\documentclass{article}

\usepackage{arxiv}
\usepackage{multirow}
\usepackage{tabularx}
\usepackage{booktabs}

\usepackage{float} 
\usepackage{adjustbox}
\usepackage{multirow, bigdelim}
\usepackage{threeparttable}
\usepackage{amsmath} 

\usepackage[table]{xcolor} 
\usepackage{colortbl} 

\usepackage[utf8]{inputenc} 
\usepackage[T1]{fontenc}    
\usepackage{hyperref}       
\usepackage{url}            
\usepackage{booktabs}       
\usepackage{amsfonts}       
\usepackage{nicefrac}       
\usepackage{microtype}      
\usepackage{lipsum}
\usepackage{graphicx}
\graphicspath{ {./images/} }

\title{Understanding the Disparities in Mathematics Performance: An Interpretability-Based Examination}

\author{
 Ismael Gomez-Talal \\
  Department of Signal Theory \\
  and Communications and Telematic \\
  Systems and Computation \\
  Madrid, Spain 28943 \\
  \texttt{ismael.gomez.talal@urjc.es} \\
   \And
 Luis Bote-Curiel \\
  Department of Signal Theory \\
  and Communications and Telematic \\
  Systems and Computation \\
  Madrid, Spain 28943 \\
  \texttt{luis.bote@urjc.es} \\
  \And
 Jose Luis Rojo-Alvarez \\
  Department of Signal Theory \\
  and Communications and Telematic \\
  Systems and Computation \\
  Madrid, Spain 28943 \\
  \texttt{joseluis.rojo@urjc.es} \\
}

\begin{document}
\maketitle
\begin{abstract}
\textit{Problem.} Educational disparities in Mathematics performance are a persistent challenge. This study aims to unravel the complex factors contributing to these disparities among students internationally, with a focus on the interpretability of the contributing factors. \textit{Methodology.} Utilizing data from the Programme for International Student Assessment (PISA), we conducted rigorous preprocessing and variable selection to prepare for applying binary classification interpretability models. These models were trained using the Stratified K-Fold technique to ensure balanced representation and assessed using six key metrics. \textit{Solution.} By applying interpretability models such as Shapley Additive Explanations (SHAP) analysis, we identified critical factors impacting student performance, including reading accessibility, critical thinking skills, gender, and geographical location. \textit{Results.} Our findings reveal significant disparities linked to resource availability, with students from lower socioeconomic backgrounds possessing fewer books and demonstrating lower performance in Mathematics. The geographical analysis highlighted regional educational disparities, with certain areas consistently underperforming in PISA assessments. Gender also emerged as a determinant, with females contributing differently to performance levels across the spectrum. \textit{Conclusion.} The study provides insights into the multifaceted determinants of student Mathematics performance and suggests potential avenues for future research to explore global interpretability models and further investigate the socioeconomic, cultural, and educational factors at play.
\end{abstract}


\section{Introduction}

One of the most critical studies of adolescent students worldwide is the Programme for International Student Assessment (PISA). This study is an international educational assessment conducted every three years on 15-year-old students, and it evaluates the degree to which they have acquired essential knowledge and skills for successful participation in society \cite{pisa2019results, pisa2018results}. The 2018 PISA assessment focuses on areas of reading, Mathematics, and Science, as well as an innovative domain and the well-being of students \cite{schleicher2019pisa}. These results are used to evaluate student performance worldwide and compare the effectiveness of education systems in participating countries. In particular, the PISA study with the Mathematics and Science results are of great importance, as these subjects are fundamental in the technological development of a country \cite{shin2021systematic, kandeel2021learners}. Within the PISA framework, Mathematics occupies a paramount position. It is a crucial indicator of students' problem-solving capabilities, logical reasoning, and quantitative aptitude. 

In the current research landscape, a palpable need exists for comprehensive exploration and analysis. Numerous studies have indeed dissected specific facets using PISA metrics, and an emerging cohort is tapping into the vast potential of Machine Learning (ML). Nevertheless, despite these advancements, there remains a glaring void. The scientific community needs a nonlinear, intuitive method that can sift through the vast array of data and discern essential variables from a broader vantage point. This method should provide deeper insights and be accessible and comprehensible for various stakeholders, from researchers to educators. As we stand on the cusp of a data-driven era, proposing and adopting such a methodology becomes not only beneficial but imperative for the progression of educational research \cite{podda2018machine, lezhnina2022combining, xiaomin2020historical}. 

Given the significance of the PISA study, our primary goal was to identify and address educational global disparities in mathematical performance among Spanish adolescent students. We used interpretability models within the Spanish educational landscape to achieve this aim. These models help us discern the impacts of key factors such as reading accessibility, critical thinking skills, and geographical contexts. Additionally, we preprocessed the PISA data for Spain strategically to focus on mathematical outcomes and to segment student performance into three distinct categories: low, medium, and high. We also evaluated various binary classification models to determine the optimal model that delineates and deciphers those interpretable variables in the study. Lastly, we investigated the factors influencing mathematical performance using advanced analytical tools, such as Shapley Additive exPlanations (SHAP) model analysis. This Interpretable ML (IML) model shows us an understanding of the relationships between these factors and students' different levels of Mathematics. Through applying interpretability models, the research sheds light on the myriad multivariate factors influencing students' Mathematics performance. Key determinants explored include access to reading materials, critical thinking skills, gender disparities, and geographic location.

In this paper, we preprocess the PISA data in Spain to understand and cover three profiles of students (low, medium, and high) based on the original levels of the official reports presented in PISA \cite{hu2023decoding, frade2021factors}. At the low proficiency level, students demonstrate a fundamental grasp of essential mathematical concepts and possess basic arithmetic skills. The medium proficiency level encompasses a more comprehensive comprehension of mathematical principles, enabling students to apply mathematical reasoning to practical situations. The high proficiency level represents the apex of mathematical prowess, reflecting an ability to engage in complex mathematical reasoning, abstract thinking, and applying advanced mathematical concepts.

The paper is organized as follows. Section II provides an overview of the related background, including the relationship between teaching quality and student achievement in Mathematics and Science, the influence of the gender gap on Mathematics achievement, and the moderating effect of parental involvement. Section III details the research methodology, outlining how predictive models of student achievement levels are developed, validated, and interpreted. This process encompasses four phases: data preprocessing, construction of the binary classification model, validation, and interpretation using SHAP analysis. Section IV then presents the complete results of the experiments conducted. It shows the performance and validation results of the ML models, clarifies the classification of significant features, and highlights the directional impact of these features on the final performance prediction. In Section V, the focus is on the discussion of the results obtained, where we elaborate on the implications of the results for students at low and high-performance levels. Section VI is the final chapter of this research. It synthesizes the study contributions, highlighting its role in advancing the understanding of educational disparities in Mathematics achievement. In addition, this section acknowledges the study limitations and offers recommendations for future research, pointing toward avenues that could broaden and enrich the scope of this investigation.

\section{Background}

This section serves as a foundation for understanding the multifaceted influences on academic achievement. Firstly, we  delve into how the quality of teaching directly correlates with student performance in Mathematics and Science. Subsequently, the spotlight shifts to the gender gap and its unique impact on Mathematics achievement, shedding light on existing disparities. Lastly, we discuss the pivotal role of parental involvement, underscoring its potential to either bolster or hinder a student's academic trajectory.

{\bf Teaching Quality Affects Mathematics/Science Achievement.} One of the studies investigated the socioeconomic status and the relationship of growth mindset to Mathematics and Science learning filtered in the country of the Philippines. The results of this study show that a growth mindset is positively related to scores in these two technical subjects. On the other hand, the student's socioeconomic background moderated this relationship, concluding that the influences of a growth mindset on Mathematics and Science learning are stronger in students with unfavorable family economic status \cite{bernardo2021socioeconomic}. This same study also found that students from economically disadvantaged backgrounds have lower levels of growth mindset compared to students from more advantaged backgrounds. It concluded that this result may be due to varying educational opportunities and resources depending on the student's economic status. 

The results of some other studies indicated that there is a complex relationship between the use of Information and Communication Technologies (ICT) and performance in Mathematics and Science. Some studies suggest that strategic use of ICT can have a positive impact on learning and performance, as these tools can provide opportunities for practice, exploration, and access to online educational resources. However, evidence was also found that excessive or inappropriate use of ICT could be associated with poor academic performance \cite{odell2020scoping}.

{\bf Gender Gap Impacts Mathematics Achievement.} 
One of the consistent findings across PISA assessments has been the gender gap in Mathematics. Traditionally, males have outperformed females in Mathematics in many participating countries. However, the magnitude of this gap varies widely among countries. In their PISA data analysis, \cite{else2010cross} examined the math scores of 15-year-old students from 40 countries. While they found that males outperformed females in most countries, they also identified several countries with no significant gender differences, even some of them where females outperformed males.

Other studies regarding Mathematics results with the 2018 PISA data have investigated whether there is a gender gap in the performance of Mathematics results globally, where they showed that the results are similar between males and females \cite{lu2023assessing}. Specifically, this study found that females outperformed in Mathematics in 11 of the 40 countries in the study. However, there is some gap in the choice of more technical careers, and they are less confident in the Science and Mathematics fields, which are known as Science, Technology, Engineering, and Mathematics (STEM) careers.

{\bf Parental Involvement Moderates Academic Achievement.} 
Parental involvement in a child's education has long been identified as a crucial factor influencing academic outcomes. While students possess innate abilities and schools offer resources and instruction, the role of parents cannot be understated. A study conducted by \cite{hill2004parent} delved into the nuances of how parental involvement impacts academic achievement. Using a diverse dataset, the researchers found that when parents are actively involved in their child's education, either by aiding with homework, attending school meetings, or engaging in educational activities at home, there was a noticeable positive impact on the student's academic performance. 
Another of the studies related to the PISA data is based on the moderated effect of parental and student involvement on their academic performance \cite{ma2022association}. The authors analyze data to determine whether students with positive teacher-student relationships tend to perform better academically and whether parental involvement influences this relationship strength.

\section{Methods}

In this section, we delineate the systematic approach undertaken in this study. Our journey begins with data preprocessing, where the PISA dataset undergoes rigorous refinement to ensure its suitability for modeling. Following this, we transition into binary classification models, setting the stage for understanding the predictive mechanisms. The training of these models is addressed with a focus on the Stratified K-Fold technique, ensuring a balanced representation of our dataset categories. The choice of the Stratified K-Fold technique in our model training process is pivotal to ensuring an equitable representation of each category within the dataset. This method is especially beneficial in handling imbalanced datasets, as it maintains the original proportion of each category across all folds, thereby facilitating a more balanced and unbiased model training and validation process. For model assessment, we employ six key metrics. The section culminates with exploring the nuances of IML, utilizing the SHAP model to demystify the relationships and impacts of individual features.

\subsection{Data Preprocesing}

\begin{figure}[t]
\centering
\includegraphics[width=0.9\columnwidth]{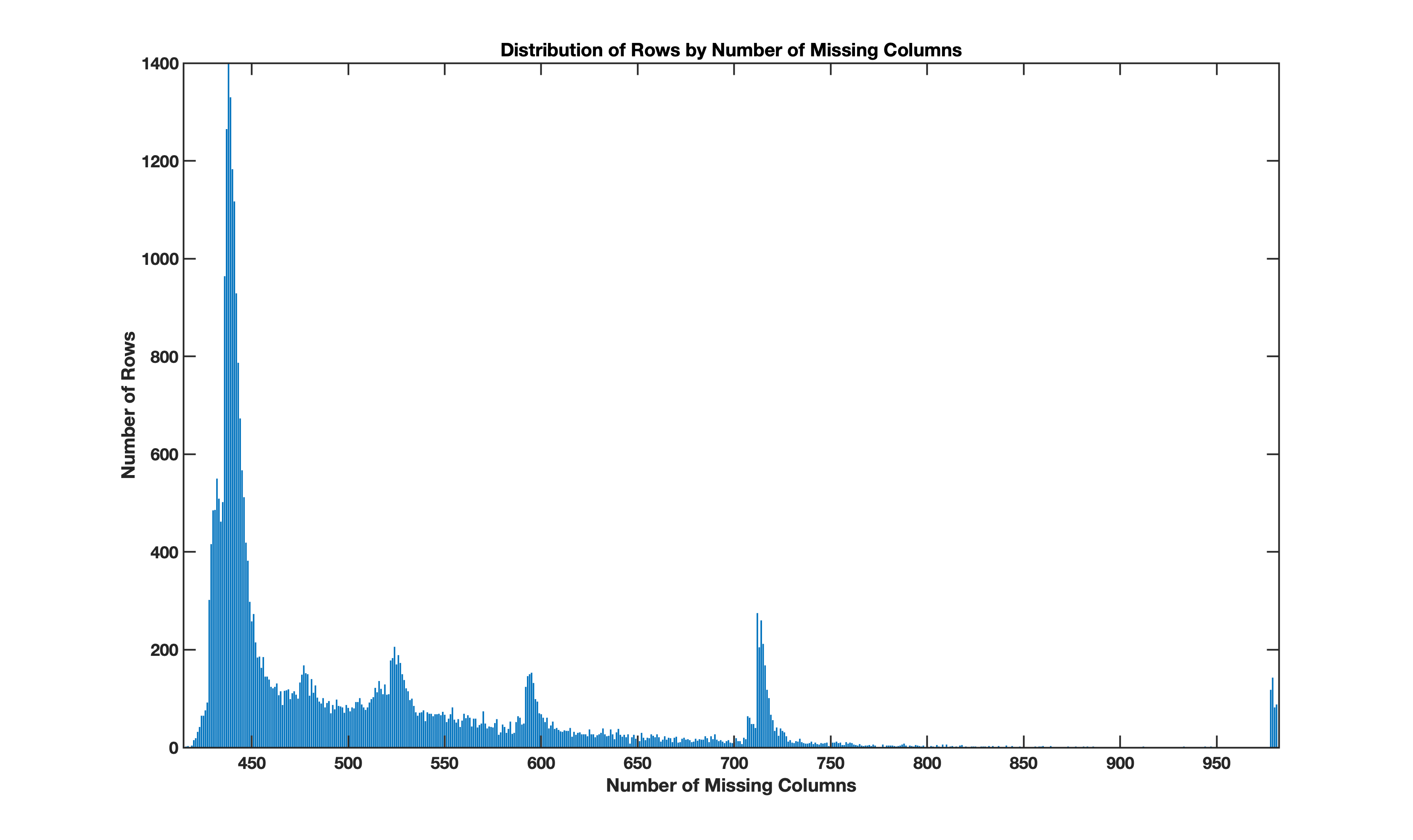} 
\caption{Distribution of rows by number of missing columns. The histogram presents the frequency of rows relative to the count of missing data points across all student records in the PISA dataset. A significant concentration of rows exhibits over 400 missing columns, which underscores the rationale for data pruning to enhance the dataset robustness for subsequent analysis.}
\label{fig_preprocessing}
\end{figure}

The dataset leveraged in this study is drawn from the most recent PISA assessment, focusing on student performance in Spain. The original dataset, sourced from the Organisation for Economic Co-operation and Development (OECD), encompasses results from 35,943 students. However, a critical step in our data preparation involved the exclusion of rows with missing values. Consequently, the analysis was conducted on a refined sample of 26,657 Spanish students, reducing the initial dataset by 9,286 entries. This elimination of incomplete records is substantiated by the distribution of the data, as depicted in Figure \ref{fig_preprocessing}. The histogram reveals a significant prevalence of rows with a high count of missing columns; notably, most of the omitted entries contained more than 400 missing values out of 1,118 total columns. Such a substantial portion of missing data could potentially skew the analysis, thereby justifying their removal to ensure the integrity and validity of the subsequent analysis. The key variables retained for examination are detailed in Table \ref{tab_variables}, where the emphasis is placed on ensuring a robust dataset conducive to a reliable investigation.

\begin{table}[t]
\centering
\caption{PISA Mathematics survey variables: descriptions and data types. This table catalogs the variables collected in the PISA, targeting the home environments and digital literacy of mathematics students. Variables range from the number of certain household items to students' responses to digital communication scenarios, each described with its respective data type. The variables include both normalized scalar values for quantifiable items and binary values for yes/no responses.}
    \begin{adjustbox}{width=0.81\textwidth, keepaspectratio}
    \begin{tabularx}{\textwidth}{@{}l>{\hsize=1.6\hsize}X>{\hsize=0.4\hsize}X@{}}
        \toprule
        \textbf{Variable} & \textbf{Description} & \textbf{Data types} \\
        \midrule
        ST012 & How many of the following items exist in your home?  &  \\
        \quad ST012Q01TA & Number of televisions & Normalized scalar \\
        \quad ST012Q02TA & Number of cars & Normalized scalar \\
        \quad ST012Q03TA & Number of rooms with bathroom & Normalized scalar \\
        \quad ST012Q05TA & Number of smartphones & Normalized scalar \\
        \quad ST012Q06TA & Number of computers & Normalized scalar \\
        \quad ST012Q07TA & Number of tablets & Normalized scalar \\
        \quad ST012Q08TA & Number of e-books & Normalized scalar \\
        \quad ST012Q09TA & Number of musical instruments & Normalized scalar \\
        
        \midrule
        \multirow{2}{*}{ST013} & How many books are in your house? Answer with 0-10, 11-25,
                                26-100, 101-200, 201-500, and more than 500 books & Normalized scalar \\
        \midrule

        ST166 & If you receive an email from a phone company claiming you have won  &  \\
        \quad ST166Q01HA & Respond to the email and ask for more information about the smartphone & Normalized scalar \\
        \quad ST166Q02HA & Verify the email address of the email & Normalized scalar \\
        \quad ST166Q03HA & Click on the form link as quickly as possible & Normalized scalar \\
        \quad ST166Q04HA & Delete the email message without clicking on the form link & Normalized scalar \\
        \quad ST166Q05HA & Check the mobile operator's website to see if it is true & Normalized scalar \\
        
        \midrule
        STRATUM & Columns representing the autonomous communities of each student & Binary value \\
        \midrule
        ST004D01T & Sex of student, answer with male and female & Binary value \\
        \midrule
        ST019AQ01T & In which country was the student born? & Binary value \\
        \midrule
        ST019BQ01T & In which country was the student's mother born?  & Binary value \\
        \midrule
        ST019CQ01T & In which country was the student's father born? & Binary value \\
        \midrule
        ST011 & Do any of the following items exist in your house? Answer yes or no &  \\
        \quad ST011Q01TA & A desk for studying & Binary value \\
        \quad ST011Q02TA & A room of your own & Binary value \\
        \quad ST011Q03TA & A quiet place to study & Binary value \\
        \quad ST011Q04TA & If you have a computer for studying & Binary value \\
        \quad ST011Q05TA & educational software & Binary value \\
        \quad ST011Q06TA & If you have internet & Binary value \\
        \quad ST011Q07TA & Classic literature at home & Binary value \\
        \quad ST011Q08TA & Poetry books & Binary value \\
        \quad ST011Q09TA & Art paintings & Binary value \\
        \quad ST011Q10TA & Books that can help you with your homework & Binary value \\
        \quad ST011Q11TA & Technical books & Binary value \\
        \quad ST011Q12TA & A dictionary & Binary value \\
        \quad ST011Q16TA & Books on art, music or design & Binary value \\
        \quad ST011Q17TA & A video camera & Binary value \\
        \quad ST011Q18TA & A tablet & Binary value \\
        \quad ST011Q19TA & Pay television & Binary value \\
        \midrule
        ST158 & Have you been taught the following tools at school? Answer yes or no &  \\
        \quad ST158Q01HA & Have you been taught keywords to be able to use search engines? & Binary value \\
        \quad ST158Q02HA & How to know if the information on the internet is reliable? & Binary value \\
        \quad ST158Q03HA & How to cross-check sources for task verification? & Binary value \\
        \quad ST158Q04HA & Do you grasp the risks of posting on socials? & Binary value \\
        \quad ST158Q05HA & How to use the brief description that appears in the search engine results? & Binary value \\
        \quad ST158Q06HA & Do you know how to detect if information is subjective or reliable? & Binary value \\
        \quad ST158Q07HA & Do you know how to detect phishing messages or spam messages? & Binary value \\
        \midrule
        Mathematics scores & Mathematics score prediction label with 3 categories (low, medium, and high) & Categorical \\
        \bottomrule
    \end{tabularx}
    \end{adjustbox}
    \label{tab_variables}
\end{table}

In the case of the score variables, preprocessing has been done to transform them into three recognized labels that define three categories (high, medium, and low scores). For this, the first step is to obtain the mean of the scores of the ten plausible values of the scores of all students. The cut-off criterion for the different levels is the one established by the OECD reports to homogenize the levels with the rest of the countries and to be able to make an equitable comparison \cite{OECD2019}. The levels with each cut-off mentioned in the OECD reports are shown in Table \ref{tab_levels}, which also shows the category considerations of student score levels with labels of low, medium, and high values. This categorization criterion to evaluate each score by reducing it to three categories allows the different types of students, according to the Mathematics results, to discern those of high/low performance and those in the average (medium). As an additional preprocessing tool, a control variable has been included to evaluate the SHAP interpretability analysis subsequently. In this case, a Gaussian noise variable has been generated pseudo-randomly. The Gaussian distribution criterion follows a mean of zero and a standard deviation of one. 

In further elaboration of our methodological approach, it is pertinent to discuss the strategic rationale behind adopting a three-category classification system, despite initially categorizing the samples into seven levels. This decision was primarily driven by the need to enhance interpretability and streamline the analysis process in dealing with complex educational data sets. Reducing the number of categories for evaluative purposes aligns with a widely accepted practice in educational research, as exemplified by the works of \cite{hu2023decoding} and by \cite{gorostiaga2016use}. These studies illustrate how simplification aids in isolating key findings and trends without compromising the depth of analysis and demonstrate the effectiveness of categorizing complex data into fewer, more manageable groups. By focusing on broader categories, our study aimed to distill the most relevant insights, ensuring that our findings are not only robust but also directly applicable and easily comprehensible in the broader context of educational research. This strategic simplification facilitates a clearer understanding of the underlying patterns and drivers in educational performance and outcomes.

In the preprocessing of our dataset, categorical variables of a binary nature were transformed using the one-hot encoding method. This approach allows us to convert categorical data into a numerical format by creating binary columns for each category, which can be processed by ML algorithms effectively. One-hot encoding eliminates the imposition of a potentially misleading ordinal relationship among categories, thereby providing a more nuanced and accurate representation of our dataset. It is important to note that while one-hot encoding was applied to most categorical variables within our dataset, the labels presented in Table \ref{tab_variables} were exempted from this process. The decision to retain the original categorical format for the labels was made to facilitate a clearer interpretation of the outcomes of our analysis. As our labels represent discrete, non-ordinal entities, one-hot encoding for these particular variables was deemed unnecessary and potentially obfuscating, given the context of our research goals.

\begin{table}[t]
\centering
\caption{Detailed distribution of Mathematical skill levels: This table presents the distribution of mathematical skill levels among the sample population. It includes the score ranges, corresponding skill levels, the number of samples in each category, and a qualitative categorization of skill levels. This breakdown provides insights into the proficiency distribution and the prevalence of various skill levels within the dataset.}
    \begin{tabular}{@{}llll@{}}
        \toprule
        \textbf{Score Ranges} & \textbf{Levels} & \textbf{Number of Samples} & \textbf{Categories} \\
        \midrule
        \text{score} $\leq 358$ & Below Level 1 & 967 & Low \\
        $358 < \text{score} \leq 420$ & Level 1 & 3080 & Low \\
        $420 < \text{score} \leq 482$ & Level 2 & 6196 & Low \\
        $482 < \text{score} \leq 545$ & Level 3 & 8341 & Medium \\
        $545 < \text{score} \leq 607$ & Level 4 & 6143 & Medium \\
        $607 < \text{score} \leq 669$ & Level 5 & 1795 & High \\
        \text{score} $> 669$ & Level 6 & 135 & High \\
        \bottomrule
    \end{tabular}
    \label{tab_levels}
\end{table}

\subsection{Binary Classification Models}

We tested eight binary classification models across various combinations of three distinct math score levels: low, medium, and high. Each model was trained using the Stratified K-fold method, ensuring data balance within each fold through the application of an undersampling technique. For optimization, every model employed differing hyperparameters, followed by the introduction of a grid search to pinpoint the model that delivered the highest performance. The evaluated models were the following ones.

{\it Logistic Regression (LR)}. This is a linear model utilized for binary classification. The tested hyperparameters included regularization (penalty) with ``L1'' (LASSO) and ``L2'' (Ridge) options and the regularization values ``C'' with selections of 0.1, 1, and 10 \cite{libreria_skl}.

{\it Decision Trees.} This is a nonlinear classification model that segregates the feature space into regions by employing a set of decision rules. The assessed hyperparameters included the splitting criterion with ``gini'' and ``entropy'' options, along with the maximum tree depth, which had options of None, 5, and 10. The ``gini'' index and the hyperparameter entropy are measures of impurity that are used to evaluate how mixed the samples of different classes are in a node. The main difference lies in the equation where for the ``gini'' index it is calculated as  \begin{math}Gini =1-\Sigma\left(p_i\right)^2\end{math} while for the entropy equation, it is calculated as  \begin{math} Entropy=-\Sigma\left(p_i * \log_ 2\left(p_i\right)\right)\end{math}, where $p_i$ is the probability that a sample belongs to class $i$ \cite{libreria_skl, yerpude2020predictive}.

{\it Random Forest.} This is an ensemble model that relies on a combination of multiple decision trees. The evaluated hyperparameters were the number of estimators named in the code as (``n\_estimators''), with choices of 50, 100, and 200, along with the maximum depth of the trees, which included options of None, 5, and 10 \cite{libreria_skl}.

{\it Gradient Boosting (GB).} This ensemble technique amalgamates multiple weak models to create a more robust model. The hyperparameters evaluated included the number of estimators (``n\_estimators'') with options of 50, 100, and 200, alongside the learning rate (``learning\_rate'') with alternatives of 0.1, 0.01, and 0.001 \cite{libreria_skl}.

{\it Support Vector Machine (SVM).} This is a linear classification model designed to identify the optimal hyperplane that separates samples of different classes. The hyperparameters evaluated were the kernel, with options of ``linear'' and ``RBF'', and the regularization parameter ``C'', which included alternatives of 0.1, 1, and 10. The linear kernel (``linear'') is employed when the data can be separated linearly, while the Radial Basis Function (RBF) kernel is used in SVM for problems involving non-linearly separable classification \cite{libreria_skl, buitinck2013api}.

{\it XGBoost.} This model employs a BG technique, with decision trees serving as the base models. The evaluated hyperparameters included the number of estimators (``n\_estimators'') with options of 50, 100, and 200, and the learning rate (learning\_rate) with alternatives of 0.1, 0.01, and 0.001 \cite{chen2016xgboost}.

{\it LightGBM.} This model represents another GB technique, utilizing decision trees as base models. The hyperparameters assessed included the number of estimators (``n\_estimators'') with choices of 50, 100, and 200, along with the learning rate (learning\_rate) with alternatives of 0.1, 0.01, and 0.001 \cite{ke2017lightgbm}.

{\it Multilayer Perceptron (MLP).} This is a feedforward neural network comprising multiple hidden layers. The evaluated hyperparameters included the size of the hidden layers, with options of (100,), (50, 50), and (100, 50, 25), and the activation function, with choices of ``relu'', ``tanh'', and ``logistic'' \cite{libreria_skl}.

\subsection{Training Binary Models via Stratified K-Fold} 

The Stratified K-Fold cross-validation technique is a pivotal methodology, particularly useful when dealing with imbalanced datasets. This method ensures a homogeneous class distribution across each fold, providing an equitable and consistent evaluation of a model performance \cite{prusty2022skcv, wong2017dependency}.

Initially, the training data (which constitute 80\% of the whole set) are sorted according to the target class and divided into five different stratified folds (since, in this problem, we have defined a k=5). Each fold contains the same proportion of target classes as the original set. At each iteration, one fold is assigned as the validation set, while the remaining folds serve as the training set. The model is then fitted to the training set and evaluated concerning the validation set \cite{zeng2000distribution}.

This process is repeated five times, treating each fold as a validation set once. It is crucial to note that the models are fitted from scratch at each iteration to avoid any influence from previously fitted models. Finally, the performance measures for each fold are averaged to obtain a robust and accurate assessment of model effectiveness \cite{purushotham2011evaluation}.

\subsection{Evaluation Models}

The metrics used in this study in each of the models are the six fundamental metrics: accuracy (ACC), recall (RC), F1 score (F1S), Precision (PR), Specificity (SP), and area under the curve (AUC) \cite{bishop2006pattern}. In the case of the first metric, it provides an overall measure of model performance by indicating the proportion of correct predictions concerning the total number of predictions made. The ACC is calculated as 
\begin{equation}
ACC =\frac{T P+T N}{T P+T N+F P+F N},
\end{equation}
where the TP value represents the true positive rate, the TN value represents the true negative rate, the FP value represents the false positive rate and the FN value represents the false negative rate.

The RC value represents the model ability to correctly identify all positive cases, being especially useful in situations where FN are more problematic than FP. Its equation 
\begin{equation}
RC =\frac{T P}{T P+F N}.
\end{equation}

The PR measures the proportion of correctly classified positive samples (true positives) among all samples classified as positive ($TP + FP$), defined as
\begin{equation}
 PR =\frac{T P}{T P+F P}.
\end{equation}

The F1-Score metric corresponds as a balanced measure between ACC and completeness and is calculated as follows,
\begin{equation}
F1 =\frac{2 \times( PR \times RC)}{PR + RC}.
\end{equation}

On the other hand, SP measures the proportion of correctly classified negative samples (true negatives) among all samples classified as negative (true negatives + false positives) as defined by the following equation
\begin{equation}
SP =\frac{T N}{T N+F P}.
\end{equation}

Finally, the AUC, which represents the area under the receiver operating characteristic (ROC) curve, is used to evaluate the totality of two parameters: the true positive rate and the false positive rate. Where the equation is defined as the integral that calculates the area as follows
\begin{equation}
AUC=\int_0^1 \mathrm{ROC}(\mathrm{t}) \mathrm{dt},
\end{equation}
where an AUC of 1 denotes a perfect classification model, while a value of 0.5 suggests a performance no better than random classification \cite{powers2020evaluation, fawcett2006introduction}.

\subsection{Interpretability in Machine Learning}

Shapley values have their origin in game theory and measure the average marginal contribution of a player in a cooperative game \cite{roth1988shapley}. In the context of ML, this player is interpreted as a characteristic or an attribute, and the cooperative game becomes the prediction task performed by the model. Shapley values, therefore, can explain the contribution of an attribute to the prediction of an ML mode \cite{merrick2020explanation}. The Shapley value of a feature for a particular query point, therefore, explains the contribution of that feature to the model prediction (the response for regression or the score of each class for classification) at the specified query point. The Shapley value corresponds to the deviation of the prediction for the query point from the average prediction due to the feature. For each query point, the sum of the Shapley values for all features corresponds to the total deviation of the prediction from the average. Mathematically, the Shapley value of the i-th feature for query point x is defined by the value function $v_x$ as
\begin{equation}
\varphi_i\left(v_x\right)=\frac{1}{M} \sum_{S \subseteq M_s\{i\}} \frac{v_x(S \cup\{i\})-v_x(S)}{(M-1) !},
\end{equation}
where $M$ is the total number of features, \begin{math}M_s\end{math} is the set of all features, $|S|$ is the cardinality of the set $S$, i.e., the number of elements in the set $S$ and $v_x(S)$ is the value function of the features in a set $S$ for query point $x$ (indicates the expected contribution of the features in $S$ to the prediction for query point $x$).

Several algorithms are provided to compute these values within the SHAP package in Python. These algorithms, often called {\it interventional algorithms}, include variants such as Kernel SHAP, Linear SHAP, and Tree SHAP.


{\bf SHAP Intervention Algorithms.} Intervention Algorithms define the value function for a set of attributes at a query point as the expected prediction for the intervention distribution, which is the joint distribution of the attributes in the complement of the set
\begin{equation}
v_x(S)=E_D[f(x_S, X_{S^c})],
\end{equation}
where $x_S$ is the query point value for the features in $S$, and $X_{S^c}$ are the features in $S^c$. The intervention algorithm evaluates the value function $v_x(S)$ at query point $x$, under the assumption that the features are not highly correlated, using the values in the data $X$ as samples from the intervention distribution $D$ for the features in $S^c$
\begin{equation}
v_x(S)=E_D\left[f\left(x_S, X_{S^c}\right)\right] \approx \frac{1}{N} \sum_{j=1}^N f\left(x_S,\left(X_{S^c}\right){ }_j\right),
\end{equation}
where $N$ is the number of observations, and $\left(X_{S^c}\right){ }_j$ contains the values of the features in $S^c$ for the $j$-th observation.

The main advantage of interventional algorithms is that they are computationally less expensive, although, in contrast, they require the assumption of feature independence and use samples outside the distribution, which may result in unrealistic observations \cite{kumar2020problems}. The main SHAP intervention algorithms include the following ones.

{\it Kernel SHAP} is a version of the Shapley algorithm that uses a kernel approximation to estimate the Shapley values. This method is particularly useful when faced with a high-dimensional feature space but can be computationally intensive \cite{lundberg2017unified}.

{\it Linear SHAP} is used when we have a linear model, and it calculates the Shapley values analytically rather than approximately. This makes Linear SHAP computationally more efficient than Kernel SHAP \cite{lundberg2017unified}.

{\it Tree SHAP} is a variant of the Shapley algorithm designed specifically for tree-based models, such as decision trees, random forests, and GB algorithms. Tree SHAP is particularly computationally efficient and can handle feature interactions explicitly \cite{lundberg2020local}.

\subsection{Research Workflow}

\begin{figure}[t]
\centering
\includegraphics[width=0.7\columnwidth]{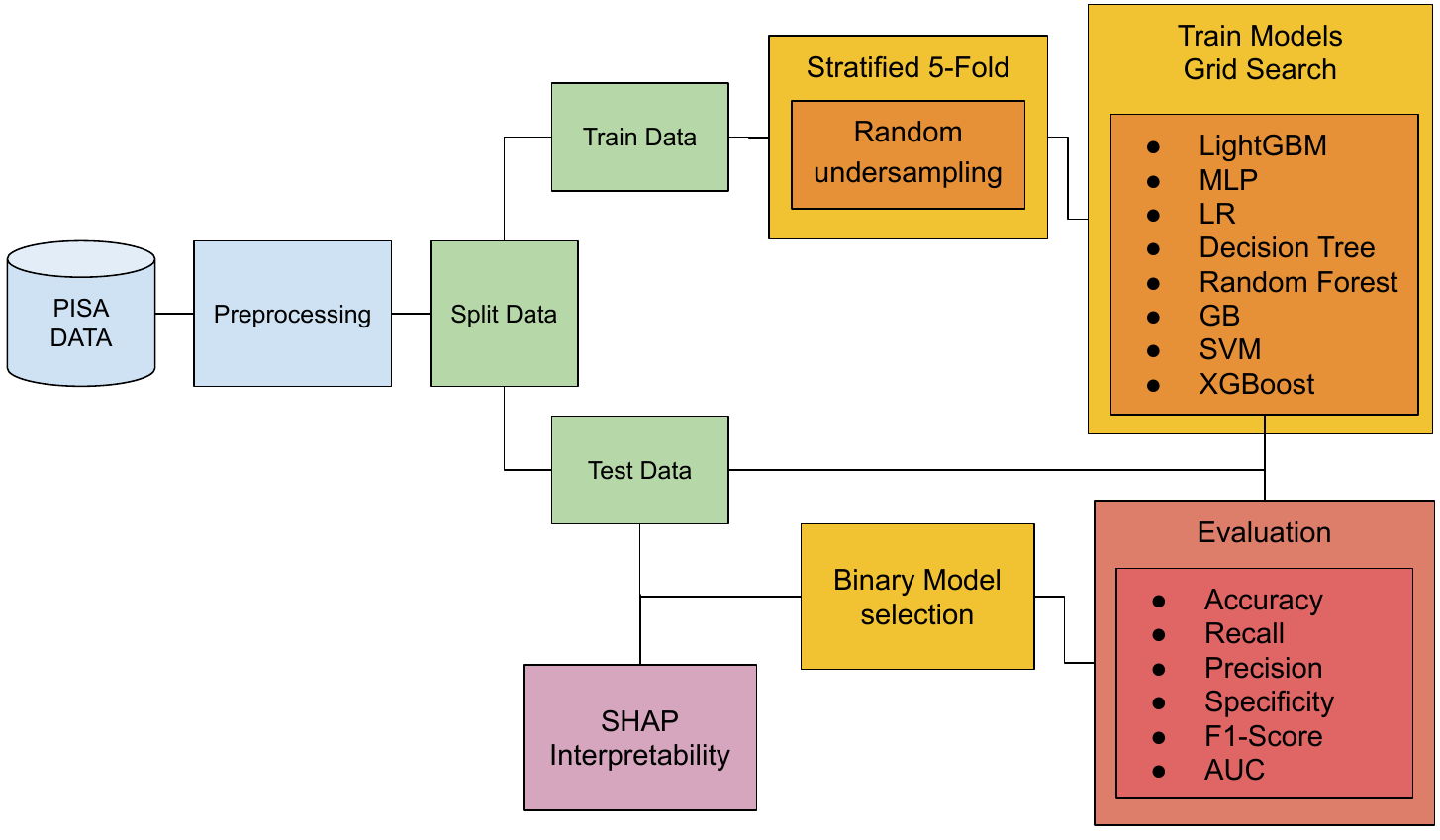} 
\caption{Comprehensive Research Workflow utilizing the PISA dataset. Starting with (1) preprocessing, (2) data splitting into training and test sets, (3) addressing class imbalance with stratified 5-fold cross-validation and undersampling, (4) training various models with grid search optimization, (5) evaluating with metrics including ACC and AUC, and (6) applying SHAP for interpretability.}
\label{fig_framework}
\end{figure}

The comprehensive analysis of the PISA dataset, as illustrated in Figure \ref{fig_framework}, commences with a meticulous preprocessing phase. This initial stage is instrumental in ensuring data quality by applying various cleaning procedures, addressing missing values, and conducting feature engineering to enhance the forthcoming modeling steps. Following preprocessing, the dataset is systematically divided into a training set and a testing set. The training set is crucial for the development and calibration of the predictive models, while the testing set is reserved for an unbiased assessment of the models' predictive prowess. This bifurcation is indispensable to ascertain the generalizability of the models beyond the observed data.

To address potential class imbalance inherent in the PISA dataset, the study incorporates stratified 5-fold cross-validation in conjunction with random undersampling techniques during model training. This stratification ensures that each cross-validation fold retains a proportionate representation of the class distributions, thereby promoting equity in model training and preventing the overrepresentation of majority classes from skewing the results. In the quest for the optimal predictive model, a suite of advanced ML algorithms is employed. These include but are not limited to, GB machines like LightGBM, MLP, logistic regression, decision trees, random forests, GB, SVM, and XGBoost. A meticulous grid search is undertaken to fine-tune the hyperparameters, thereby facilitating the selection of a binary model that excels in classification ACC. The rigor of the modeling process culminates in a thorough evaluation using a diverse set of metrics: ACC, RC, PR, SP, F1S, and AUC of the ROC curve are computed to provide a holistic view of model performance. These metrics collectively offer a nuanced understanding of the model abilities to predict accurately while balancing the trade-offs between various types of prediction errors. Enhancing the robustness of the analysis, SHAP interpretability is harnessed to demystify the model decision-making process. This interpretability framework elucidates the contribution of each feature to the predictive outcomes, fostering a deeper understanding and ensuring that the model decisions are transparent and justifiable.

\section{Experiments and Results}

In this section, we present three experiments based on binary comparisons of the different levels of Mathematics scores: (1) the comparison of the low level concerning the medium level, (2) the study of the high level and the medium level, and (3) the comparison of the low level and the high level. In this section, we compare the eight models to obtain the best model in each study and then gut the knowledge of the best model in each case to learn which variables contribute most to the prediction using the SHAP model.

\subsection{Comparing Models Across Different Levels}

This Subsection compares the metrics of the eight models used in this study that have been trained using the Stratified 5-Fold method. The trained data corresponds to 80\% of the student data set of the pairs of levels (Low-Medium, High-Medium, and Low-High). The rest of the data (20\%) corresponds to the test data balanced by the undersampling method to obtain the metrics of the eight models, compare them homogeneously, and ensure no class imbalance.

{\bf Low and Medium Level Study.} In this experiment, filtering the data with the students who have obtained a grade in Mathematics with a low and medium level have been introduced in the training of the Stratified 5-Fold model (with 80 \% of the data set), and the metrics have been obtained with 20 \% of test data and that the model has not employed in the training. The compilation of the metrics is shown in the Table \ref{tab_model_hyperparameters_low_medium}.

The Table \ref{tab_model_hyperparameters_high_medium} meticulously delineates the optimal hyperparameters for various ML models across. This framework is instrumental in guiding the selection of the most effective model and hyperparameter combinations for each specific scenario. In the current context, focused on the lo-medium performance, a comprehensive analysis of the model performance is conducted using a range of metrics, including ACC, RC, PR, SP, F1S, and AUC. Each model efficacy is evaluated to ascertain which configurations yield the highest performance in classifying student achievement levels.

Particularly noteworthy in this evaluation is the SVM model. In the Low-Medium comparison, the SVM, especially with the RBF kernel and a regularization parameter (C) set to 1, emerges as the leading model (which you can see in Table \ref{tab_all_hyperparameters}). It distinguishes itself not only by tying in the number of top metrics but also by achieving the highest AUC score, a crucial indicator of the model ability to balance true positive rate and false positive rate effectively. This superior AUC score underscores the SVM robustness in accurately predicting student performance levels, making it an optimal choice for this specific study case. In addition to SVM, Table \ref{tab_all_model_hyperparameters} also highlights the strengths of other models like GB, which excels in different scenarios with its top-tier performance across various metrics. For instance, in the High-Medium and Low-High scenarios, GB demonstrates its prowess by leading in multiple metrics, offering a comprehensive view of its effectiveness in differentiating student performance levels.

{\bf High and Medium Level Study.} In this experiment, the data used in the training and test are the sets of students at medium and high levels in Mathematics. The Table \ref{tab_model_hyperparameters_high_medium} serves as a crucial reference in this analysis, presenting a detailed overview of performance metrics across different models. Among these, the GB model (referenced in Table \ref{tab_all_model_hyperparameters}) stands out for its superior performance in the High-Medium case. This model, particularly with 50 estimators and a learning rate 0.01, excels in three key metrics: ACC, PR, and SP. These findings suggest that GB is adept at distinguishing between higher and medium levels of student achievement.

\begin{table}[t]
\centering
\caption{Model hyperparameters and optimal values for different experiments. This table displays the best-performing hyperparameters for all models across three experimental scenarios: Low-Medium, High-Medium, and Low-High. Each model key hyperparameters are listed with their optimal values, offering a quick reference for model tuning in diverse comparative studies.}
\label{tab_all_hyperparameters}
\begin{tabular}{@{}lllll@{}}
\toprule
\textbf{Model} & \textbf{Hyperparameter} & \textbf{Low-Medium} & \textbf{High-Medium} & \textbf{Low-High} \\
\midrule
\multirow{2}{*}{Logistic Regression} & Penalty: L1, L2 & L1 & L1 & L1 \\
                                      & C: 0.1, 1, 10 & 10 & 10 & 10 \\
\midrule
\multirow{2}{*}{Decision Tree} & Criterion: gini, entropy & gini & entropy & entropy \\
                                & Max Depth: None, 5, 10 & 5 & 5 & 5 \\
\midrule
\multirow{2}{*}{Random Forest} & N. Estimators: 50, 100, 200 & 200 & 200 & 100 \\
                                & Max Depth: None, 5, 10 & 10 & 10 & None \\
\midrule
\multirow{2}{*}{Gradient Boosting} & N. Estimators: 50, 100, 200 & 200 & 50 & 200 \\
                                    & Learning Rate: 0.1, 0.01, 0.001 & 0.1 & 0.01 & 0.1 \\
\midrule
\multirow{2}{*}{SVM} & Kernel: linear, RBF & RBF & RBF & RBF \\
                      & C: 0.1, 1, 10 & 1 & 10 & 10 \\
\midrule
\multirow{2}{*}{XGBoost} & N. Estimators: 50, 100, 200 & 50 & 50 & 200 \\
                          & Learning Rate: 0.1, 0.01, 0.001 & 0.1 & 0.1 & 0.1 \\
\midrule
\multirow{2}{*}{MLP} & Hidden Layer Sizes: (100), (50, 50), (100, 50, 25) & (100) & (100) & (100) \\
                      & Activation: relu, tanh, logistic & tanh & tanh & tanh \\
\midrule
\multirow{2}{*}{LightGBM} & N. Estimators: 50, 100, 200 & 50 & 200 & 50 \\
                           & Learning Rate: 0.1, 0.01, 0.001 & 0.1 & 0.01 & 0.1 \\
\bottomrule
\end{tabular}
\end{table}

{\bf Low- and High-Level Study.} In this third experiment, the collection of model metrics is performed, but in this case, trained data covering students with a low and high level in Mathematics. Table \ref{tab_model_hyperparameters_low_high} compiles the metrics where it is highlighted that all the methods present outstanding performances to be able to classify both classes, which will allow us to understand (through the interpretability of the model) the differentiation and the keys of the students with a high and low level in Mathematics. As delineated in Table \ref{tab_model_hyperparameters_low_high}, an array of performance metrics is employed to thoroughly assess each model capability. The GB model, specifically with 200 estimators and a learning rate of 0.1, emerges as the standout performer in this scenario (referenced in Table \ref{tab_all_model_hyperparameters}). This configuration excels across four metrics (ACC, PR, SP, and F1S), indicating its superior ability to effectively differentiate between low and high student performance levels.

This comprehensive evaluation process, as presented in the Low-High study case, is instrumental in guiding educators and data analysts toward the most appropriate models and hyperparameters. By leveraging these insights, educational institutions can adopt a more nuanced and effective approach to student performance analysis, thereby enhancing the overall quality of educational assessment strategies.

\begin{table}[t]
  \centering
  \caption{Consolidated performance metrics of various ML models across three study cases: Low-Medium, High-Medium, and Low-High. The metrics provide insight into the models' abilities to classify between different levels of student performance.}
  \label{tab_all_model_hyperparameters}
  \begin{adjustbox}{width=1.0\textwidth, keepaspectratio}
  \begin{tabular}{lcccccc|cccccc|cccccc}
    \toprule
    & \multicolumn{6}{c}{\textbf{Low-Medium (I)}} & \multicolumn{6}{c}{\textbf{High-Medium (II)}} & \multicolumn{6}{c}{\textbf{Low-High (III)}} \\
    \cmidrule(lr){2-7} \cmidrule(lr){8-13} \cmidrule(lr){14-19}
    \textbf{Model} & \textbf{ACC} & \textbf{RC} & \textbf{PR} & \textbf{SP} & \textbf{F1S} & \textbf{AUC} & \textbf{ACC} & \textbf{RC} & \textbf{PR} & \textbf{SP} & \textbf{F1S} & \textbf{AUC} & \textbf{ACC} & \textbf{RC} & \textbf{PR} & \textbf{SP} & \textbf{F1S} & \textbf{AUC} \\
    \midrule
    LR & .7211 & .7238 & .7836 & .7174 & .7525 & .7940 & .6874 & .6632 & .2222 & .6907 & \bf{.3329} & .7340 & .8546 & .8756 & .6563 & .5248 & .8507 & .9335 \\
    Dec. Tree & .6767 & .6123 & .7884 & \bf{.7676} & .6893 & .7576 & .6746 & .6243 & .2070 & .6813 & .3109 & .6868 & .8464 & .7979 & .6222 & .5099 & .8555 & .8824 \\
    Rand. Forest  & .7126 & .7169 & .7755 & .7066 & .7451 & .7886 & .6783 & .6761 & .2189 & .6786 & .3307 & .7267 &  .8604 & .8653 & .6627 & .5370 & .8594 & .9264 \\
    \textbf{GB} (II, III)\textsuperscript{**} &  .7274 & .7248 & .7921 & .7310 & .7570 & \bf{.8023} &  \bf{.7069} & .6010 & \bf{.2230} & \bf{.7210} & .3253 & .7112  &  \bf{.8645} & .8627 & \bf{.6687} & \bf{.5459} & \bf{.8648} & .9340  \\
    \textbf{SVM} (I)\textsuperscript{*} & \bf{.7282} & .7259 & \bf{.7926} & .7315 & .7578 & .7998 &  .6859 & .6528 & .2193 & .6903 & .3283 & .7286 & .8628 & .8705 & .6680 & .5419 & .8614 & \bf{.9371} \\
    XGBoost & .7234 & .7252 & .7860 & .7208 & .7543 & .7961 & .6853 & .6528 & .2189 & .6896 & .3279 & .7230 & .8542 & .8601 & .6516 & .5245 & .8531 & .9287 \\
    MLP & .7234 & \bf{.7528} & .7698 & .6817 & \bf{.7612} & .7920 &  .6554 & \bf{.7227} & .2141 & .6465 & .3303 & \bf{.7368} & .8567 & \bf{.8834} & .6615 & .5287 & .8516 & \bf{.9371} \\
    LightGBM & .7209 & .7152 & .7887 & .7291 & .7501 & .7965 &  .6819 & .6528 & .2168 & .6858 & .3255 & .7213 & .8517 & .8472 & .6443 & .5199 & .8526 & .9299 \\
    \bottomrule
  \end{tabular}
  \end{adjustbox}
\begin{tablenotes}
\footnotesize
\item \textsuperscript{*}Note: SVM is particularly effective in the Low-Medium comparison of student performance, being tied in the number of top metrics but leading with a superior AUC score.
 \item \textsuperscript{**}Note: GB excels in the High-Medium case with leading performance in three metrics (ACC, PR and SP) and outperforms in the Low-High scenario across all four metrics (ACC, RC, PR and SP).
\end{tablenotes}
\end{table}

\subsection{Research on Model Interpretability}

In this subsection, we go deeper into the interpretability of the best-performing models (depending on the case study of students' levels) to understand the variables that most affect or contribute more significantly to predicting the different levels in Mathematics. As in the previous subsection, this subsection is divided into three studies (Low-Medium, High-Medium, and Low-High), selecting the best model in each case and exposing the best model in each case.

The SHAP values allow us to obtain which variables provide the highest to the lowest contribution in the predictions of 20 \% of the test data. However, these test values do not detail which variables affect each level. Therefore, in this study, we propose to be able to select students at the extremes of each of the levels and analyze locally and individually how the variables that affect these extreme predictions affect them. Table \ref{table_value_shap} shows the compilation of the indices together with the sum of the SHAP contributions of each variable ordered from highest to lowest SHAP value together with the student's index to locate the student in the data matrix and to be able to select him/her in the SHAP interpretability study. 

{\bf SHAP Study of Low-Medium Level Students.} In this experiment, we analyze the extreme students corresponding to the first experiment (choosing a student with a low and a medium level). In Table \ref{table_value_shap} (left), we have a student with the lowest SHAP contribution, corresponding to student 2537, and student 3629 with the highest SHAP contribution (corresponding to a medium level).

\begin{figure}[t]
\centering
\begin{tabular}{cc}
\includegraphics[width=0.45\columnwidth]{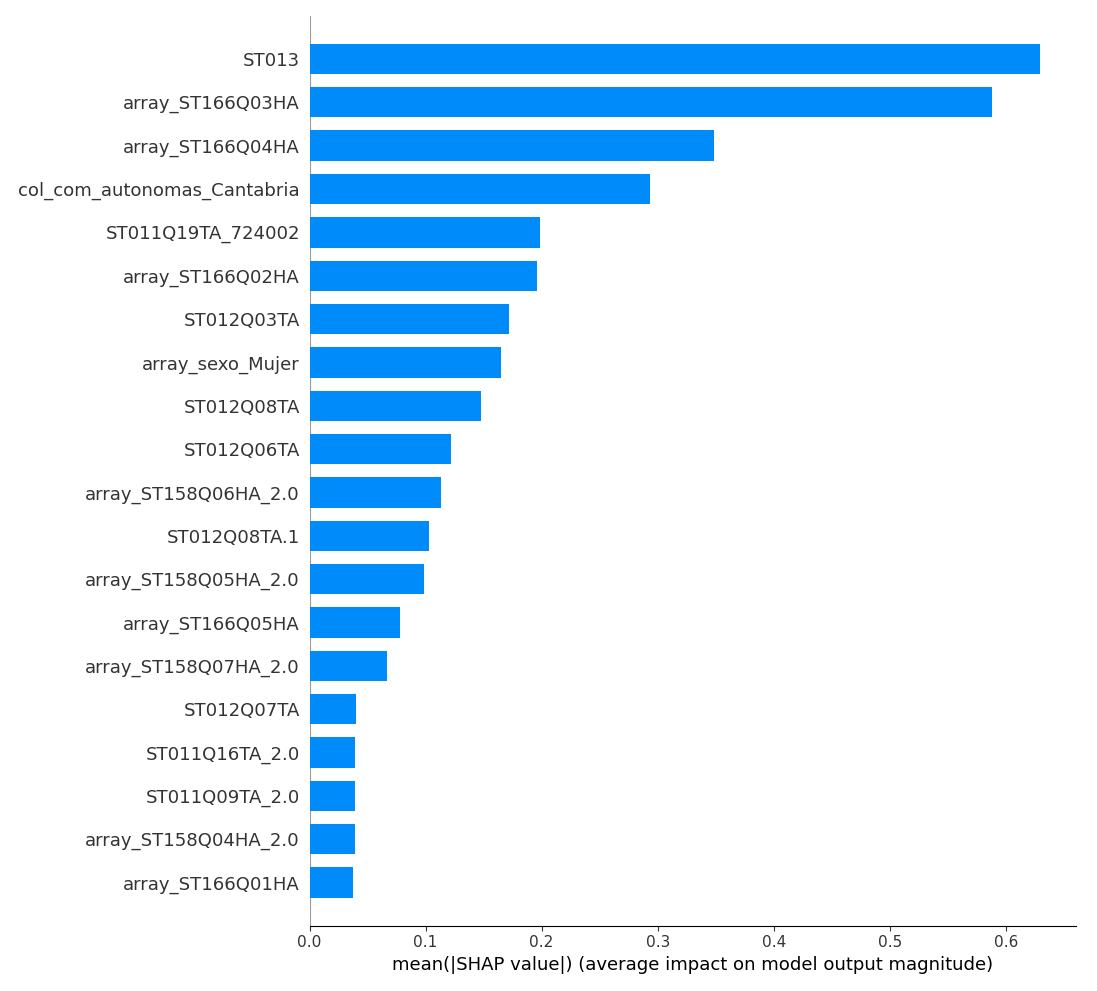} & 
\includegraphics[width=0.45\columnwidth]{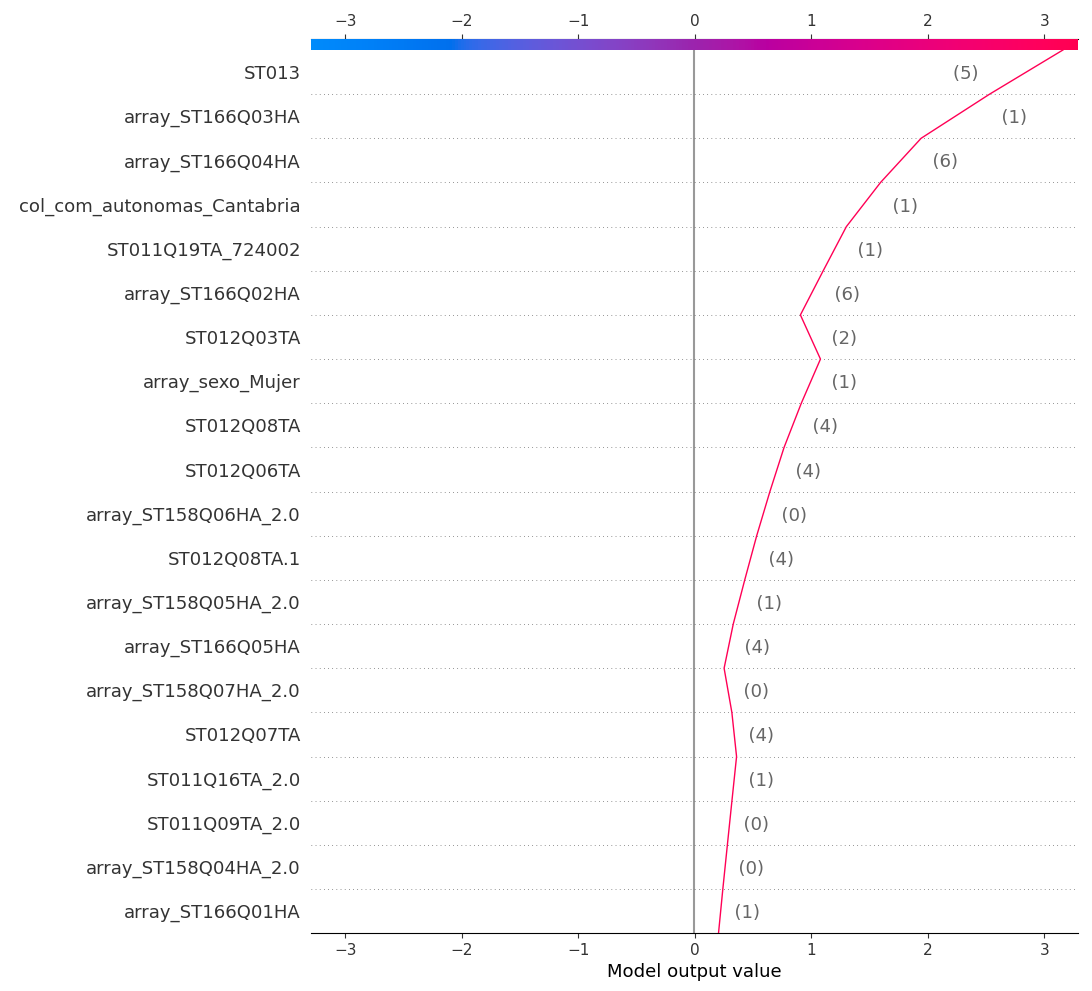} \\ 
 (a) & (b) \\
\includegraphics[width = 0.45\columnwidth]{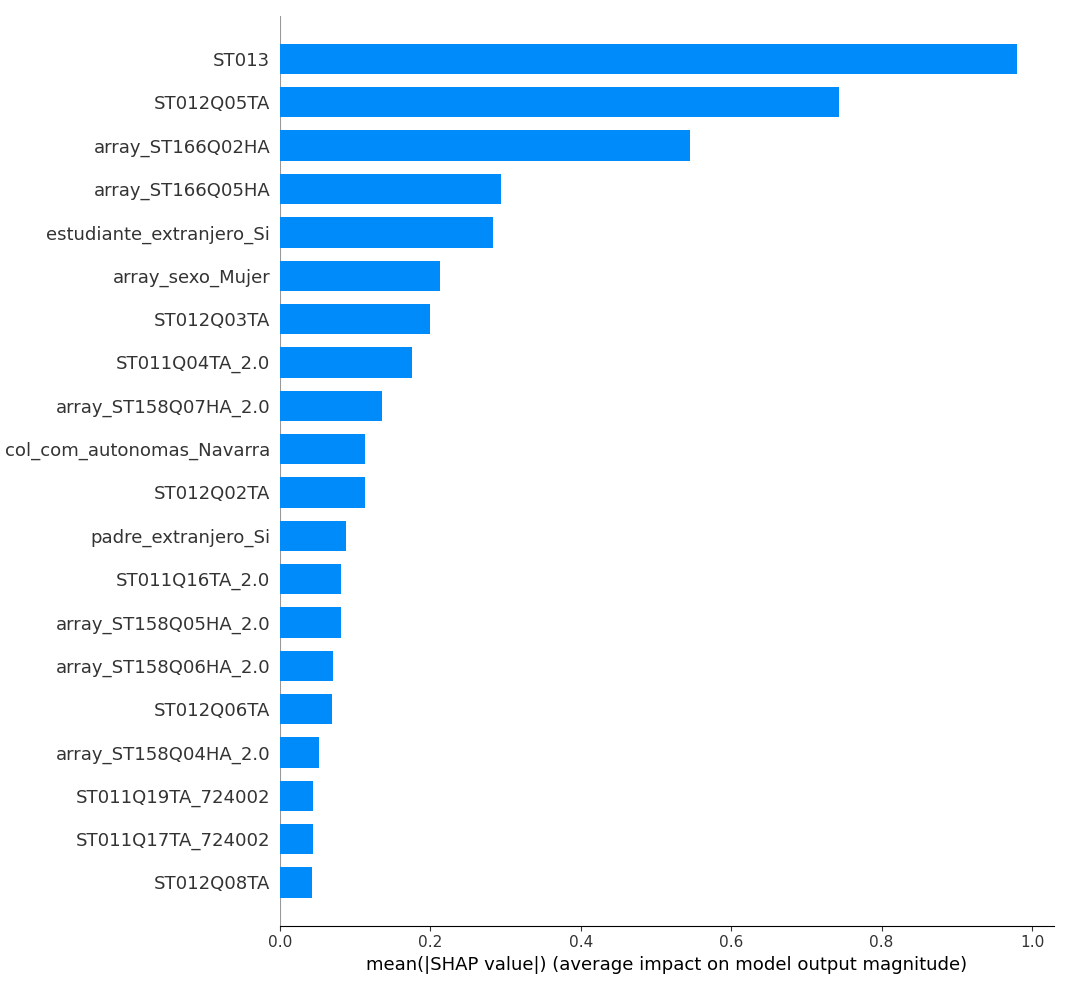} & 
\includegraphics[width = 0.45\columnwidth]{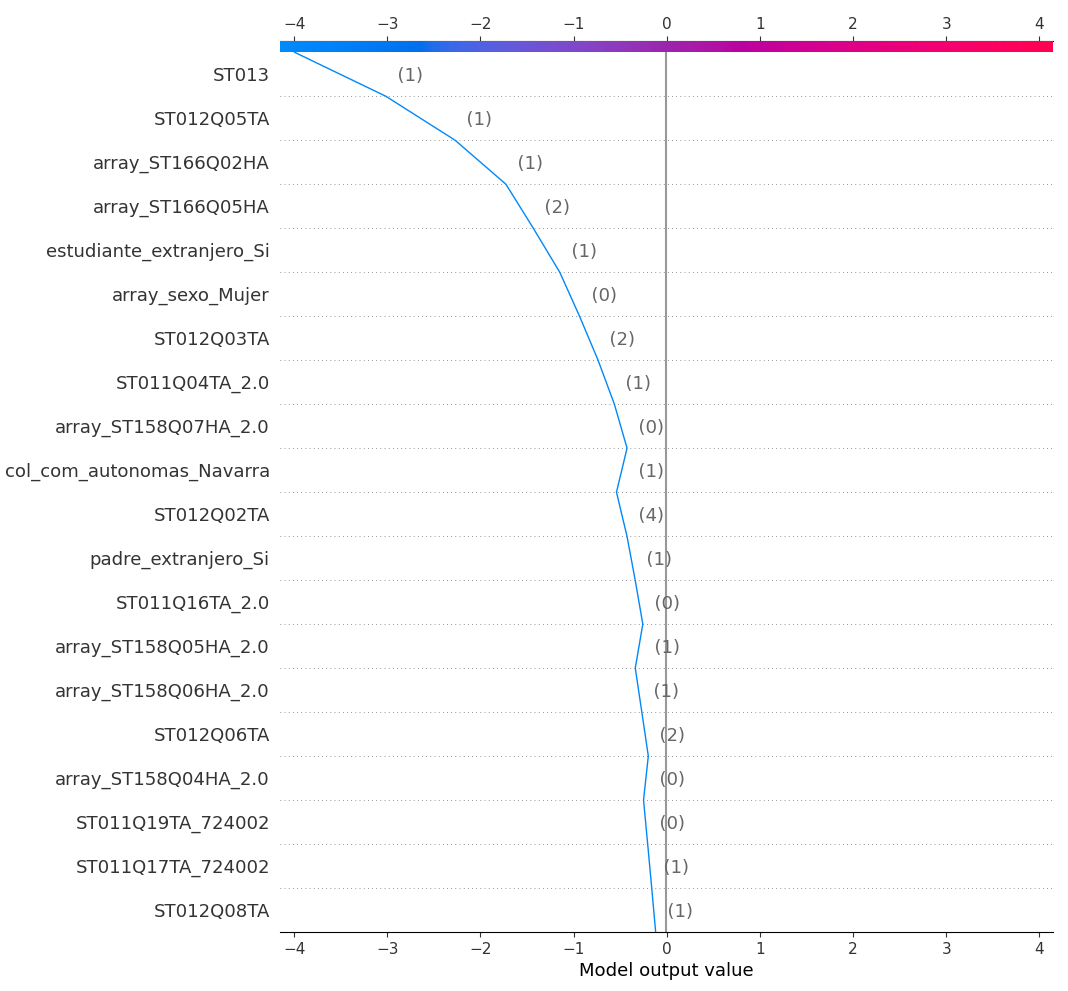} \\ 
  (c) & (d) \\
\end{tabular}
\caption{SHAP analysis comparing students with low and medium levels in Mathematics. The mean absolute SHAP value of student 3629 (medium level) is represented as summary plot (a) and decision plot (b) compared with student 2537 (low level), showing summary plot (c) and decision plot (d).}
\label{fig_shap_exp_1}
\end{figure}

When selecting student 3629 with the medium level, it is observed that the mean values of the SHAP contributions of the variables of this student provide that the most determinant variables are: (1) ST013, (2) ST166Q03HA,  and (3) ST166Q04HA. Figure \ref{fig_shap_exp_1} (a) shows the contribution of the variables mentioned in a histogram summarizing the mean absolute values. However, this figure does not concisely show which specific variables or values of these variables affect middle-level students. Therefore, Figure \ref{fig_shap_exp_1} (b) shows in a decision plot that the variables: (1) variable ST013 has a value of 5 (where the student has 201-500 books at home), (2) variable ST166Q03HA is active (the response being that it is not appropriate to respond to a phishing attack), and (3) variable ST166Q04HA with a value of 6 (where it is considered very appropriate to delete the email message without clicking on the form link in a phishing attack).

In the opposite case, when selecting student 2537 (low level), Figure \ref{fig_shap_exp_1} (c) shows that the variables that most affect the prediction of this student are: (1) the variable ST013, (2) the variable ST012Q05TA, (3) the variable ST166Q02HA and (4) the variable ST166005HA. The values of these variables that affect the mentioned variables are shown in Figure \ref{fig_shap_exp_1} (d) where it is observed that the variables: (1) the variable ST013 (indicates that the student has between 0-10 books at home), (2) the variable ST012Q05TA with a response of 1 (indicates that the student has 0 smartphones at home) (3) ST166002HA with a response of 1 out of 6 where verifying the email of the phishing attack is not entirely appropriate and (4) the variable ST166005HA with a response of 2/6 where it is considered inappropriate to verify if the operator is doing any campaign in the alleged phishing attack.

{\bf  SHAP Study of High-Medium Level Students.} In this second experiment, the model used corresponds to a GB. As a result of this and using SHAP, we obtain the contribution values of the SHAP values to be able to discern between the most extreme students. The values of the SHAP contributions of this second experiment are shown in Table \ref{table_value_shap} (center). The lowest contribution corresponds to student 656 (with a highly medium level close to the low level) and another student 626, with the highest SHAP contribution (with a high level).

\begin{figure}[t]
\centering
\begin{tabular}{cc}
\includegraphics[width=0.45\columnwidth]{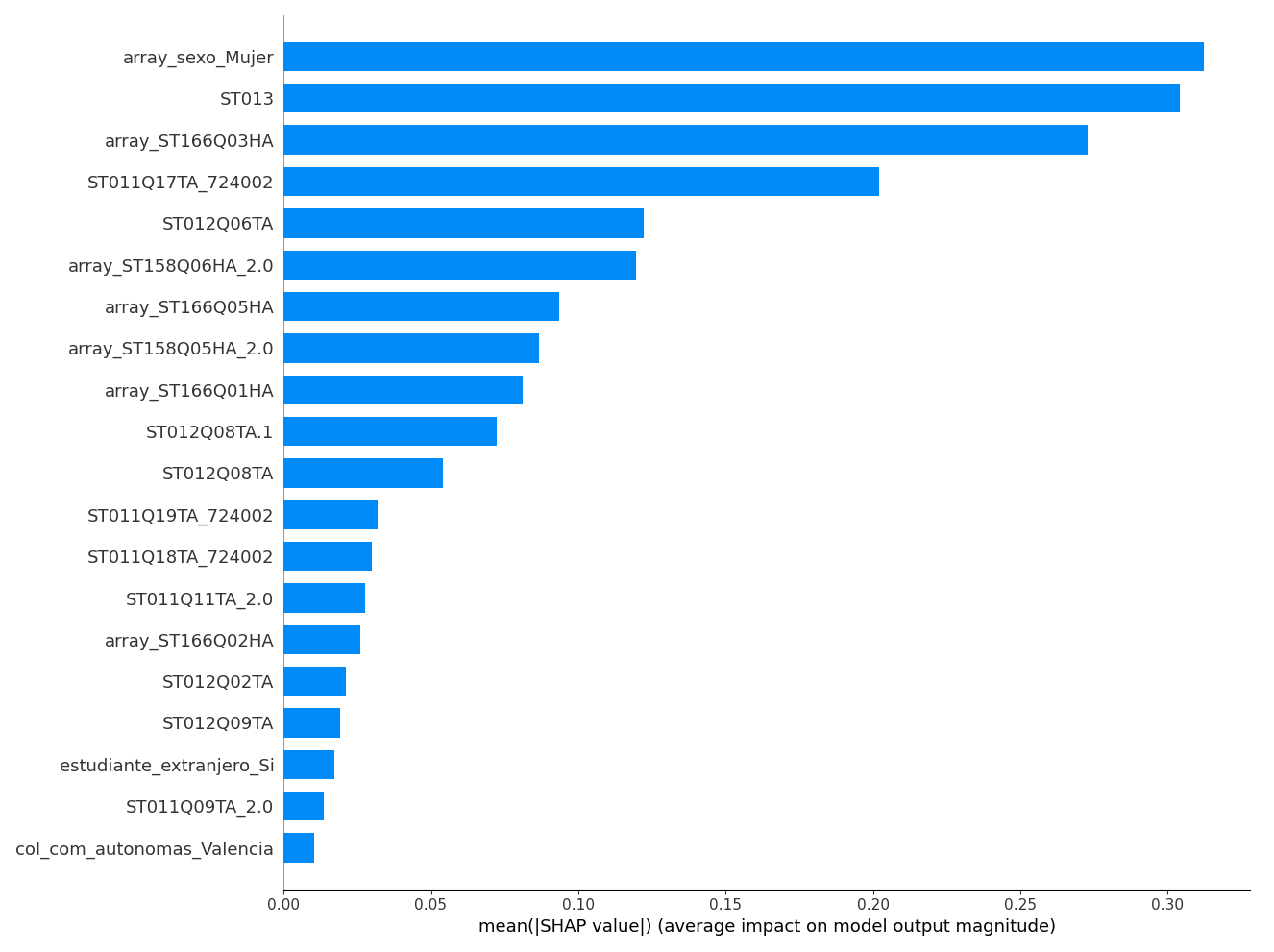} & 
\includegraphics[width=0.45\columnwidth]{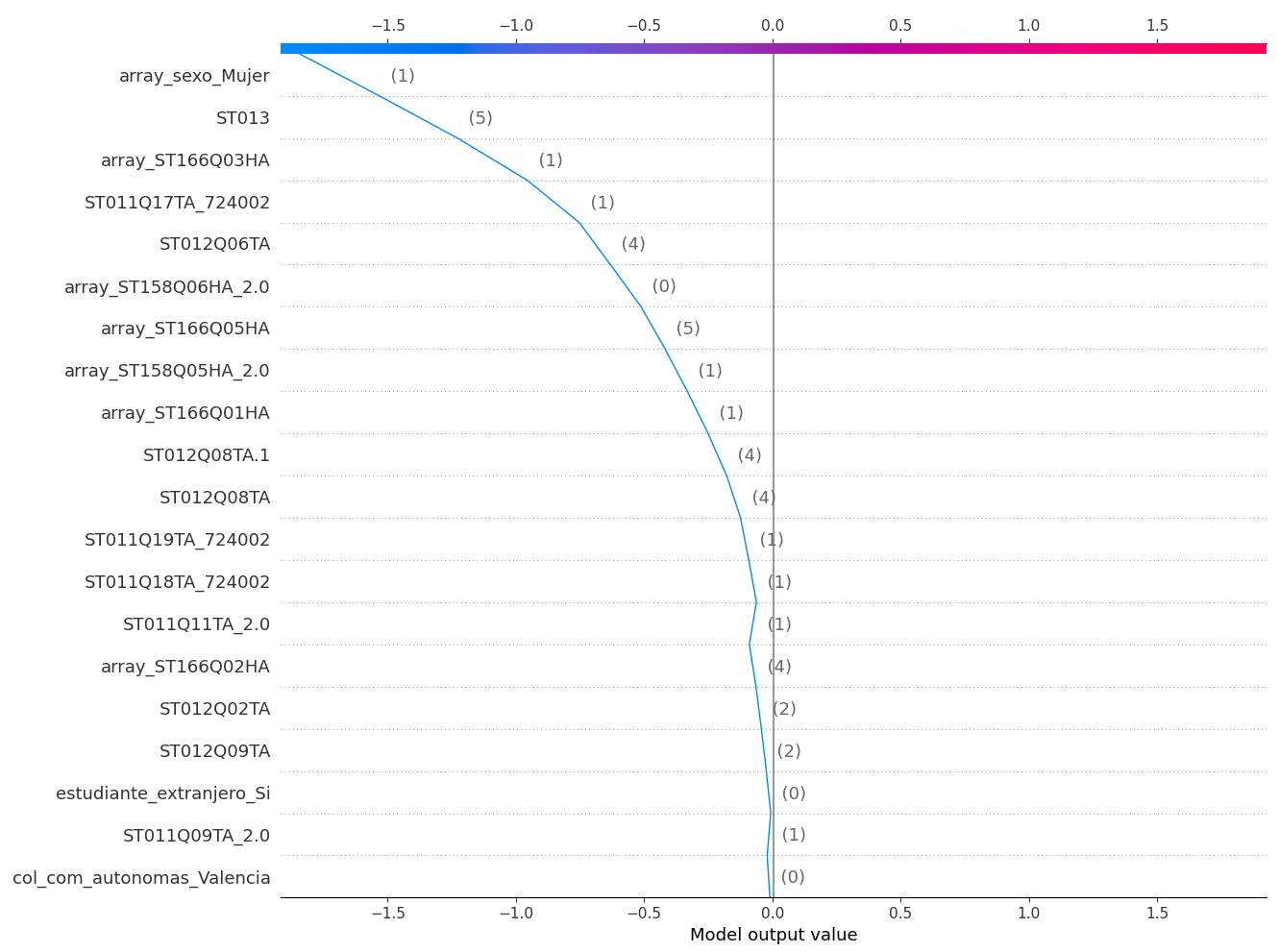} \\ 
 (a) & (b) \\
\includegraphics[width = 0.45\columnwidth]{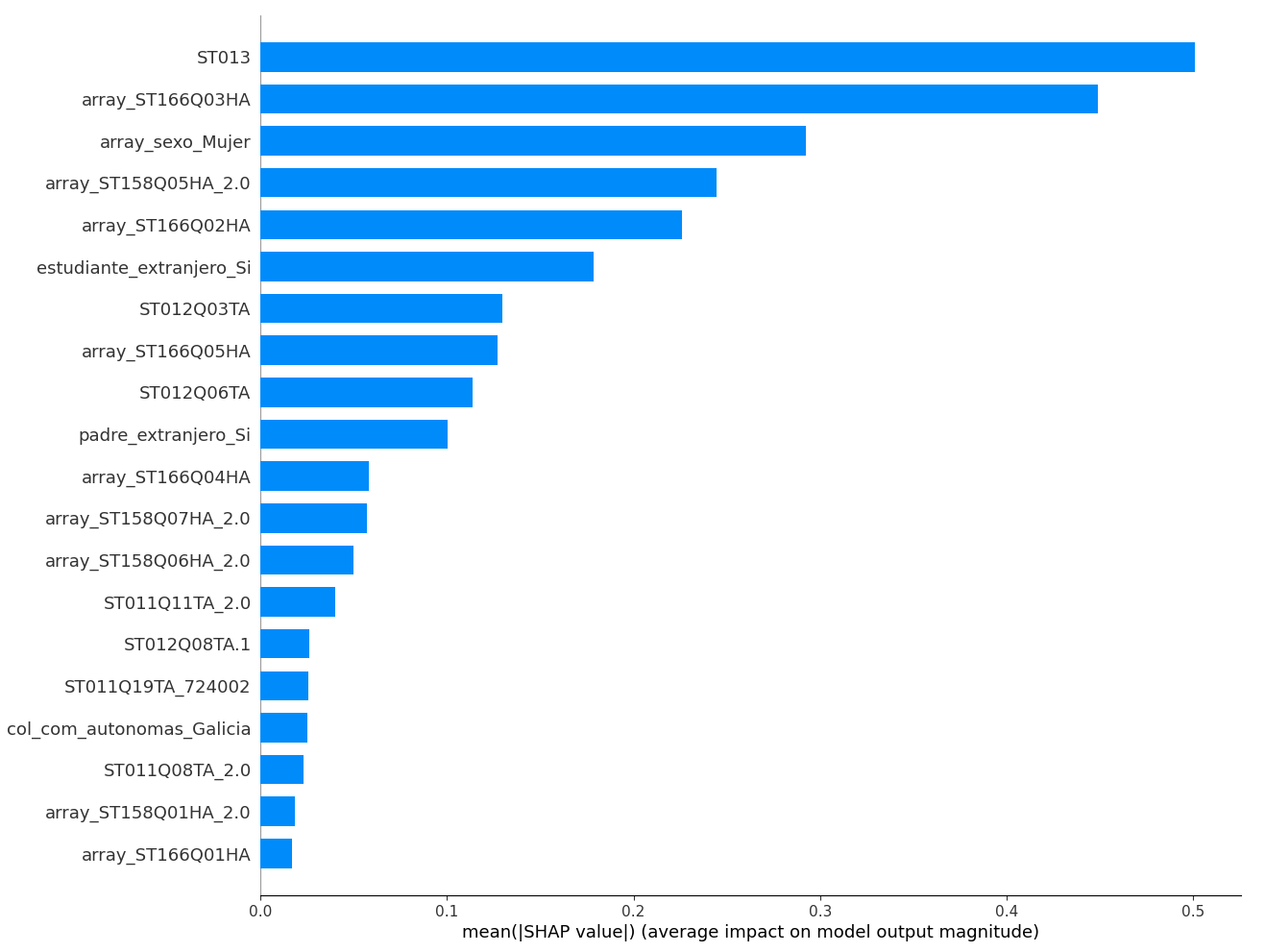} & 
\includegraphics[width = 0.45\columnwidth]{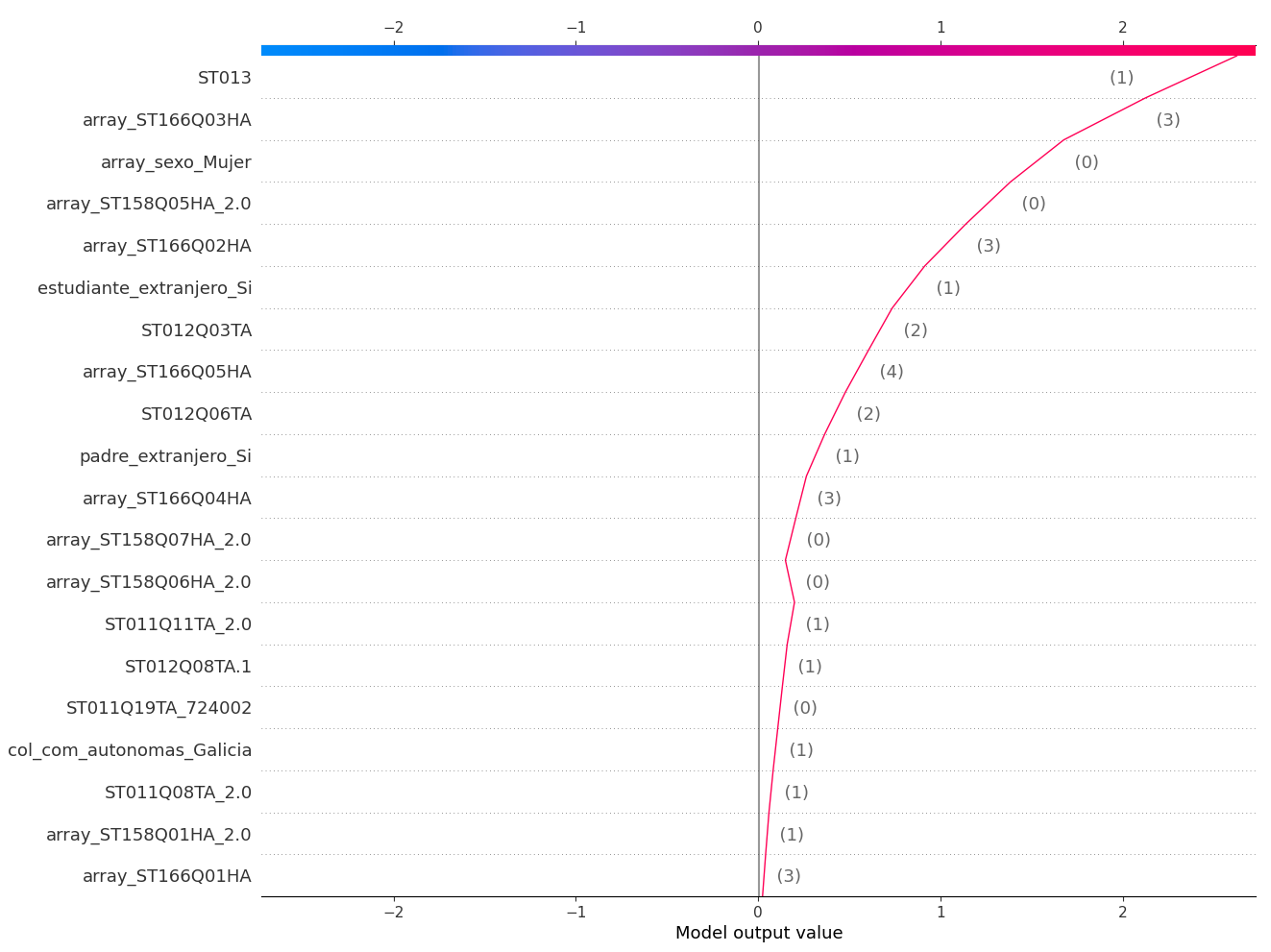} \\ 
  (c) & (d) \\
\end{tabular}
\caption{SHAP analysis comparing students with high and medium levels in Mathematics. The mean absolute SHAP value of student 626 (high level) is represented as summary plot (a) and decision plot (b) compared with student 656 (medium level), showing summary plot (c) and decision plot (d).}
\label{fig_shap_exp_2}
\end{figure}

In the case of the student high (corresponding to index 626), Figure \ref{fig_shap_exp_2} (a) shows that through this GB model for this student, the most influential variables are: (1) the variable sex Female, (2) the variable ST013, and (3) the variable ST166Q03HA. To understand the options and to deepen the interpretability of these variables, Figure \ref{fig_shap_exp_2} (b) shows the values obtained in the previous variables where the following stand out: (1) the variable of the student's sex Female is essential at this high level, (2) the variable ST013 with a value of 5 (indicating that the student has 201-500 books at home), and (3) the variable ST166003HA with a value of 1 (where it is considered that it is not appropriate to respond to a phishing attack). 

In the case of student 656, the highest contribution corresponds to a student with an average level (but close to the low-level zone). Figure \ref{fig_shap_exp_2} (c) shows that the essential variables in the prediction are: (1) the variable ST013, (2) the variable ST166Q03HA, and (3) the variable of sex Female. To be able to go deeper into the interpretability in Figure \ref{fig_shap_exp_2} (d) of the values of these variables, the following stand out: (1) the variable ST013 with a value of 1 (which indicates that the student has between 0-10 books at home), (2) the sex variable is not female, and (3) click on the form link (phishing attack) as quickly as possible.

{\bf SHAP Study of Low-High Level Students.} In the third experiment, we selected two students from the most extreme levels, where using the SHAP contribution values, we selected student 885 with a lower contribution (corresponding to the low level) and on the other hand, student 1382 with a higher SHAP contribution (corresponding to a student of a high level in Mathematics). The values of the SHAP contributions of this third experiment are shown in Table \ref{table_value_shap} (right).

\begin{figure}[t]
\centering
\begin{tabular}{cc}
\includegraphics[width=0.45\columnwidth]{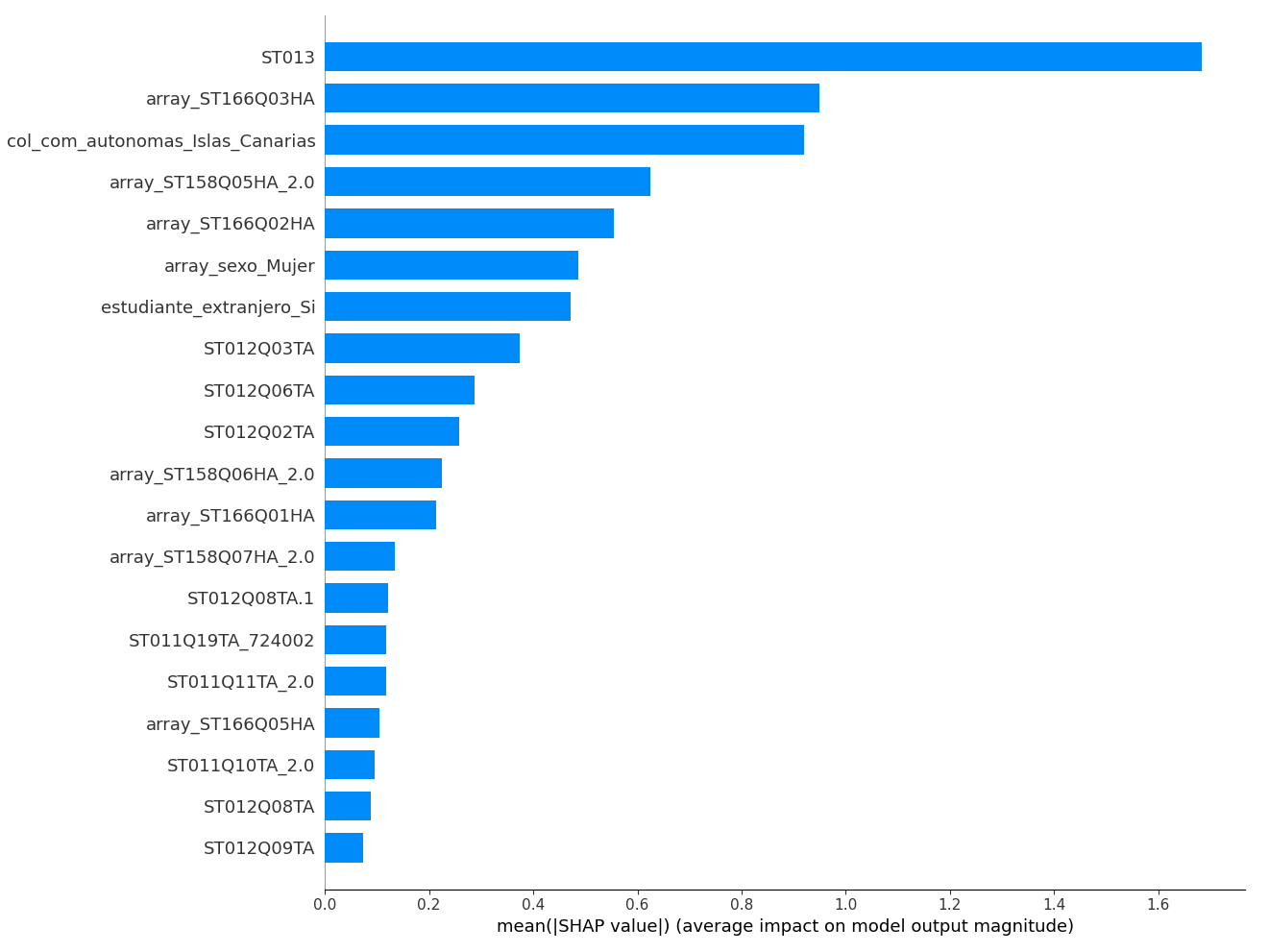} & 
\includegraphics[width=0.45\columnwidth]{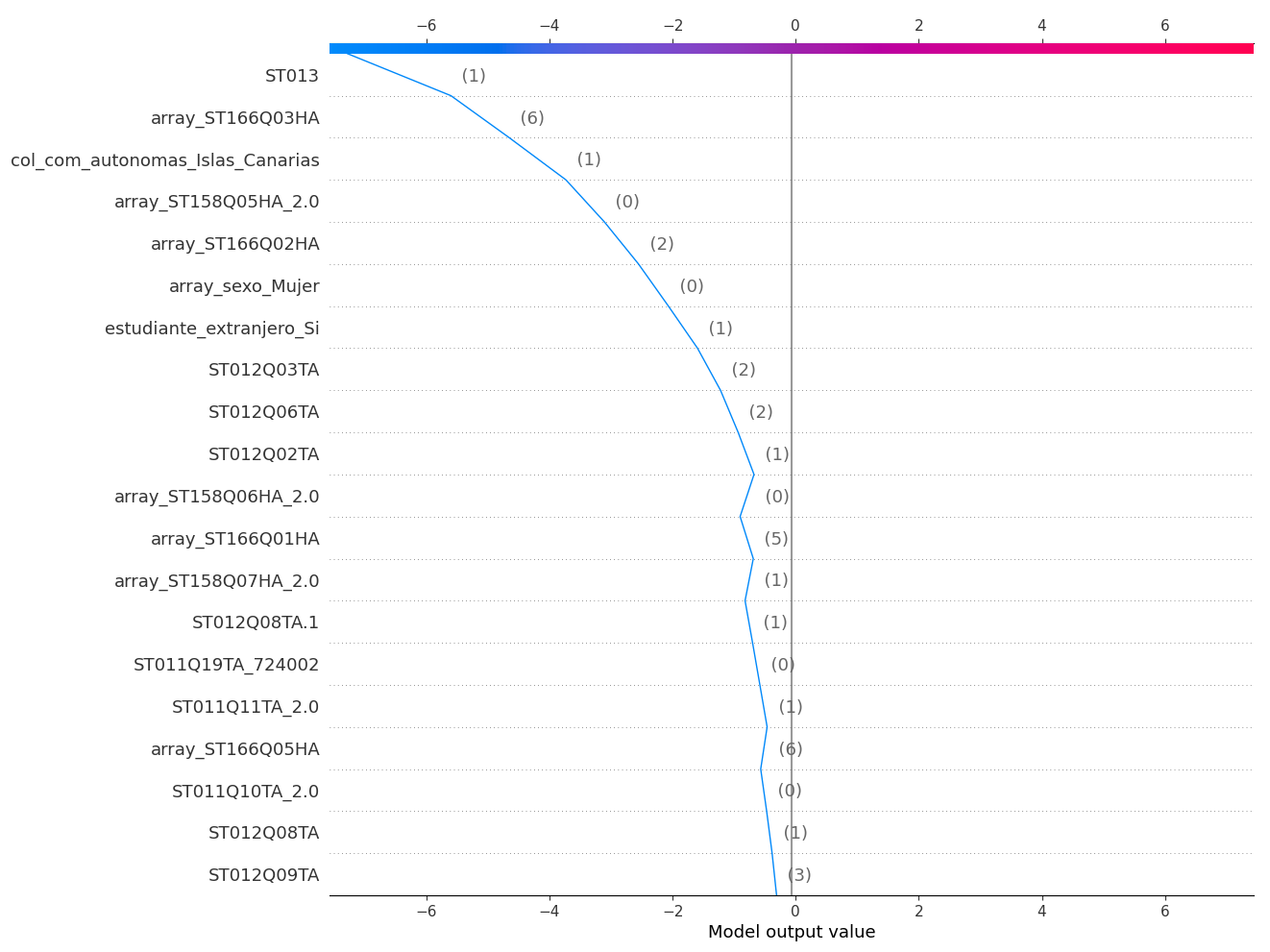} \\ 
 (a) & (b) \\
\includegraphics[width = 0.45\columnwidth]{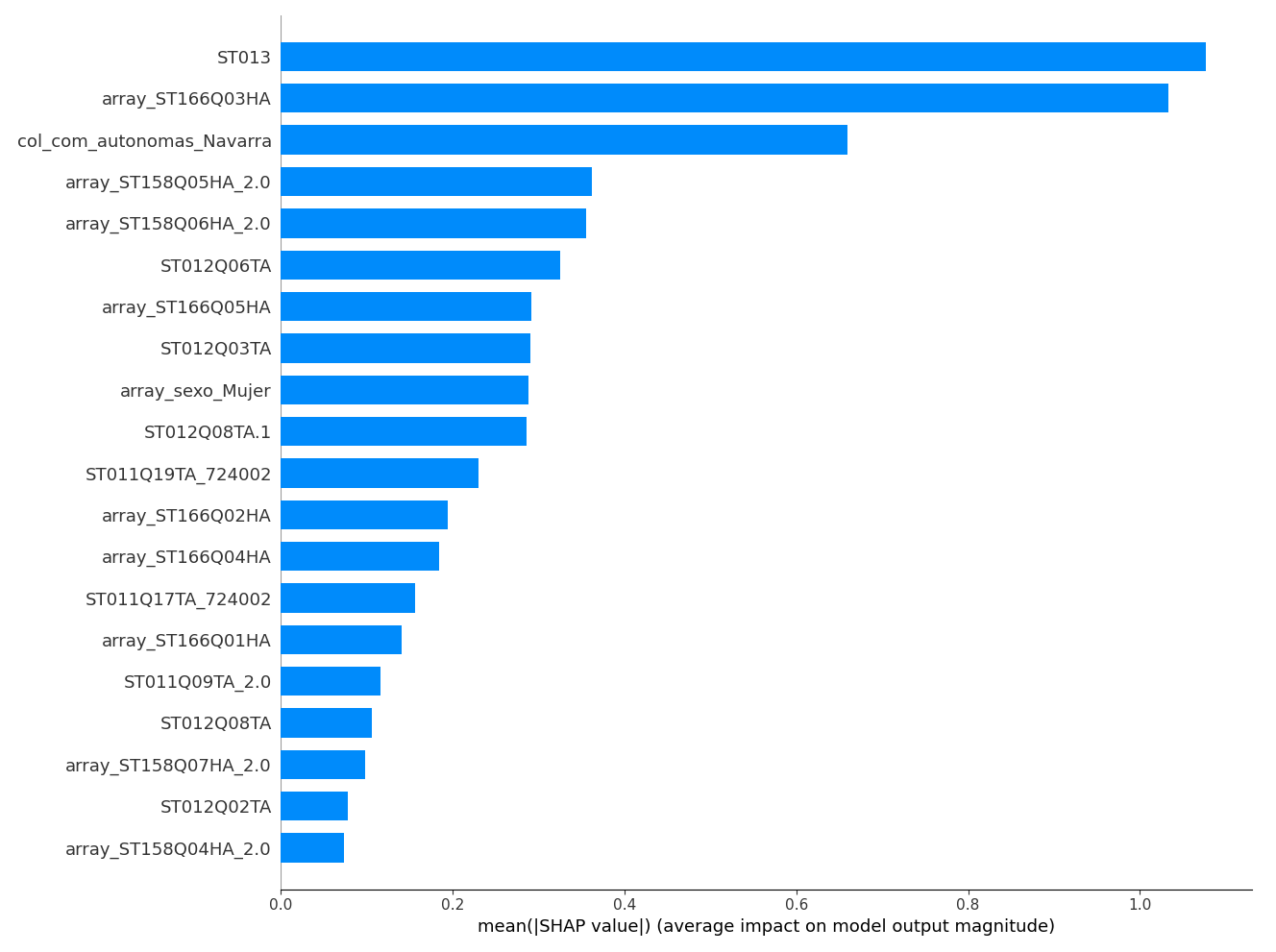} & 
\includegraphics[width = 0.45\columnwidth]{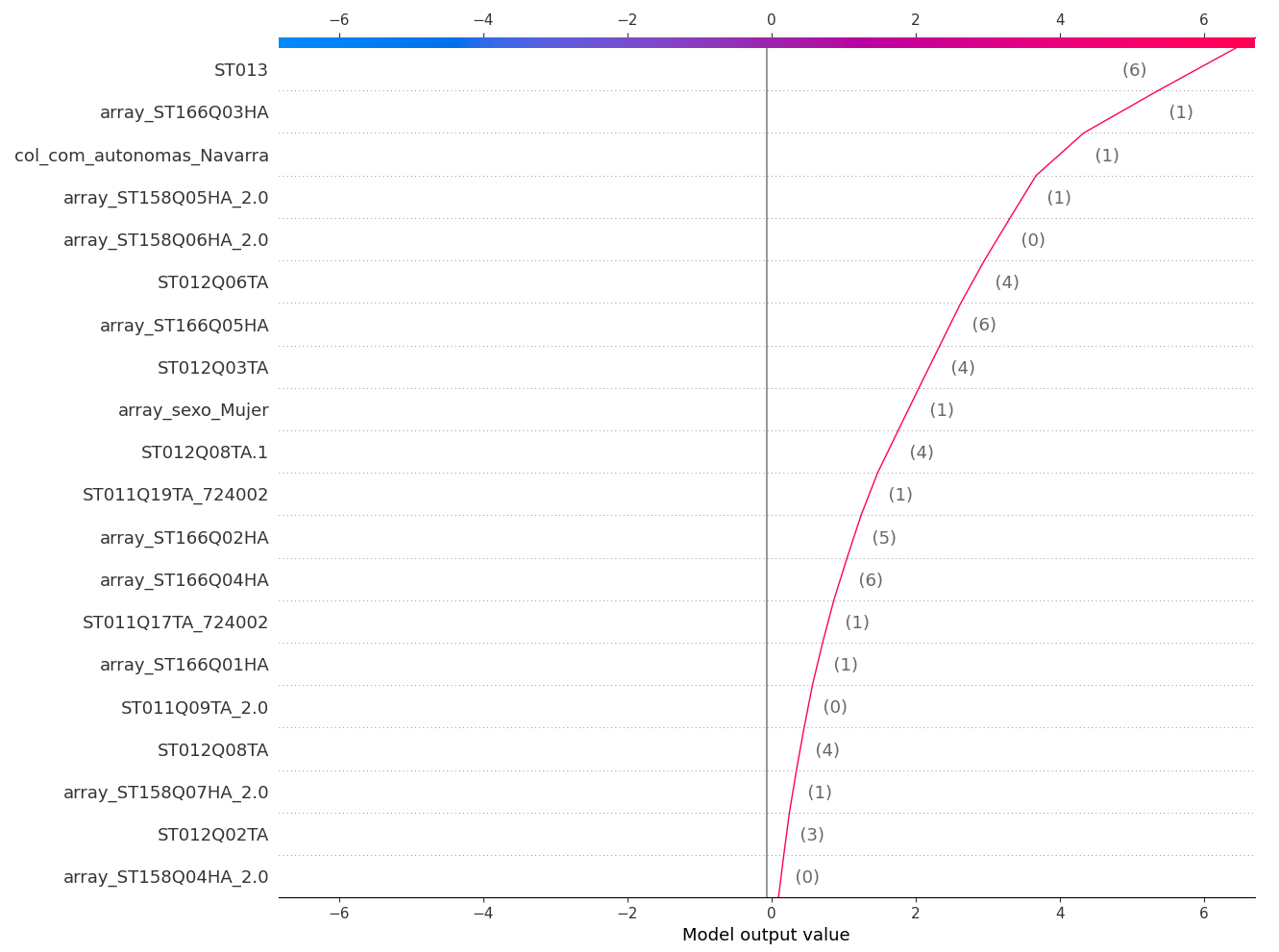} \\ 
  (c) & (d) \\
\end{tabular}
\caption{SHAP analysis comparing students with low and high levels in Mathematics. The mean absolute SHAP value of student 885 (low level) is represented as summary plot (a) and decision plot (b) compared with student 1382 (high level), showing summary plot (c) and decision plot (d).}
\label{fig_shap_exp_3}
\end{figure}

Starting with student 885 with a low level, the variables that contribute most to the prediction are (1) the ST013 variable, (2) the ST166Q03HA variable, and (3) the variable if the student studies in the Canary Islands. The contribution values of the SHAP values of the above variables are displayed in Figure \ref{fig_shap_exp_3} (a). Going deeper into the interpretability of the values selected by this low-level student in Mathematics, Figure \ref{fig_shap_exp_3} (b) shows the values of each of these variables where the following stand out: (1) the variable ST013 indicates selection ``01'' (indicating that the student owns between 0-10 books at home), and (3) the Canary Islands is active.

In the second case, corresponding to student 1382 (with a high level) in the SHAP model after training the GB model, Figure \ref{fig_shap_exp_3} (c) shows that the variables that most influence prediction are: (1) the ST013 variable, (2) the ST166003HA variable, and (3) the autonomous community of Navarra variable. Figure \ref{fig_shap_exp_3} (d) shows the values of each of these variables, where the following stand out: (1) in the variable ST013 (which indicates that the student has more 500 books at home), (2) in the variable ST166003HA, the value of ``01'' (which means that the student considers it inappropriate to click as quickly as possible in a possible phishing attack), and (3) the variable of the autonomous community of Navarra is active.

\begin{table}[t]
\centering
\caption{Set of tables compiling the SHAP values of the three experiments (experiment with low-middle levels on the left, high-middle in the middle, and Low-High on the right) ordered from highest to lowest, along with the student's index.}
    \begin{minipage}{0.3\linewidth}
        \centering
        \begin{tabular}{cc}
        \toprule
        \textbf{Index} & \textbf{ SHAP} \\
        \midrule
        \underline{\bf{3629}} & \underline{\bf{3.1651}} \\
        3114 & 3.0867 \\
        4004 & 3.0430 \\
        \vdots & \vdots \\
        \vdots & \vdots \\
        \vdots & \vdots \\
        359 & -3.6936 \\
        1012 & -3.9345 \\
        \underline{\bf{2537}} & \underline{\bf{-3.9892}} \\
        \bottomrule
        \end{tabular}
    \end{minipage}%
    \hspace{0.02\linewidth}
    \begin{minipage}{0.3\linewidth}
        \centering
        \begin{tabular}{cc}
        \toprule
        \textbf{Index} & \textbf{ SHAP} \\
        \midrule
        \underline{\bf{626}} & \underline{\bf{2.6225}} \\
        714 & 2.3465 \\
        1335 & 2.3282 \\
        \vdots & \vdots \\
        \vdots & \vdots \\
        \vdots & \vdots \\
        1224 & -1.6110 \\
        3033 & -1.6765 \\
        \underline{\bf{656}} & \underline{\bf{-1.8490}} \\
        \bottomrule
        \end{tabular}
    \end{minipage}%
    \hspace{0.02\linewidth}
    \begin{minipage}{0.3\linewidth}
        \centering
        \begin{tabular}{cc}
        \toprule
        \textbf{Index} & \textbf{ SHAP} \\
        \midrule
        \underline{\bf{1382}} & \underline{\bf{6.5067}} \\
        1669 &  5.8863 \\
        479 & 5.7600 \\
        \vdots & \vdots \\
        \vdots & \vdots \\
        \vdots & \vdots \\
        2354 & -7.1437 \\
        2378 & -7.1736 \\
        \underline{\bf{885}} & \underline{\bf{-7.2152}} \\
        \bottomrule
        \end{tabular}
    \end{minipage}
    \label{table_value_shap}
\end{table}

\section{Discussion}

The presented results reflect an in-depth discussion regarding the different student profiles obtained through Mathematics scores. The SHAP analysis defines the variables from highest to lowest contribution to the predictions obtained. According to these variables and the profiles obtained, we can classify the selected students into the levels shown in Table \ref{tab_dis}. The discussions in this section are defined by the profiles obtained for students with low and high Mathematics scores and the common variables that determine students with a medium level on the borderline between the low and high levels.

\begin{table}[t]
  \centering
  \caption{Compilation of the metrics of the best models in the binary classification problem with student scores at the low and high level.}
  \label{tab_dis}
\begin{tabular}{lc}
\toprule
\textbf{Level} & \textbf{Students} \\
\midrule
\cellcolor{red!50} High & \cellcolor{red!50} 626, 1382 \\
\cellcolor{orange!50} Medium-High & \cellcolor{orange!50} 3629 \\
\cellcolor{orange!25} Medium-Low & \cellcolor{orange!25} 656 \\
\cellcolor{blue!25} Low & \cellcolor{blue!25} 2537, 885 \\
\bottomrule
\end{tabular}
\end{table}


{\bf Discussion on Students with Low Test Level.} In this area of our discussion, the most contributing variables to the different models obtained from students 656, 2537, and 885 are compiled, with particular attention to student 885, since the SHAP values obtained from this student belong to a model with better performance than the previous ones. Therefore, the following ones stand out when discussing the most influential variables from the most to the least important.

\textbf{Variable ST013}. This variable corresponds to the question asked to students regarding the number of books they own at home. This variable appears with greater importance in the results obtained, with particular attention to students with a low level where the answers appear with a value of ``01'' (they own between 0-10 books) and ``02'' (they own between 11-25 books). This response indicates that students with low levels in Mathematics do not possess or cannot access reading at home.

\textbf{Variable ST166}. This variable is surprising since the purpose of this questionnaire is to evaluate the student's sensitivity to a possible phishing attack, where a scenario is proposed to the student where he/she has supposedly won a cell phone and different strategies and a range of values where these actions are appropriate are proposed. Among the different strategies, the different responses among students with a low level stand out:
\begin{enumerate}
    \item \textbf{ST166Q05HA}. The proposed strategy is to check the cell phone operator's website (where they supposedly regulate cell phones). This strategy, in principle, in the face of a possible phishing attack, could be considered entirely appropriate. However, what is surprising is that among students with a low level in Mathematics, the answers correspond to the fact that this type of action is not at all or not very appropriate (with answers of ``01'' and ``02'').
    \item \textbf{ST166Q02HA}. On this occasion, the proposed action is to verify the email address of the email where we have supposedly won a cell phone. As in the previous strategy, in this case, it is proposed to the student whether this action is appropriate in the face of a possible phishing attack (where at first, we could consider that it is entirely appropriate). The answers with a low level in Mathematics emphasize that this action could be more appropriate (with answers of ``01'' and ``02'').
\end{enumerate}

\textbf{Student's Gender}. In all cases, the gender of the student is not female since the binary variable is deactivated (with a value of 0). This is a determinant for the higher levels in Mathematics, which we will go into more detail later.

\textbf{Student's Autonomous Community}. In this case, the autonomous city of Ceuta and students from the Canary Islands stand out, respectively, from most to least important. This information is supported by the PISA report 2018, where the percentages of exceeding levels by autonomous communities are indicated. The autonomous communities of Castilla y León, Navarra, Basque Country, Galicia, Cantabria, Aragón, and La Rioja have between 58\% and 61\% of students achieving at least Level 3 in Mathematics. On the opposite end, the Canary Islands (40\%), Andalucía (44\%), and Extremadura (45\%) have the lowest proportions of students reaching at least Level 3. The autonomous cities of Ceuta (20\%) and Melilla (29\%) have concerning proportions of students achieving at least Level 3 in mathematical competency (where we have our cut-off from the low to the medium level in our work) \cite{cebrian2019pisa}.

{\bf Discussion on Students with High Test Level.} In the case of high-level students, the variables obtained from student 1382 stand out the most, followed by students 626 and 3629. The variables that most influence the prediction of the models are the following ones.

\textbf{Variable ST166Q03HA}. In this case, within the problem of the phishing attack scenario, the proposed action is to click as fast as possible on the ad where we have supposedly won a cell phone. The response of students with a high level in Mathematics is that this action is inappropriate (with a response of ``01''). These types of students are {\it a priori} less sensitive to phishing attacks.

\textbf{ST013}. Students with high levels answered this questionnaire with values between ``04'' and ``06'', which indicates that they have at least 100 books at home, something relatively significant compared to students with a low level, where they stand out with the ability to have less than 25 books at home.

\textbf{Student's Gender}. The sex of students with a high level in Mathematics generally has a more significant contribution to predicting these levels when they are females. This does not indicate that the highest gender significance is in females. Instead, the prediction model infers this gender distinction in students with a high level. The 2018 PISA reports define this gender distinction at high mathematics levels, while the percentages are similar at low levels. In Spain, 8\% of males reach the highest levels (5 and 6) on the mathematics scale, compared to only 5.5\% of females, a trend that is repeated in the average of OECD countries, but with higher figures: 12.2\% of males compared to 9.5\% of females. However, at the lowest levels (below Level 2), the figures are very similar in Spain (24.8\% of females and 24.6\% of males). In the average of OECD countries (24\% of males and 23.9\% of females), these percentages indicate that approximately 25\% of males and females in Spain do not reach Level 2 in mathematics, which stands at 24 \% on average in OECD countries \cite{cebrian2019pisa}.

\textbf{Student's Autonomous Community}. In the case of students with a high test level, the autonomous community of Navarra stands out above all, although the local analysis of this student is always carried out. This is in line with the PISA reports in Spain. In the case of La Rioja (9.0\%), Navarra (9.3\%) and Castilla y León (9.5\%) show the highest percentages of students who are at Level 5 of performance in Mathematics, and with respect to Level 6 in Mathematics, La Rioja (2.4\%) and Navarra (2.1\%) stand out again \cite{cebrian2019pisa}.

\textbf{Variable ST166Q05HA}. In this case, the action performed concerning the questionnaire ST166 corresponds to consult the mobile operator's website, and most of the answers of the students with a high level in Mathematics is ``06'' which they consider this action to be very appropriate. Therefore, apparently, students at these levels are less sensitive to phishing attacks.

\textbf{Variable ST166Q02HA}. Similarly, this action where the verification of the sender's email of the possible phishing attack is considered an appropriate strategy for the possibility of this scenario (answered with a value of ``06'').

In our discussion, we synthesize our findings with the related work, drawing upon several studies highlighting the multifactorial nature of academic achievement. Our analysis dovetails with the work of \cite{bernardo2021socioeconomic}, which emphasizes the critical interplay between socioeconomic status and growth mindset within the Philippines' educational context. This intersection aligns with our observations on how economic factors shape student outcomes.

Moreover, our investigation into the gender gap in Mathematics achievement finds resonance with the cross-national study by \cite{else2010cross} and the global assessment by \cite{lu2023assessing}. These studies bring to light the variable degrees of gender disparities across different regions, reinforcing our conclusions regarding the impact of gender on academic choices and performance, with a particular emphasis on STEM fields. Our findings contribute to this nuanced discourse, suggesting that while strides have been made, gender-based educational differences persist and continue to shape the academic landscape.

Taken together, our study weaves these strands (socioeconomic factors, gender dynamics, and familial engagement) into a cohesive narrative, shedding light on the complex tapestry of influences that underpin academic success in Mathematics. This comprehensive approach not only provides a deeper understanding of the determinants of educational outcomes but also underscores the contributions of our research to ongoing scholarly conversations.

\section{Conclusion and Future Works}

This study, through models generated from representative students, has examined the multivariate factors contributing to students' mathematical performance, from those with low to the highest levels. 

In low-performing students, the limited amount of books at home suggests a correlation between lack of access to reading and low Mathematics performance. In addition, their response to a possible phishing attack scenario indicates a lower level of logical and analytical thinking. At the geographic level, the concentration of low-performing students in the autonomous city of Ceuta and the Canary Islands is consistent with the 2018 PISA reports, underscoring the persistence of regional disparities in education.

In contrast, high-achieving students in Mathematics possess significantly more books at home, indicating greater access to reading materials. These students also demonstrate superior critical thinking skills in the phishing scenario, reflecting their competence to manage and evaluate information effectively. The role of gender in Mathematics performance is also evident in this group, with females contributing significantly to the prediction of these levels. 
Regarding regionality, the autonomous community of La Rioja and Navarra stands out in having a higher percentage of students with a high Mathematics level.

Therefore, this study provides valuable information on how variables, such as access to reading, critical thinking skills, gender, and geographic region, may influence students' Mathematics achievement. However, the limitation of the SHAP model in the local analysis of students may affect the need for more information on other variables at the global level. Therefore, as future research, it is interesting to evaluate this issue using global interpretability models. In the future, it would be beneficial to explore further the association between these variables and their potential impact on other aspects of students' academic performance.

As future research avenues, exploring global interpretability models to overcome limitations observed in the SHAP model during local student analyses appears compelling. Scholars might consider extending the current study findings by employing these global interpretability models to comprehend better the multifaceted variables affecting student performance on a broader scale.

Additionally, while this investigation has highlighted the importance of reading accessibility, logical reasoning, gender, and regional influences on Mathematics achievements, there is potential to delve deeper. For instance, it might be intriguing to analyze how the interplay between these factors could influence student performances in other academic subjects, thus providing a more comprehensive educational landscape.

Furthermore, given the sharp regional disparities highlighted, such as the prominence of high-performing students in Navarra and the concentration of low-performing ones in places like Ceuta and the Canary Islands, there is an avenue for researchers to examine the socioeconomic and infrastructural reasons behind such disparities. This could pave the way for targeted educational interventions tailored to specific regions.

Moreover, the significant gender-based insights derived from this study invite further exploration. Could there be underlying sociocultural reasons behind these observations? Such questions are ripe for future research endeavors.

By understanding and addressing these educational disparities, especially in Mathematics performance, we have the potential to craft more effective and personalized educational policies and interventions. This study has laid the groundwork, but the horizon ahead offers numerous opportunities for deeper dives and wider explorations. We anticipate that this investigation serves as a launchpad, inspiring future endeavors in this crucial academic domain.

\section*{Declaration of competing interest}
The authors declare that they have no known competing financial interests or personal relationships that could have appeared to influence the work reported in this paper.

\section*{Acknowledgments}
Partially funded by the Autonomous Community of Madrid (ELLIS Madrid Node). Also partially supported by project PID2022-140786NB-C32/AEI/10.13039/501100011033 (LATENTIA) from the Spanish Ministry of Science and Innovation.

\bibliographystyle{unsrt}  
\bibliography{references}  

\newpage

\appendix
\section{Appendix}

\subsection{Low and Medium Level Study}
In the context of predicting student performance, Table \ref{tab_model_hyperparameters_low_medium} presents an exhaustive analysis of eight ML models, each evaluated under a range of hyperparameters. This evaluation focuses on distinguishing between low and medium performance levels among students.

\textbf{Logistic Regression} demonstrates notable performance in terms of ACC and F1S, especially with ``Penalty: L1, C: 10'', achieving an ACC of 0.7211 and an F1S of 0.7525. This indicates considerable efficacy in binary classification, possibly due to the L1 regularization, which favors the selection of relevant features for the model.

\textbf{Decision Tree}, particularly with ``Criterion: gini, Max Depth: 10'', show a balance in PR and SP, but moderate ACC, which could suggest overfitting. Limited depth might be crucial in avoiding this phenomenon.

\textbf{Random Forest}. By increasing the number of estimators, as in ``n\_estimators: 200, Max Depth: None'', the Random Forest model significantly improves, reflecting an ACC of 0.7139 and an AUC of 0.7872. This demonstrates the ensemble’s effectiveness in generalization, reducing the overfitting common in individual trees.

\textbf{Gradient Boosting}, with ``n\_estimators: 200, learning\_rate: 0.1'', stands out for its high ACC (0.7274) and F1S (0.7570), indicating its efficacy in sequential model corrections and in handling various data distributions.

\textbf{SVM} with ``Kernel: RBF, C: 1'' exhibits high PR (0.7926) and F1S (0.7578), highlighting its capability to create complex boundaries between classes.

\textbf{XGBoost and LightGBM} Both models, with their respective optimal hyperparameters, demonstrate high levels of Accuracy and AUC, indicating their suitability for large datasets due to their scalability and speed.

\textbf{MLP}, especially with ``HLS: (100,), Activation: tanh'', achieves high RC and F1S, indicating its effectiveness in capturing non-linear relationships in the data.

\subsection{High and Medium Level Study}

Table \ref{tab_model_hyperparameters_high_medium} presents an in-depth analysis of eight ML models, each assessed across a range of hyperparameters. The objective is to differentiate between high and medium performance levels among students, a crucial task in educational data analysis.

\textbf{Logistic Regression}. This model, particularly with ``Penalty: L1, C: 10'', achieves an impressive ACC (0.6874) and the highest PR (0.2222) among its configurations. The results suggest that while LR is reasonably effective in classifying high and medium performers, the challenge of distinguishing these two groups is more complex, reflected in the relatively lower performance metrics compared to lower-level differentiation.

\textbf{Decision Tree}, especially with ``Criterion: entropy, Max Depth: 5'', shows an improved balance between PR (0.2070) and SP (0.6813). This configuration indicates a modest capability to differentiate between high and medium performers, likely due to the decision tree simplicity and interpretability.

\textbf{Random Forest}. In the ensemble category, this model with ''n\_estimators: 100, Max Depth: None'' registers the highest ACC (0.6947) and PR (0.2200), suggesting its effectiveness in handling more nuanced distinctions in student performance levels, likely due to its robustness against overfitting and its ability to aggregate multiple decision trees.

\textbf{Gradient Boosting}. Demonstrating strong performance, this model with ``n\_estimators: 50, learning\_rate: 0.01'' achieves high ACC (0.7069). This model success can be attributed to its sequential approach to correcting errors from previous trees, making it particularly effective for complex classification tasks like distinguishing high and medium performers.

\textbf{SVM} with a radial basis function kernel (``Kernel: RBF, C: 10'') shows substantial PR (0.2193), indicating its strength in classifying high-performance students accurately, albeit with a trade-off in RC.

\textbf{XGBoost}, particularly with ``n\_estimators: 50, learning\_rate: 0.1'', stands out with its balanced ACC (0.6853) and F1S (0.3279). This highlights XGBoost capability to handle sophisticated patterns within the data, which is beneficial for differentiating nuanced performance levels.

\textbf{MLP}, especially with ``HLS: (100,), Activation: tanh'', achieves a significant RC (0.7227) and F1S (0.3303). This indicates its strength in identifying high performers but also suggests the potential for false positives, a common issue in neural network-based models.

\textbf{LightGBM} with ``n\_estimators: 200, learning\_rate: 0.01'', this model achieves a notable balance with high PR (0.2168) and F1S (0.3255), reflecting its efficiency and effectiveness in handling large datasets and complex classification tasks.

\subsection{Low- and High-Level Study}
In Table \ref{tab_model_hyperparameters_low_high}, we delve into the evaluation of eight ML models, each assessed across diverse hyperparameters. The goal is to discern between low and high-performance levels among students, a significant challenge in educational data analysis.

\textbf{Logistic Regression}. Exhibiting notable efficacy, especially with ``Penalty: L1, C: 10'', achieves the highest ACC (0.8546) and F1S (0.8507). This model strength lies in its ability to balance RC and PR essential for distinguishing between extreme performance levels.

\textbf{Decision Tree}, particularly with ``Criterion: entropy, Max Depth: 5'', shows considerable effectiveness, indicated by a high F1S (0.8555) and SP (0.5099). This suggests its capability to differentiate performance levels, though its simpler structure may lead to challenges in more nuanced cases.

\textbf{Random Forest}. Standing out in the ensemble category, this model with ``n\_estimators: 100, Max Depth: None'' records the highest ACC (0.8604) and F1S (0.8594). This underscores the model robustness and its ability to handle complex classification tasks, such as distinguishing varied performance levels in educational data.

\textbf{Gradient Boosting} with ``n\_estimators: 200, learning\_rate: 0.1'', this model achieves excellent results, including the highest AUC (0.9354). Its sequential approach to model building allows it to effectively address the classification challenges inherent in distinguishing between low and high student performance levels.

\textbf{SVM} with a radial basis function kernel (``Kernel: RBF, C: 10'') demonstrates strong PR (0.6680) and F1S (0.8614), indicative of its strength in accurately classifying high-performing students while maintaining a balance with RC.

\textbf{XGBoost}, particularly with ``n\_estimators: 200, learning\_rate: 0.1'', shows a balanced performance with high ACC (0.8542) and F1S (0.8531). Its capability to manage complex patterns within the data makes it a suitable choice for this challenging classification task.

\textbf{MLP}, especially with ``HLS: (100,), Activation: tanh'', achieves a significant RC (0.8834) and F1S (0.8516), reflecting its effectiveness in identifying high performers, although it may require careful tuning to prevent overfitting.

\textbf{LightGBM}, especially with n\_estimators: 50, learning\_rate: 0.1, excels in ACC (0.8517) and PR (0.6443), illustrating its efficiency in handling large datasets and complex tasks like distinguishing between low and high student performance levels.

\subsection{Confusion Matrices}

In a comparative analysis of ML algorithms for binary classification tasks, various models were assessed based on their confusion matrices at distinct classification thresholds. As depicted in Table \ref{tab_model_hyperparameters_low_medium}, LR exhibited a relatively balanced performance across all thresholds, with a slight inclination towards higher false positives in the Low-High threshold scenario. Decision Trees, while demonstrating robustness in the Low-Medium and High-Medium thresholds, showed a propensity for increased false negatives in the Low-High.

Random Forest and GB methods showed similar patterns, with the latter having a marginally higher count of true positives in most cases. SVM and XGBoost both presented competitive true positive rates, but SVM displayed a slightly elevated rate of false negatives. The MLP model, a type of neural network, demonstrated high true positive rates but was also accompanied by a notable number of false negatives, particularly in the Low-High threshold.

Finally, LightGBM showed a balanced distribution of predictive outcomes, with a moderate increase in false negatives in the Low-High threshold. This table allows for a detailed comparison of model performance, highlighting the trade-offs between detecting true positives and avoiding false positives across different models and threshold settings.

Upon closer examination of the confusion matrices presented in Table \ref{tab_model_hyperparameters_low_medium}, it is pertinent to acknowledge the potential impact of data imbalance on the classification outcomes. The skewness in the distribution of test data classes may contribute to the observed discrepancy in false negatives and false positives across the models. This phenomenon underscores the importance of considering the underlying data distribution when training and evaluating ML models. The prevalence of one class over another can lead to a model overfitting to the majority class, thereby diminishing its predictive performance on the minority class, as evidenced by the disproportionate false negative rates. 

\begin{table}[h]
  \centering
  \caption{This table presents a comprehensive overview of eight ML models evaluated across a range of hyperparameters to distinguish between low and medium performance levels among students. }
  \begin{adjustbox}{width=0.80\textwidth, keepaspectratio}
  \label{tab_model_hyperparameters_low_medium}
  \begin{tabular}{llcccccc}
    \toprule
    \textbf{Model} & \textbf{Hyperparameters} & \textbf{ACC} & \textbf{RC} & \textbf{PR} & \textbf{SP} & \textbf{F1S} & \textbf{AUC} \\
    \midrule
    \multirow{6}{*}{Logistic Regression} 
      & Penalty: L1, C: 0.1 & 0.7189 & 0.7169 & \bf{0.7846} & \bf{0.7218} & 0.7492 & 0.7929 \\
      & Penalty: L1, C: 1 & 0.7201 & 0.7228 & 0.7828 & 0.7164 & 0.7516 & 0.7938 \\
      & \textbf{Penalty: L1, C: 10} & \bf{0.7211} & \bf{0.7238} & 0.7836 & 0.7174 & \bf{0.7525} & \bf{0.7940} \\
      & Penalty: L2, C: 0.1 & 0.7187 & 0.7210 & 0.7818 & 0.7154 & 0.7502 & 0.7935 \\
      & Penalty: L2, C: 1 & 0.7207 & \bf{0.7238} & 0.7830 & 0.7164 & 0.7522 & \bf{0.7940} \\
      & Penalty: L2, C: 10 & 0.7209 & \bf{0.7238} & 0.7833 & 0.7169 & 0.7524 & \bf{0.7940} \\
    \midrule
    \multirow{6}{*}{Decision Tree}
      & Criterion: gini, Max Depth: None & 0.6148 & 0.6130 & 0.6937 & 0.6173 & 0.6509 & 0.6152 \\
      & \textbf{Criterion: gini, Max Depth: 5} & 0.6767 & 0.6123 & \bf{0.7884} & \bf{0.7676} & 0.6893 & \bf{0.7576} \\
      & Criterion: gini, Max Depth: 10 & 0.6785 & \bf{0.6983} & 0.7385 & 0.6505 & 0.7178 & 0.7226 \\
      & Criterion: entropy, Max Depth: None & 0.6231 & 0.6220 & 0.7008 & 0.6246 & 0.6591 & 0.6233 \\
      & \textbf{Criterion: entropy, Max Depth: 5} & 0.6767 & 0.6123 & \bf{0.7884} & \bf{0.7676} & 0.6893 & \bf{0.7576} \\
      & Criterion: entropy, Max Depth: 10 & \bf{0.6837} & 0.6924 & 0.7487 & 0.6715 & \bf{0.7195} & 0.7278 \\
    \midrule
    \multirow{9}{*}{Random Forest}
      & n\_estimators: 50, Max Depth: None & 0.7076 & 0.6934 & \bf{0.7826} & \bf{0.7276} & 0.7353 & 0.7813 \\
      & n\_estimators: 50, Max Depth: 5 & 0.7064 & 0.7162 & 0.7670 & 0.6925 & 0.7408 & 0.7773 \\
      & n\_estimators: 50, Max Depth: 10 & 0.7131 & 0.7166 & 0.7763 & 0.7081 & \bf{0.7452} & 0.7862 \\
      & n\_estimators: 100, Max Depth: None & 0.6997 & 0.6889 & 0.7736 & 0.7149 & 0.7288 & 0.7798 \\
      & n\_estimators: 100, Max Depth: 5 & 0.7060 & 0.7190 & 0.7649 & 0.6876 & 0.7412 & 0.7777 \\
      & n\_estimators: 100, Max Depth: 10 & 0.7086 & 0.7145 & 0.7712 & 0.7003 & 0.7418 & 0.7881 \\
      & n\_estimators: 200, Max Depth: None & \bf{0.7139} & 0.7100 & 0.7815 & 0.7193 & 0.7440 & 0.7872 \\
      & n\_estimators: 200, Max Depth: 5 & 0.7064 & \bf{0.7210} & 0.7643 & 0.6856 & 0.7420 & 0.7754 \\
      & \textbf{n\_estimators: 200, Max Depth: 10} & 0.7126 & 0.7169 & 0.7755 & 0.7066 & 0.7451 & \bf{0.7886} \\
    \midrule
    \multirow{9}{*}{Gradient Boosting}
      & n\_estimators: 50, learning\_rate: 0.1 & 0.7191 & 0.7079 & 0.7906 & 0.7349 & 0.7470 & 0.7940 \\
      & n\_estimators: 50, learning\_rate: 0.01 & 0.6674 & 0.6831 & 0.7313 & 0.6451 & 0.7064 & 0.7524 \\
      & n\_estimators: 50, learning\_rate: 0.001 & 0.6637 & 0.6910 & 0.7227 & 0.6251 & 0.7065 & 0.7262 \\
      & n\_estimators: 100, learning\_rate: 0.1 & 0.7266 & 0.7190 & \bf{0.7947} & \bf{0.7374} & 0.7549 & 0.8000 \\
      & n\_estimators: 100, learning\_rate: 0.01 & 0.6829 & 0.6741 & 0.7578 & 0.6954 & 0.7135 & 0.7627 \\
      & n\_estimators: 100, learning\_rate: 0.001 & 0.6637 & 0.6910 & 0.7227 & 0.6251 & 0.7065 & 0.7292 \\
      & \textbf{n\_estimators: 200, learning\_rate: 0.1} & \bf{0.7274} & \bf{0.7248} & 0.7921 & 0.7310 & \bf{0.7570} & \bf{0.8023} \\
      & n\_estimators: 200, learning\_rate: 0.01 & 0.7080 & 0.6924 & 0.7839 & 0.7301 & 0.7353 & 0.7790 \\
      & n\_estimators: 200, learning\_rate: 0.001 & 0.6637 & 0.6910 & 0.7227 & 0.6251 & 0.7065 & 0.7319 \\
    \midrule
    \multirow{6}{*}{SVM}
      & Kernel: linear, C: 0.1 & 0.7187 & 0.7121 & 0.7874 & 0.7281 & 0.7478 & 0.7933 \\
      & Kernel: linear, C: 1 & 0.7185 & 0.7124 & 0.7868 & 0.7271 & 0.7478 & 0.7935 \\
      & Kernel: linear, C: 10 & 0.7189 & 0.7124 & 0.7874 & 0.7281 & 0.7480 & 0.7935 \\
      & Kernel: RBF, C: 0.1 & 0.7207 & 0.7124 & 0.7901 & \bf{0.7325} & 0.7493 & 0.7941 \\
      & \textbf{Kernel: RBF, C: 1 }& \bf{0.7282} & 0.7259 & \bf{0.7926} & 0.7315 & 0.7578 & \bf{0.7998} \\
      & Kernel: RBF, C: 10 & 0.7266 & \bf{0.7348} & 0.7847 & 0.7149 & \bf{0.7590} & 0.7961 \\
    \midrule
    \multirow{9}{*}{XGBoost}
      & \textbf{n\_estimators: 50, learning\_rate: 0.1 }& \bf{0.7234} & \bf{0.7252} & \bf{0.7860} & \bf{0.7208} & \bf{0.7543} & \bf{0.7961} \\
      & n\_estimators: 50, learning\_rate: 0.01 & 0.7013 & 0.6979 & 0.7705 & 0.7061 & 0.7324 & 0.7708 \\
      & n\_estimators: 50, learning\_rate: 0.001 & 0.6965 & 0.7003 & 0.7622 & 0.6910 & 0.7299 & 0.7610 \\
      & n\_estimators: 100, learning\_rate: 0.1 & 0.7185 & 0.7217 & 0.7810 & 0.7140 & 0.7502 & 0.7958 \\
      & n\_estimators: 100, learning\_rate: 0.01 & 0.7035 & 0.6955 & 0.7752 & 0.7149 & 0.7332 & 0.7779 \\
      & n\_estimators: 100, learning\_rate: 0.001 & 0.6979 & 0.7041 & 0.7620 & 0.6891 & 0.7319 & 0.7628 \\
      & n\_estimators: 200, learning\_rate: 0.1 & 0.7179 & 0.7231 & 0.7793 & 0.7105 & 0.7502 & 0.7902 \\
      & n\_estimators: 200, learning\_rate: 0.01 & 0.7159 & 0.7141 & 0.7819 & 0.7183 & 0.7465 & 0.7877 \\
      & n\_estimators: 200, learning\_rate: 0.001 & 0.6997 & 0.7176 & 0.7571 & 0.6744 & 0.7368 & 0.7648 \\
    \midrule
    \multirow{9}{*}{MLP}
      & HLS: (100,), Activation: relu & 0.6876 & 0.6848 & 0.7584 & 0.6915 & 0.7197 & 0.7475 \\
      & HLS: (100,), Activation: logistic & 0.6928 & 0.7324 & 0.7404 & 0.6368 & 0.7364 & 0.7487 \\
      & \textbf{HLS: (100,), Activation: tanh} & \bf{0.7234} & \bf{0.7528} & 0.7698 & 0.6817 & \bf{0.7612} & \bf{0.7920} \\
      & HLS: (50, 50), Activation: relu & 0.6793 & 0.7145 & 0.7317 & 0.6295 & 0.7230 & 0.7364 \\
      & HLS: (50, 50), Activation: logistic & 0.6688 & 0.6824 & 0.7335 & 0.6495 & 0.7070 & 0.7291 \\
      & HLS: (50, 50), Activation: tanh & 0.7135 & 0.7176 & \bf{0.7763} & \bf{0.7076} & 0.7458 & 0.7873 \\
      & HLS: (100, 50, 25), Activation: relu & 0.6391 & 0.6154 & 0.7265 & 0.6725 & 0.6664 & 0.6934 \\
      & HLS: (100, 50, 25), Activation: logistic & 0.6415 & 0.6475 & 0.7138 & 0.6329 & 0.6790 & 0.6962 \\
      & HLS: (100, 50, 25), Activation: tanh & 0.6989 & 0.7079 & 0.7613 & 0.6861 & 0.7336 & 0.7698 \\
    \midrule
    \multirow{9}{*}{LightGBM}
      & \textbf{n\_estimators: 50, learning\_rate: 0.1} & \bf{0.7209} & 0.7152 & \bf{0.7887} & 0.7291 & \bf{0.7501} & \bf{0.7965} \\
      & n\_estimators: 50, learning\_rate: 0.01 & 0.6967 & 0.6637 & 0.7852 & \bf{0.7432} & 0.7194 & 0.7694 \\
      & n\_estimators: 50, learning\_rate: 0.001 & 0.6924 & 0.6665 & 0.7767 & 0.7291 & 0.7174 & 0.7615 \\
      & n\_estimators: 100, learning\_rate: 0.1 & 0.7187 & \bf{0.7179} & 0.7837 & 0.7198 & 0.7494 & 0.7955 \\
      & n\_estimators: 100, learning\_rate: 0.01 & 0.7033 & 0.6820 & 0.7835 & 0.7335 & 0.7292 & 0.7767 \\
      & n\_estimators: 100, learning\_rate: 0.001 & 0.6906 & 0.6575 & 0.7797 & 0.7374 & 0.7134 & 0.7611 \\
      & n\_estimators: 200, learning\_rate: 0.1 & 0.7175 & 0.7162 & 0.7830 & 0.7193 & 0.7481 & 0.7926 \\
      & n\_estimators: 200, learning\_rate: 0.01 & 0.7100 & 0.6955 & 0.7849 & 0.7306 & 0.7375 & 0.7873 \\
      & n\_estimators: 200, learning\_rate: 0.001 & 0.6902 & 0.6568 & 0.7795 & 0.7374 & 0.7130 & 0.7624 \\
    \bottomrule
  \end{tabular}
  \end{adjustbox}
\end{table}

\begin{table}[h]
  \centering
  \caption{This table presents a comprehensive overview of eight ML models evaluated across a range of hyperparameters to distinguish between high and medium performance levels among students.}
  \begin{adjustbox}{width=0.80\textwidth, keepaspectratio}
  \label{tab_model_hyperparameters_high_medium}
  \begin{tabular}{llcccccc}
    \toprule
    \textbf{Model} & \textbf{Hyperparameters} & \textbf{ACC} & \textbf{RC} & \textbf{PR} & \textbf{SP} & \textbf{F1S} & \textbf{AUC} \\
    \midrule
    \multirow{6}{*}{Logistic Regression} 
      & Penalty: L1, C: 0.1 & 0.6670 & 0.6658 & 0.2104 & 0.6672 & 0.3198 & 0.7229 \\
      & Penalty: L1, C: 1 & 0.6826 & 0.6632 & 0.2191 & 0.6851 & 0.3294 & 0.7328 \\
      & \textbf{Penalty: L1, C: 10} & \bf{0.6874} & 0.6632 & \bf{0.2222} & \bf{0.6907} & \bf{0.3329} & \bf{0.7340} \\
      & Penalty: L2, C: 0.1 & 0.6756 & \bf{0.6683} & 0.2159 & 0.6765 & 0.3263 & 0.7289 \\
      & Penalty: L2, C: 1 & 0.6838 & \bf{0.6683} & 0.2208 & 0.6858 & 0.3320 & 0.7327 \\
      & Penalty: L2, C: 10 & 0.6865 & 0.6580 & 0.2206 & 0.6903 & 0.3305 & 0.7339 \\
    \midrule
    \multirow{6}{*}{Decision Tree}
      & Criterion: gini, Max Depth: None & 0.5769 & 0.5440 & 0.1475 & 0.5812 & 0.2321 & 0.5626 \\
      & Criterion: gini, Max Depth: 5 & 0.6643 & 0.6269 & 0.2016 & 0.6693 & 0.3051 & 0.6828 \\
      & Criterion: gini, Max Depth: 10 & 0.5799 & \bf{0.6424} & 0.1665 & 0.5716 & 0.2645 & 0.5902 \\
      & Criterion: entropy, Max Depth: None & 0.5945 & 0.5621 & 0.1573 & 0.5988 & 0.2458 & 0.5805 \\
      & \textbf{Criterion: entropy, Max Depth: 5} & \bf{0.6746} & 0.6243 & \bf{0.2070} & \bf{0.6813} & \bf{0.3109} & \bf{0.6868} \\
      & Criterion: entropy, Max Depth: 10 & 0.6192 & 0.6217 & 0.1785 & 0.6189 & 0.2774 & 0.6266 \\
    \midrule
    \multirow{9}{*}{Random Forest}
      & n\_estimators: 50, Max Depth: None & 0.6899 & 0.6243 & 0.2163 & 0.6986 & 0.3213 & 0.7187 \\
      & n\_estimators: 50, Max Depth: 5 & 0.6762 & 0.6373 & 0.2104 & 0.6813 & 0.3163 & 0.7140 \\
      & n\_estimators: 50, Max Depth: 10 & 0.6698 & 0.6580 & 0.2106 & 0.6713 & 0.3190 & 0.7162 \\
      & n\_estimators: 100, Max Depth: None & \bf{0.6947} & 0.6269 & \bf{0.2200} & \bf{0.7038} & 0.3257 & 0.7223 \\
      & n\_estimators: 100, Max Depth: 5 & 0.6819 & 0.6632 & 0.2188 & 0.6845 & 0.3290 & 0.7192 \\
      & n\_estimators: 100, Max Depth: 10 & 0.6756 & 0.6735 & 0.2168 & 0.6758 & 0.3280 & 0.7217 \\
      & n\_estimators: 200, Max Depth: None & 0.6862 & 0.6373 & 0.2165 & 0.6927 & 0.3232 & 0.7208 \\
      & n\_estimators: 200, Max Depth: 5 & 0.6643 & 0.6735 & 0.2103 & 0.6630 & 0.3205 & 0.7143 \\
      & \textbf{n\_estimators: 200, Max Depth: 10} & 0.6783 & \bf{0.6761} & 0.2189 & 0.6786 & \bf{0.3307} & \bf{0.7267} \\
    \midrule
    \multirow{9}{*}{Gradient Boosting}
      & n\_estimators: 50, learning\_rate: 0.1 & 0.6890 & 0.6373 & 0.2182 & 0.6958 & 0.3251 & \bf{0.7285} \\
      & \textbf{n\_estimators: 50, learning\_rate: 0.01} & \bf{0.7069} & 0.6010 & \bf{0.2230} & \bf{0.7210} & 0.3253 & 0.7112 \\
      & n\_estimators: 50, learning\_rate: 0.001 & 0.6405 & \bf{0.6632} & 0.1960 & 0.6375 & 0.3026 & 0.6973 \\
      & n\_estimators: 100, learning\_rate: 0.1 & 0.6883 & 0.6424 & 0.2188 & 0.6945 & \bf{0.3265} & 0.7273 \\
      & n\_estimators: 100, learning\_rate: 0.01 & 0.6676 & 0.6476 & 0.2074 & 0.6703 & 0.3142 & 0.7155 \\
      & n\_estimators: 100, learning\_rate: 0.001 & 0.6661 & 0.6580 & 0.2085 & 0.6672 & 0.3167 & 0.7006 \\
      & n\_estimators: 200, learning\_rate: 0.1 & 0.6850 & 0.6424 & 0.2167 & 0.6907 & 0.3241 & 0.7242 \\
      & n\_estimators: 200, learning\_rate: 0.01 & 0.6816 & 0.6450 & 0.2152 & 0.6865 & 0.3227 & 0.7249 \\
      & n\_estimators: 200, learning\_rate: 0.001 & 0.6661 & 0.6580 & 0.2085 & 0.6672 & 0.3167 & 0.7050 \\
    \midrule

    \multirow{6}{*}{SVM}
      & Kernel: linear, C: 0.1 & 0.6789 & 0.6683 & 0.2179 & 0.6803 & 0.3286 & 0.7337 \\
      & Kernel: linear, C: 1 & 0.6777 & 0.6735 & 0.2181 & 0.6782 & \bf{0.3295} & 0.7337 \\
      & Kernel: linear, C: 10 & 0.6783 & 0.6709 & 0.2180 & 0.6793 & 0.3290 & \bf{0.7341} \\
      & Kernel: RBF, C: 0.1 & 0.6414 & \bf{0.6943} & 0.2019 & 0.6344 & 0.3129 & 0.7153 \\
      & Kernel: RBF, C: 1 & 0.6695 & 0.6658 & 0.2118 & 0.6700 & 0.3214 & 0.7325 \\
      & \textbf{Kernel: RBF, C: 10} & \bf{0.6859} & 0.6528 & \bf{0.2193} & \bf{0.6903} & 0.3283 & 0.7286 \\
    \midrule
    \multirow{9}{*}{XGBoost}
      & \textbf{n\_estimators: 50, learning\_rate: 0.1} & \bf{0.6853} & \bf{0.6528} & \bf{0.2189} & 0.6896 & \bf{0.3279} & \bf{0.7230} \\
      & n\_estimators: 50, learning\_rate: 0.01 & 0.6692 & 0.6269 & 0.2043 & 0.6748 & 0.3082 & 0.7091 \\
      & n\_estimators: 50, learning\_rate: 0.001 & 0.6408 & 0.6036 & 0.1850 & 0.6458 & 0.2832 & 0.6663 \\
      & n\_estimators: 100, learning\_rate: 0.1 & 0.6847 & 0.6347 & 0.2151 & \bf{0.6914} & 0.3213 & 0.7150 \\
      & n\_estimators: 100, learning\_rate: 0.01 & 0.6704 & 0.6217 & 0.2040 & 0.6769 & 0.3072 & 0.7130 \\
      & n\_estimators: 100, learning\_rate: 0.001 & 0.6527 & 0.6010 & 0.1904 & 0.6596 & 0.2892 & 0.6832 \\
      & n\_estimators: 200, learning\_rate: 0.1 & 0.6731 & 0.6269 & 0.2066 & 0.6793 & 0.3108 & 0.7055 \\
      & n\_estimators: 200, learning\_rate: 0.01 & 0.6707 & 0.6398 & 0.2077 & 0.6748 & 0.3136 & 0.7203 \\
      & n\_estimators: 200, learning\_rate: 0.001 & 0.6625 & 0.6243 & 0.2001 & 0.6675 & 0.3031 & 0.6975 \\
    \midrule
    \multirow{9}{*}{MLP}
      & HLS: (100,), Activation: relu & 0.5881 & 0.6632 & 0.1732 & 0.5781 & 0.2746 & 0.6714 \\
      & HLS: (100,), Activation: logistic & \bf{0.6954} & 0.5803 & 0.2109 & \bf{0.7107} & 0.3093 & 0.6877 \\
      & \textbf{HLS: (100,), Activation: tanh} & 0.6554 & 0.7227 & 0.2141 & 0.6465 & \bf{0.3303} & \bf{0.7368} \\
      & HLS: (50, 50), Activation: relu & 0.5851 & 0.6010 & 0.1611 & 0.5830 & 0.2541 & 0.6327 \\
      & HLS: (50, 50), Activation: logistic & 0.6347 & 0.5699 & 0.1755 & 0.6434 & 0.2684 & 0.6489 \\
      & HLS: (50, 50), Activation: tanh & 0.6890 & 0.6502 & \bf{0.2207} & 0.6941 & 0.3296 & 0.7327 \\
      & HLS: (100, 50, 25), Activation: relu & 0.6296 & 0.5673 & 0.1727 & 0.6379 & 0.2648 & 0.6374 \\
      & HLS: (100, 50, 25), Activation: logistic & 0.6305 & 0.5647 & 0.1726 & 0.6392 & 0.2644 & 0.6331 \\
      & HLS: (100, 50, 25), Activation: tanh & 0.6073 & \bf{0.7616} & 0.1971 & 0.5868 & 0.3132 & 0.7290 \\
    \midrule
    \multirow{9}{*}{LightGBM}
      & n\_estimators: 50, learning\_rate: 0.1 & 0.6804 & 0.6424 & 0.2139 & 0.6855 & 0.3210 & 0.7158 \\
      & n\_estimators: 50, learning\_rate: 0.01 & \bf{0.6865} & 0.6139 & 0.2121 & 0.6962 & 0.3153 & 0.7139 \\
      & n\_estimators: 50, learning\_rate: 0.001 & 0.6716 & 0.5829 & 0.1970 & 0.6834 & 0.2945 & 0.6852 \\
      & n\_estimators: 100, learning\_rate: 0.1 & 0.6807 & 0.6269 & 0.2111 & 0.6879 & 0.3159 & 0.7121 \\
      & n\_estimators: 100, learning\_rate: 0.01 & 0.6850 & 0.6398 & 0.2162 & 0.6910 & 0.3232 & 0.7188 \\
      & n\_estimators: 100, learning\_rate: 0.001 & 0.6783 & 0.5958 & 0.2035 & 0.6893 & 0.3034 & 0.6934 \\
      & n\_estimators: 200, learning\_rate: 0.1 & 0.6743 & 0.6450 & 0.2108 & 0.6782 & 0.3178 & 0.7052 \\
      & \textbf{n\_estimators: 200, learning\_rate: 0.01} & 0.6819 & \bf{0.6528} & \bf{0.2168} & 0.6858 & \bf{0.3255} & \bf{0.7213} \\
      & n\_estimators: 200, learning\_rate: 0.001 & 0.7014 & 0.5673 & 0.2122 & \bf{0.7193} & 0.3088 & 0.7004 \\
    \bottomrule
  \end{tabular}
  \end{adjustbox}
\end{table}

\begin{table}[h]
  \centering
  \caption{This table presents a comprehensive overview of eight ML models evaluated across a range of hyperparameters to distinguish between low and high performance levels among students.}
  \begin{adjustbox}{width=0.80\textwidth, keepaspectratio}
  \label{tab_model_hyperparameters_low_high}
  \begin{tabular}{llcccccc}
    \toprule
    \textbf{Model} & \textbf{Hyperparameters} & \textbf{ACC} & \textbf{RC} & \textbf{PR} & \textbf{SP} & \textbf{F1S} & \textbf{AUC} \\
    \midrule
    \multirow{6}{*}{Logistic Regression} & Penalty: L1, C: 0.1 & 0.8427 & 0.8601 & 0.6342 & 0.5023 & 0.8394 & 0.9295 \\
                                          & Penalty: L1, C: 1 & 0.8526 & 0.8731 & 0.6525 & 0.5209 & 0.8487 & 0.9336 \\
                                          & \textbf{Penalty: L1, C: 10} & \bf{0.8546} & \bf{0.8756} & \bf{0.6563} & \bf{0.5248} & \bf{0.8507} & 0.9335 \\
                                          & Penalty: L2, C: 0.1 & 0.8435 & 0.8653 & 0.6368 & 0.5038 & 0.8394 & 0.9306 \\
                                          & Penalty: L2, C: 1 & 0.8505 & 0.8653 & 0.6473 & 0.5170 & 0.8477 & 0.9336 \\
                                          & Penalty: L2, C: 10 & 0.8538 & \bf{0.8756} & 0.6550 & 0.5232 & 0.8497 & \bf{0.9337} \\
    \midrule
    \multirow{6}{*}{Decision Tree} & Criterion: gini, Max Depth: None & 0.7581 & 0.7798 & 0.5055 & 0.3739 & 0.7540 & 0.7669 \\
                                    & Criterion: gini, Max Depth: 5 & 0.8263 & \bf{0.8264} & 0.6013 & 0.4726 & 0.8263 & 0.8785 \\
                                    & Criterion: gini, Max Depth: 10 & 0.7848 & 0.8031 & 0.5419 & 0.4090 & 0.7814 & 0.7869 \\
                                    & Criterion: entropy, Max Depth: None & 0.7700 & 0.7850 & 0.5197 & 0.3885 & 0.7672 & 0.7761 \\
                                    & \textbf{Criterion: entropy, Max Depth: 5} & \bf{0.8464} & 0.7979 & \bf{0.6222} & \bf{0.5099} & \bf{0.8555} & \bf{0.8824} \\
                                    & Criterion: entropy, Max Depth: 10 & 0.7893 & 0.8005 & 0.5464 & 0.4148 & 0.7872 & 0.8192 \\
    \midrule
    \multirow{9}{*}{Random Forest} & n\_estimators: 50, Max Depth: None & 0.8522 & 0.8394 & 0.6429 & 0.5209 & 0.8546 & 0.9222 \\
                               & n\_estimators: 50, Max Depth: 5 & 0.8119 & 0.8472 & 0.5881 & 0.4504 & 0.8053 & 0.9112 \\
                               & n\_estimators: 50, Max Depth: 10 & 0.8497 & 0.8627 & 0.6453 & 0.5155 & 0.8472 & 0.9237 \\
                               & \textbf{n\_estimators: 100, Max Depth: None} & \bf{0.8604} & \bf{0.8653} & \bf{0.6627} & \bf{0.5370} & \bf{0.8594} & 0.9264 \\
                               & n\_estimators: 100, Max Depth: 5 & 0.8378 & 0.8523 & 0.6249 & 0.4933 & 0.8350 & 0.9150 \\
                               & n\_estimators: 100, Max Depth: 10 & 0.8435 & 0.8549 & 0.6340 & 0.5038 & 0.8414 & 0.9267 \\
                               & n\_estimators: 200, Max Depth: None & 0.8591 & 0.8575 & 0.6587 & 0.5347 & \bf{0.8594} & \bf{0.9272} \\
                               & n\_estimators: 200, Max Depth: 5 & 0.8361 & 0.8420 & 0.6196 & 0.4902 & 0.8350 & 0.9177 \\
                               & n\_estimators: 200, Max Depth: 10 & 0.8530 & 0.8575 & 0.6490 & 0.5221 & 0.8521 & 0.9268 \\
    \midrule
    \multirow{9}{*}{Gradient Boosting} & n\_estimators: 50, learning\_rate: 0.1 & 0.8546 & 0.8497 & 0.6495 & 0.5256 & 0.8555 & 0.9295 \\
                                       & n\_estimators: 50, learning\_rate: 0.01 & 0.8078 & 0.8290 & 0.5776 & 0.4432 & 0.8038 & 0.8889 \\
                                       & n\_estimators: 50, learning\_rate: 0.001 & 0.7322 & \bf{0.8731} & 0.5083 & 0.3585 & 0.7057 & 0.8663 \\
                                       & n\_estimators: 100, learning\_rate: 0.1 & 0.8600 & \bf{0.8731} & 0.6640 & 0.5358 & 0.8575 & \bf{0.9354} \\
                                       & n\_estimators: 100, learning\_rate: 0.01 & 0.8287 & 0.8264 & 0.6047 & 0.4768 & 0.8292 & 0.8963 \\
                                       & n\_estimators: 100, learning\_rate: 0.001 & 0.7322 & \bf{0.8731} & 0.5083 & 0.3585 & 0.7057 & 0.8663 \\
                                       & \textbf{n\_estimators: 200, learning\_rate: 0.1} & \bf{0.8645} & 0.8627 & \bf{0.6687} & \bf{0.5459} & \bf{0.8648} & 0.9340 \\
                                       & n\_estimators: 200, learning\_rate: 0.01 & 0.8398 & 0.8290 & 0.6214 & 0.4969 & 0.8419 & 0.9106 \\
                                       & n\_estimators: 200, learning\_rate: 0.001 & 0.7561 & 0.8679 & 0.5301 & 0.3815 & 0.7350 & 0.8800 \\
    \midrule
    \multirow{6}{*}{SVM} & Kernel: linear, C: 0.1 & 0.8575 & 0.8679 & 0.6588 & 0.5309 & 0.8555 & 0.9329 \\
                         & Kernel: linear, C: 1 & 0.8542 & 0.8731 & 0.6550 & 0.5241 & 0.8507 & 0.9331 \\
                         & Kernel: linear, C: 10 & 0.8530 & \bf{0.8756} & 0.6538 & 0.5216 & 0.8487 & 0.9333 \\
                         & Kernel: RBF, C: 0.1 & 0.8394 & 0.8549 & 0.6280 & 0.4962 & 0.8365 & 0.9223 \\
                         & Kernel: RBF, C: 1 & 0.8563 & 0.8679 & 0.6569 & 0.5284 & 0.8541 & 0.9356 \\
                         & \textbf{Kernel: RBF, C: 10} & \bf{0.8628} & 0.8705 & \bf{0.6680} & \bf{0.5419} & \bf{0.8614} & \bf{0.9371} \\
    \midrule
    \multirow{9}{*}{XGBoost} & n\_estimators: 50, learning\_rate: 0.1 & 0.8497 & 0.8575 & 0.6440 & 0.5156 & 0.8482 & 0.9281 \\
                             & n\_estimators: 50, learning\_rate: 0.01 & 0.8234 & 0.8394 & 0.6011 & 0.4682 & 0.8204 & 0.8987 \\
                             & n\_estimators: 50, learning\_rate: 0.001 & 0.7975 & 0.8290 & 0.5649 & 0.4284 & 0.7916 & 0.8829 \\
                             & n\_estimators: 100, learning\_rate: 0.1 & 0.8530 & \bf{0.8601} & 0.6497 & 0.5220 & 0.8516 & \bf{0.9301} \\
                             & n\_estimators: 100, learning\_rate: 0.01 & 0.8308 & 0.8368 & 0.6106 & 0.4807 & 0.8297 & 0.9054 \\
                             & n\_estimators: 100, learning\_rate: 0.001 & 0.7992 & 0.8472 & 0.5722 & 0.4320 & 0.7901 & 0.8876 \\
                             & \textbf{n\_estimators: 200, learning\_rate: 0.1} & \bf{0.8542} & \bf{0.8601} & \bf{0.6516} & \bf{0.5245} & \bf{0.8531} & 0.9287 \\
                             & n\_estimators: 200, learning\_rate: 0.01 & 0.8402 & 0.8420 & 0.6256 & 0.4977 & 0.8399 & 0.9150 \\
                             & n\_estimators: 200, learning\_rate: 0.001 & 0.8037 & 0.8497 & 0.5785 & 0.4385 & 0.7950 & 0.8934 \\
    \midrule
    \multirow{9}{*}{MLP} & HLS: (100,), Activation: relu & 0.8181 & 0.8601 & 0.5998 & 0.4605 & 0.8102 & 0.9215 \\
                         & HLS: (100,), Activation: logistic & 0.8464 & 0.8601 & 0.6397 & 0.5092 & 0.8438 & 0.9303 \\
                         & \textbf{HLS: (100,), Activation: tanh} & 0.8567 & \bf{0.8834} & \bf{0.6615} & 0.5287 & 0.8516 & \bf{0.9371} \\
                         & HLS: (50, 50), Activation: relu & 0.8099 & 0.8497 & 0.5862 & 0.4475 & 0.8023 & 0.9015 \\
                         & HLS: (50, 50), Activation: logistic & 0.8353 & 0.8420 & 0.6185 & 0.4887 & 0.8341 & 0.9104 \\
                         & HLS: (50, 50), Activation: tanh & 0.8591 & 0.8679 & 0.6614 & 0.5343 & 0.8575 & 0.9348 \\
                         & HLS: (100, 50, 25), Activation: relu & 0.8160 & 0.8446 & 0.5927 & 0.4566 & 0.8106 & 0.9019 \\
                         & HLS: (100, 50, 25), Activation: logistic & 0.8172 & 0.8368 & 0.5921 & 0.4582 & 0.8136 & 0.9062 \\
                         & HLS: (100, 50, 25), Activation: tanh & \bf{0.8616} & 0.8472 & 0.6599 & \bf{0.5405} & \bf{0.8643} & 0.9322 \\
    \midrule
    \multirow{9}{*}{LightGBM} & \textbf{n\_estimators: 50, learning\_rate: 0.1} & \bf{0.8517} & 0.8472 & \bf{0.6443} & \bf{0.5199} & \bf{0.8526} & 0.9299 \\
                              & n\_estimators: 50, learning\_rate: 0.01 & 0.8292 & 0.8394 & 0.6090 & 0.4779 & 0.8272 & 0.8993 \\
                              & n\_estimators: 50, learning\_rate: 0.001 & 0.8230 & 0.8083 & 0.5915 & 0.4664 & 0.8258 & 0.8856 \\
                              & n\_estimators: 100, learning\_rate: 0.1 & 0.8493 & 0.8472 & 0.6405 & 0.5150 & 0.8497 & \bf{0.9318} \\
                              & n\_estimators: 100, learning\_rate: 0.01 & 0.8349 & 0.8420 & 0.6179 & 0.4880 & 0.8336 & 0.9045 \\
                              & n\_estimators: 100, learning\_rate: 0.001 & 0.8193 & 0.8420 & 0.5963 & 0.4616 & 0.8150 & 0.8879 \\
                              & n\_estimators: 200, learning\_rate: 0.1 & 0.8476 & \bf{0.8497} & 0.6388 & 0.5117 & 0.8472 & 0.9305 \\
                              & n\_estimators: 200, learning\_rate: 0.01 & 0.8439 & 0.8342 & 0.6289 & 0.5047 & 0.8458 & 0.9151 \\
                              & n\_estimators: 200, learning\_rate: 0.001 & 0.8172 & 0.8420 & 0.5936 & 0.4584 & 0.8126 & 0.8931 \\
    \bottomrule
  \end{tabular}
  \end{adjustbox}
\end{table}

\begin{table}[h]
  \centering
  \caption{Comparative analysis of ML models using confusion matrices across three performance scenarios: Low-Medium, High-Medium, and Low-High student achievement levels.}
  \label{tab_confusion_matrices}
  \begin{tabular}{|l|c|c|c|}
    \toprule
    \textbf{Model} & \textbf{Low-Medium} & \textbf{High-Medium} & \textbf{Low-High} \\
    \midrule
    Logistic Regression & \includegraphics[width=0.17\textwidth]{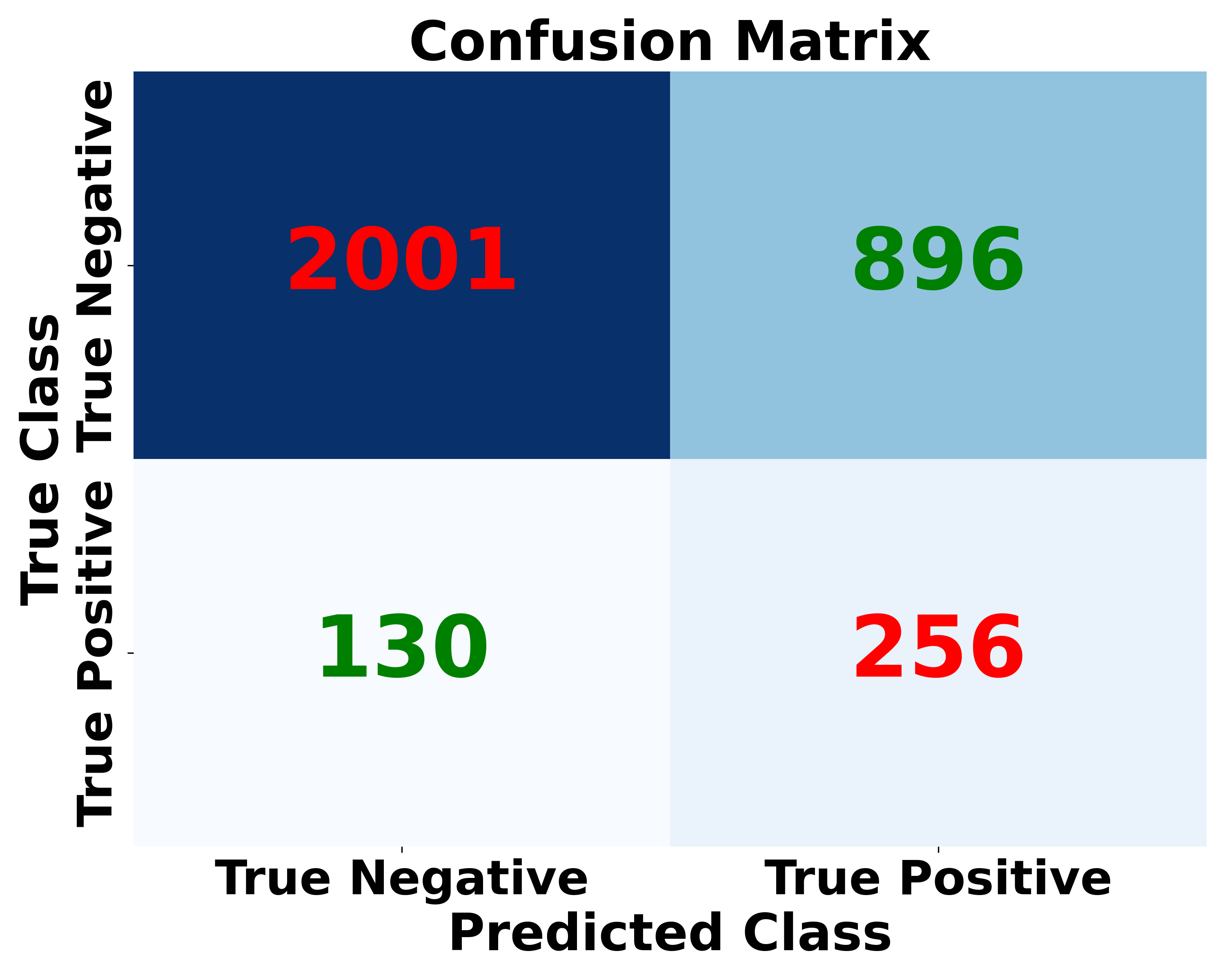} & \includegraphics[width=0.17\textwidth]{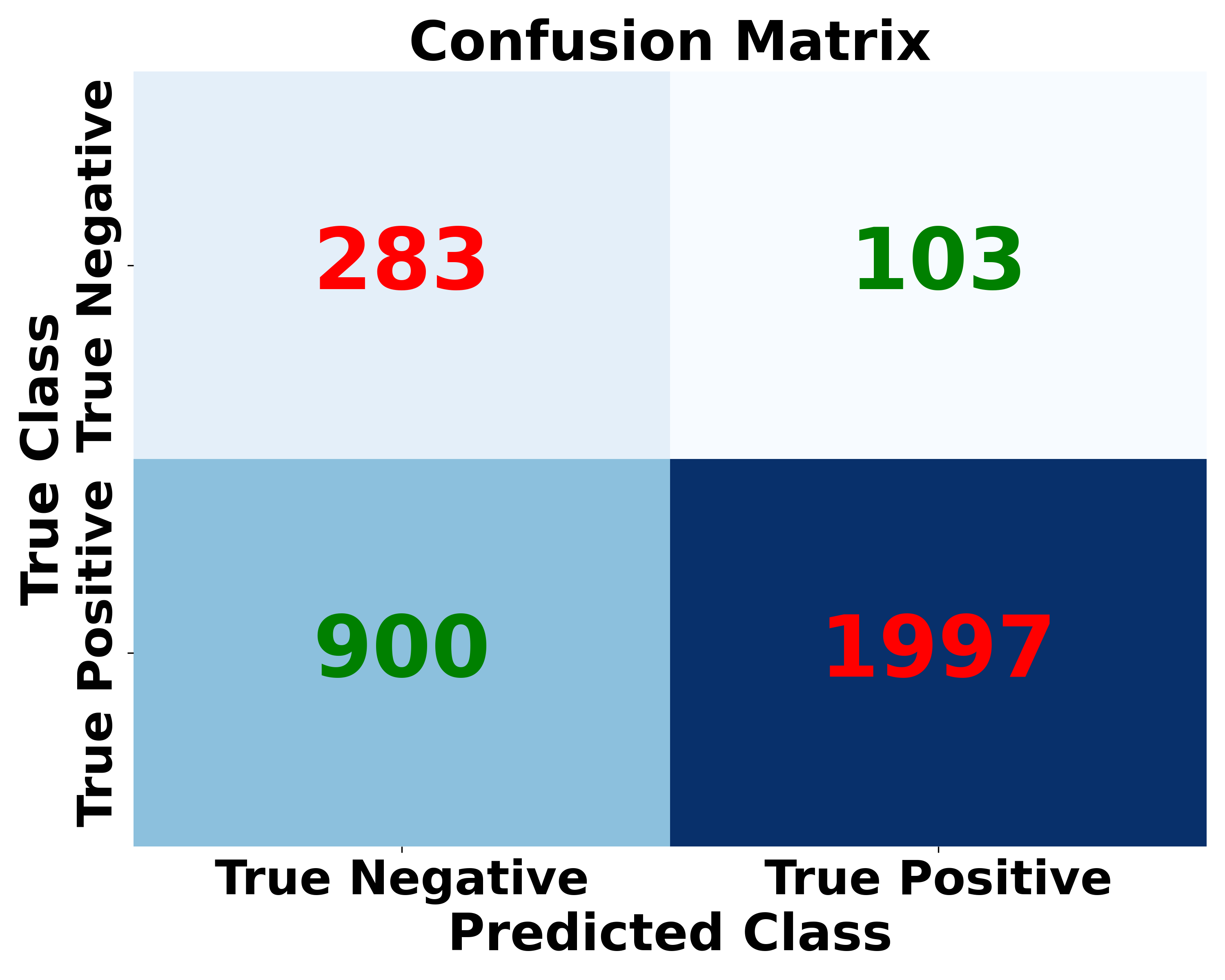} & \includegraphics[width=0.17\textwidth]{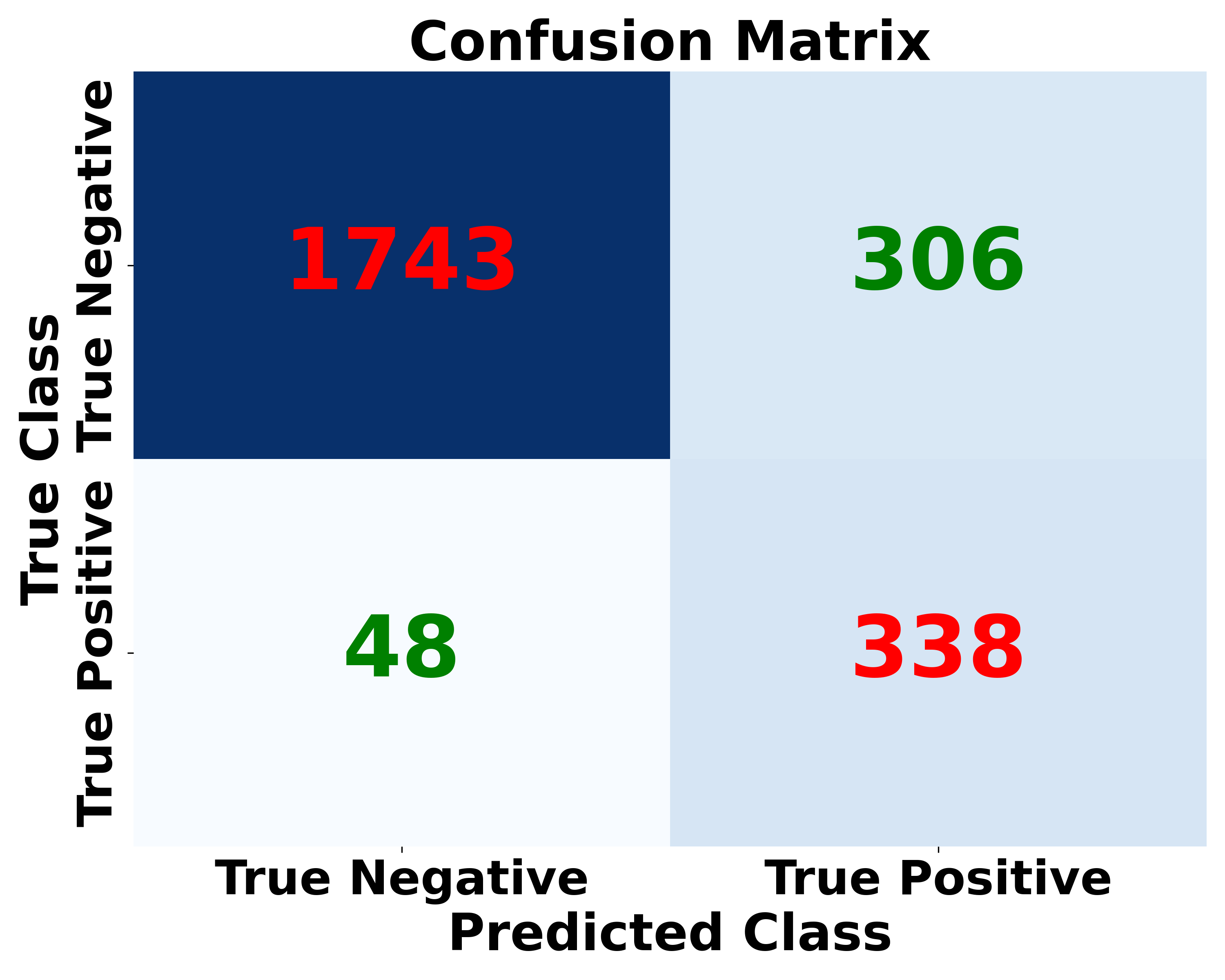} \\
    \midrule
    Decision Tree & \includegraphics[width=0.17\textwidth]{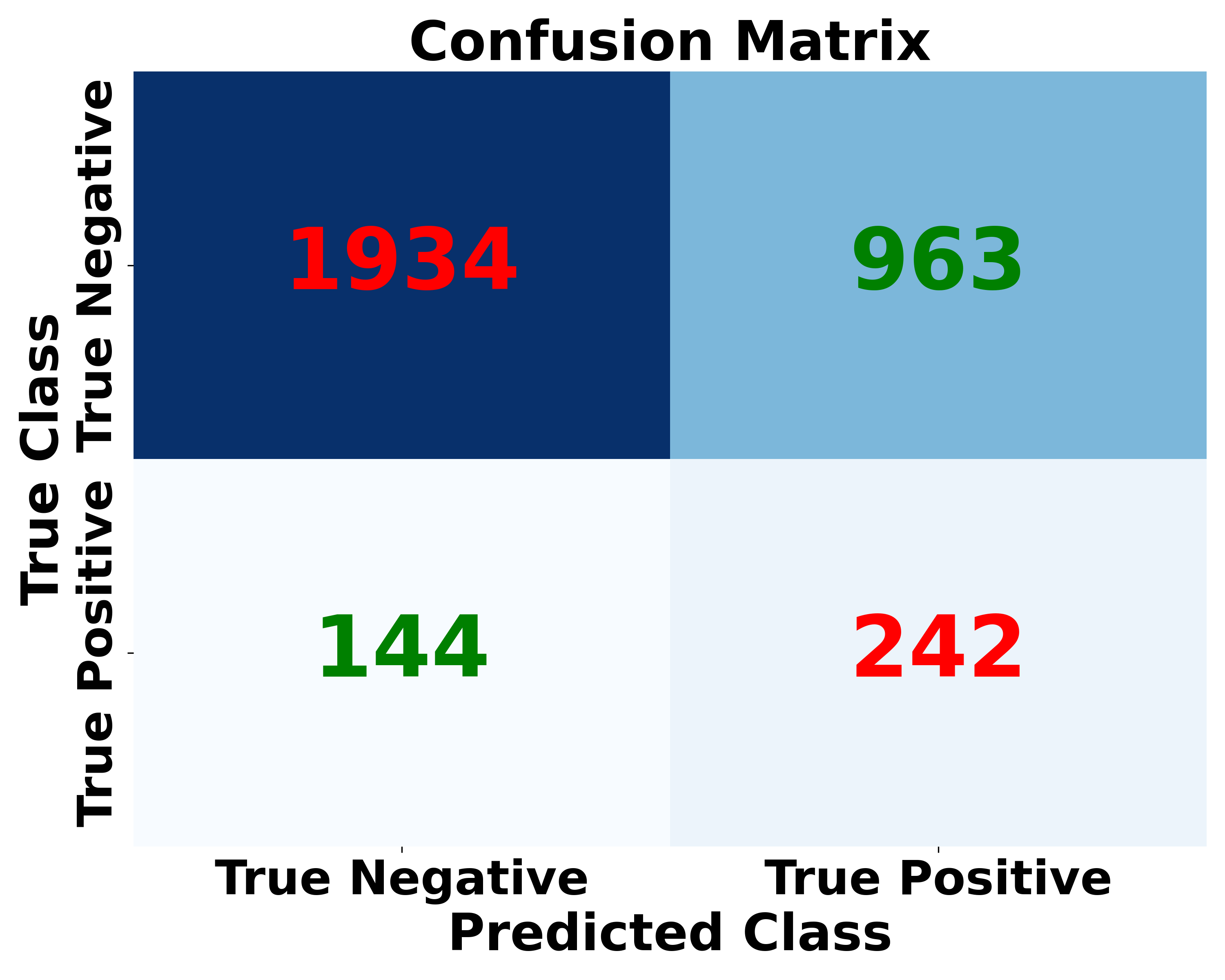} & \includegraphics[width=0.17\textwidth]{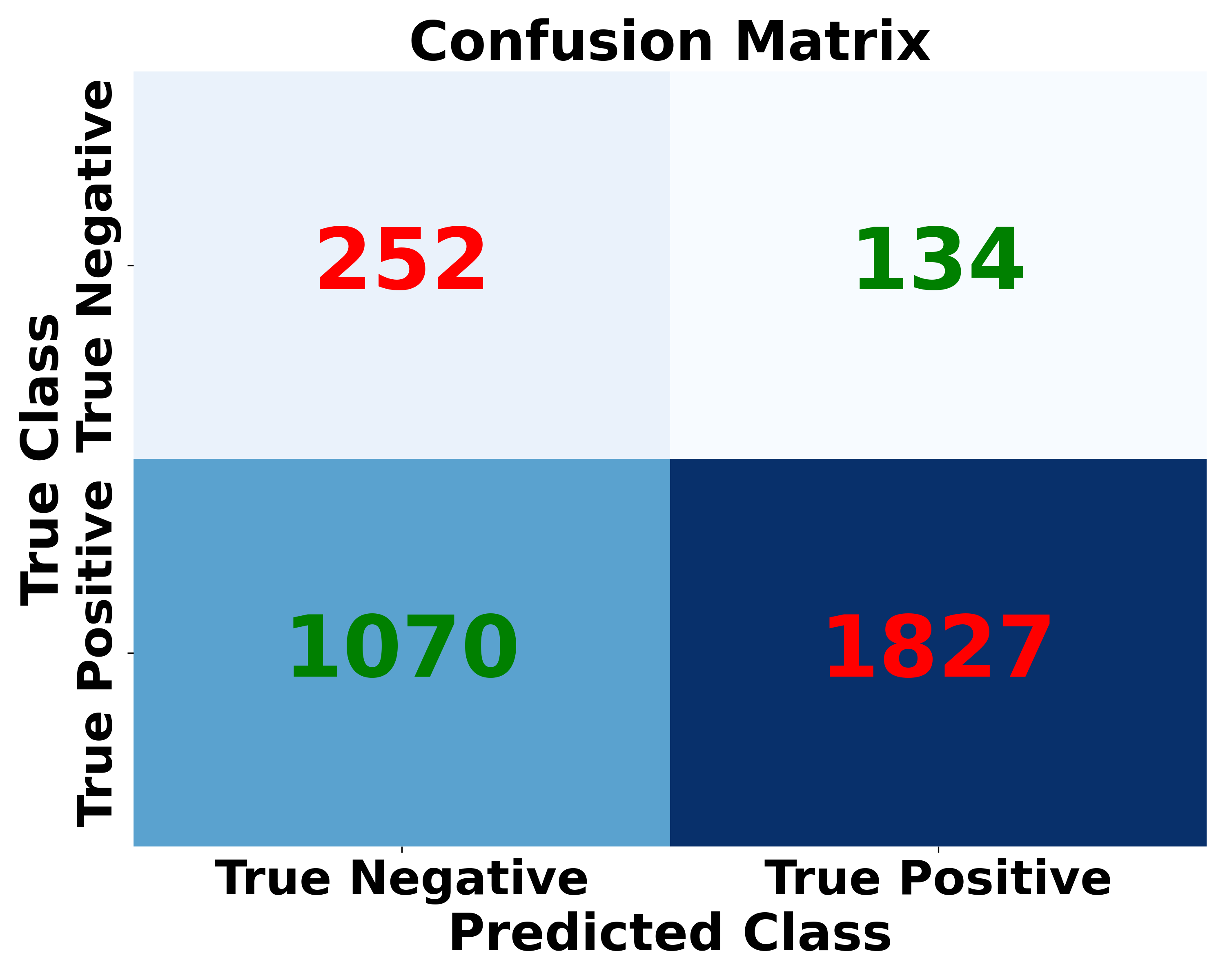} & \includegraphics[width=0.17\textwidth]{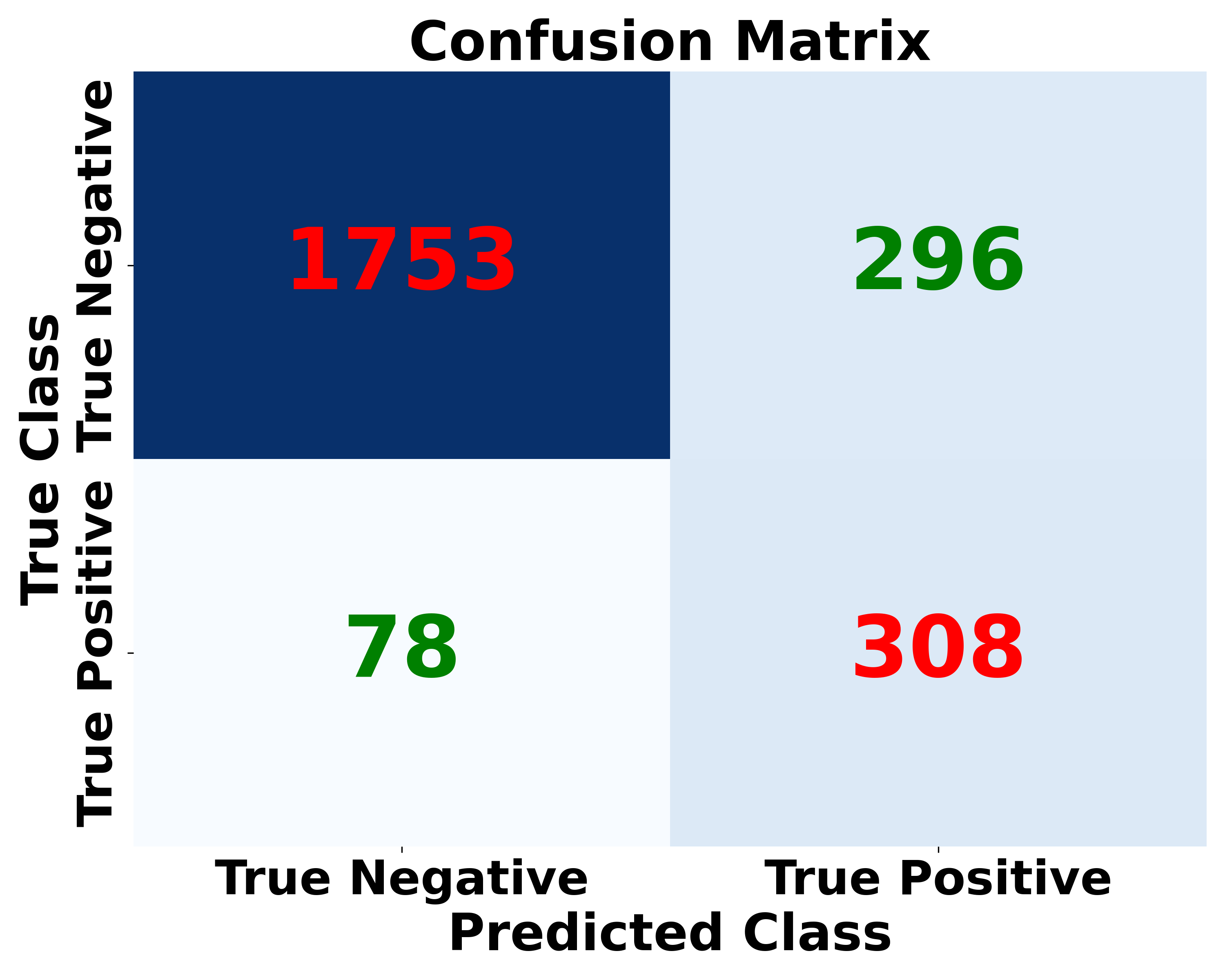} \\
    \midrule
    Random Forest & \includegraphics[width=0.17\textwidth]{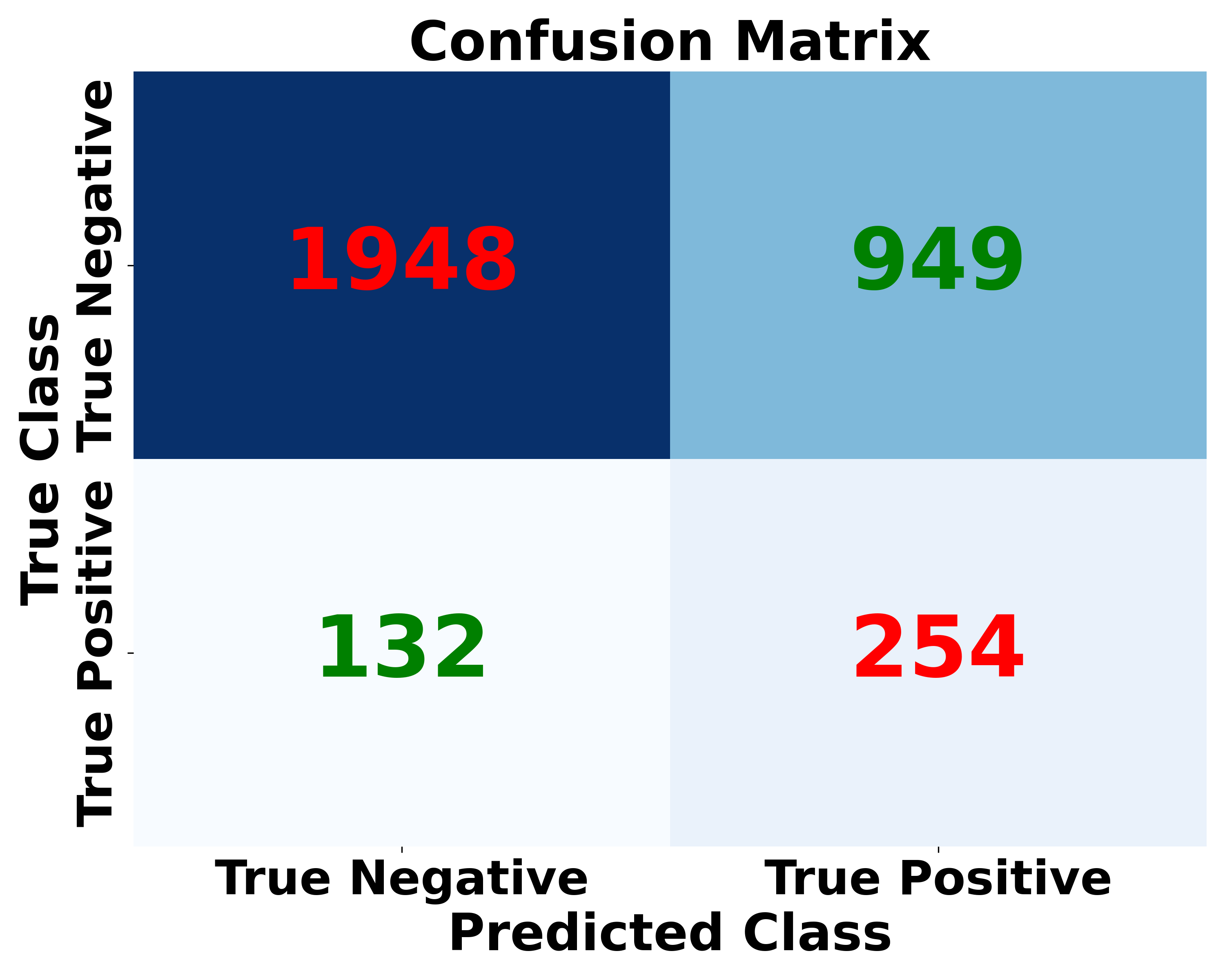} & \includegraphics[width=0.17\textwidth]{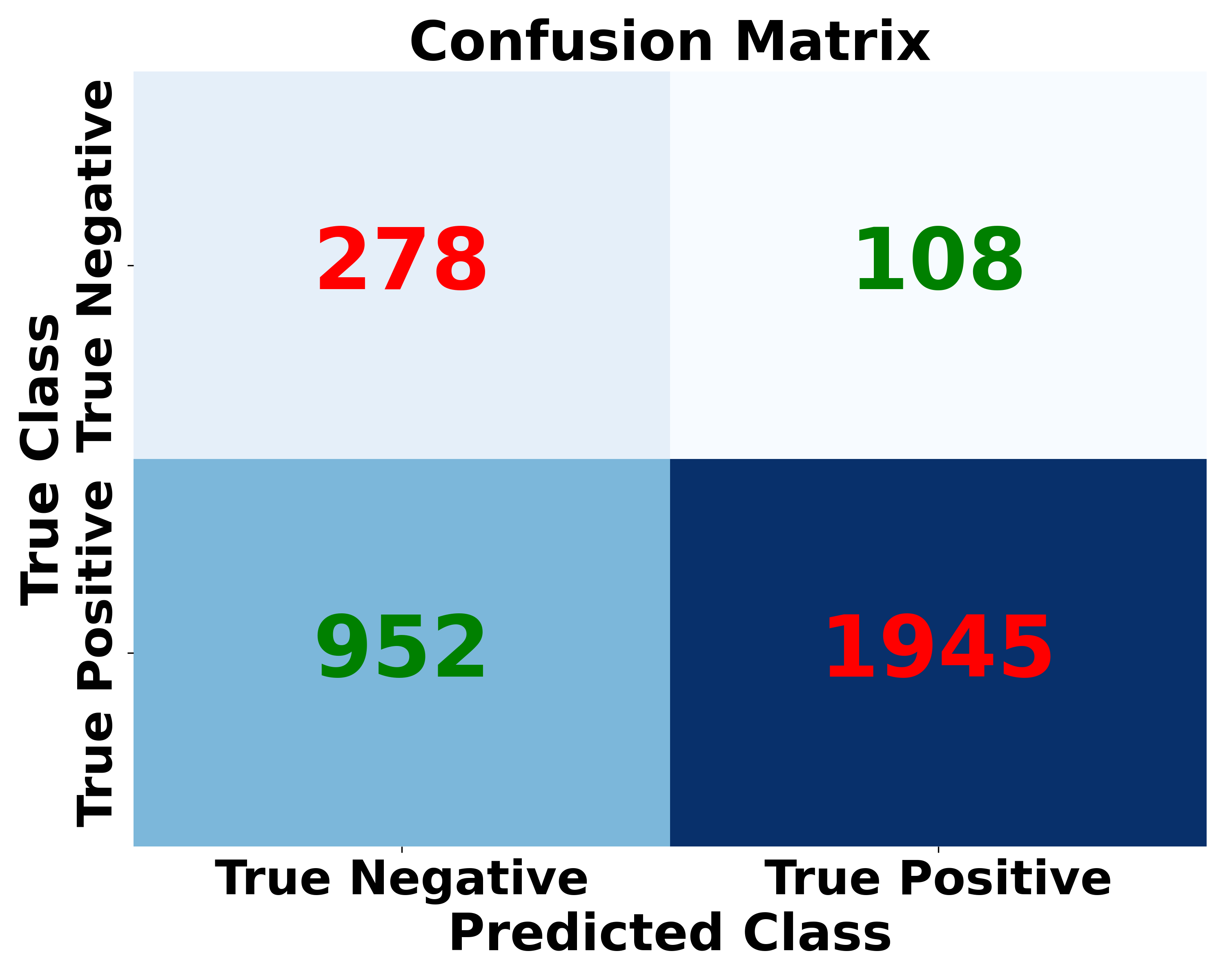} & \includegraphics[width=0.17\textwidth]{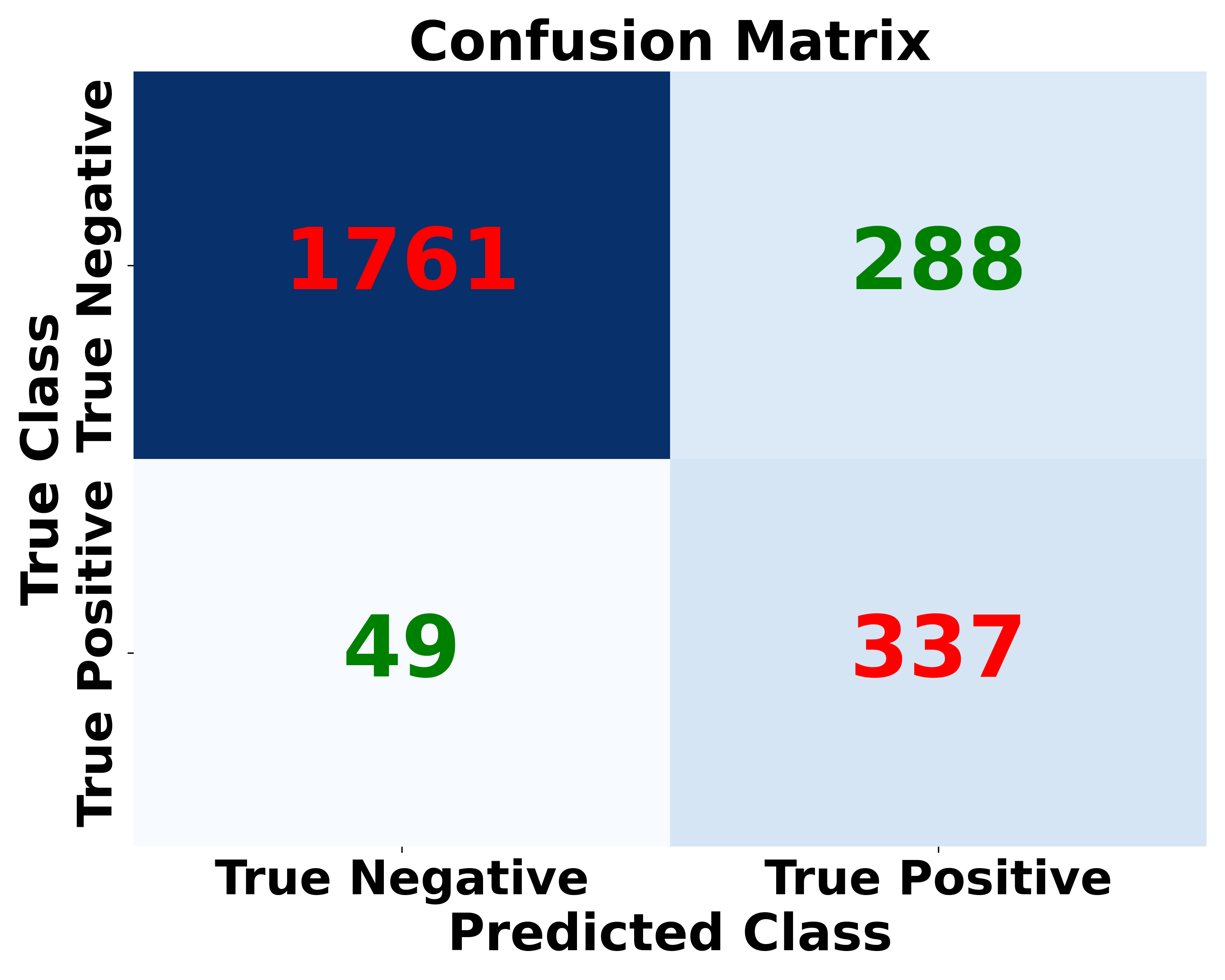} \\
    \midrule
    Gradient Boosting & \includegraphics[width=0.17\textwidth]{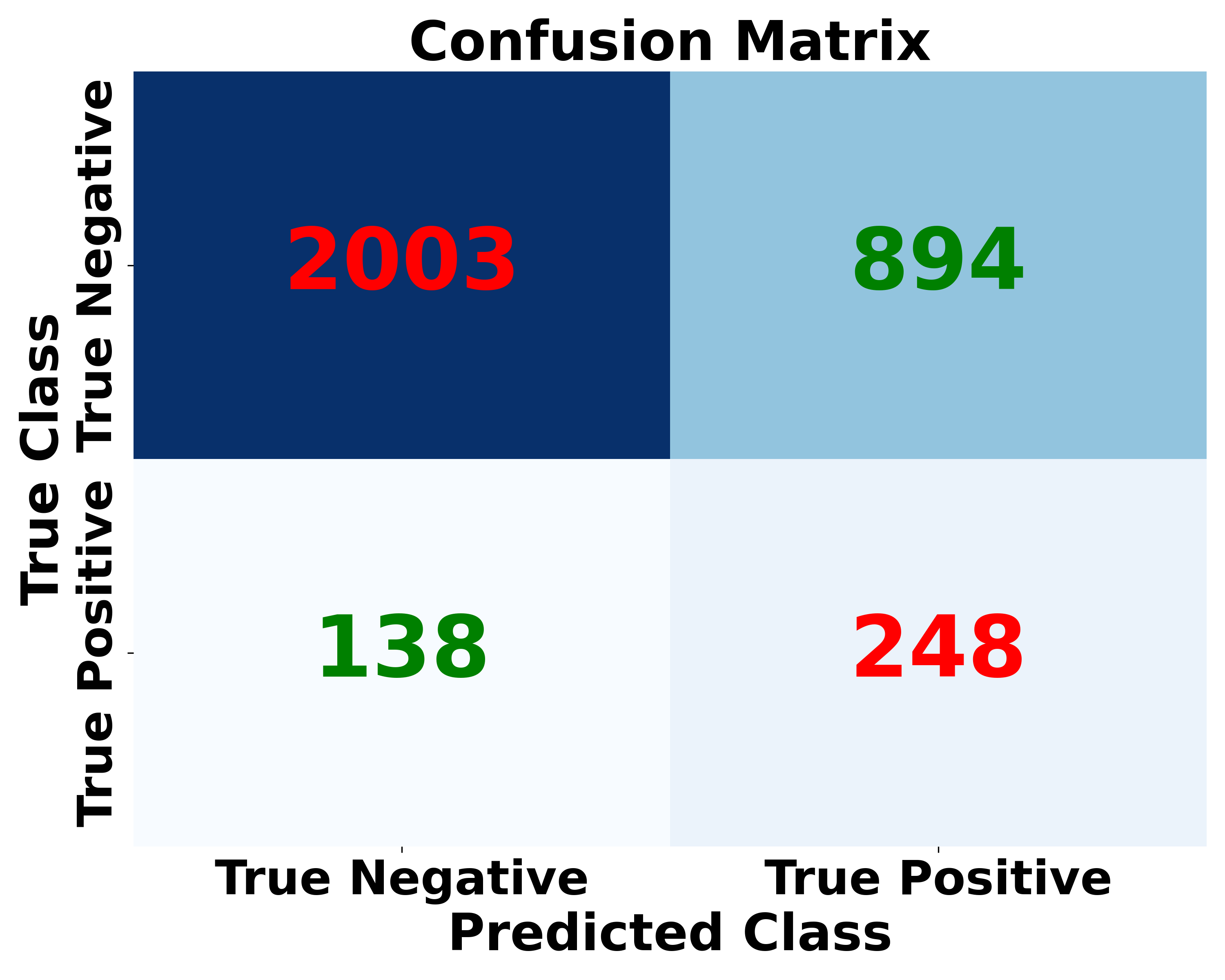} & \includegraphics[width=0.17\textwidth]{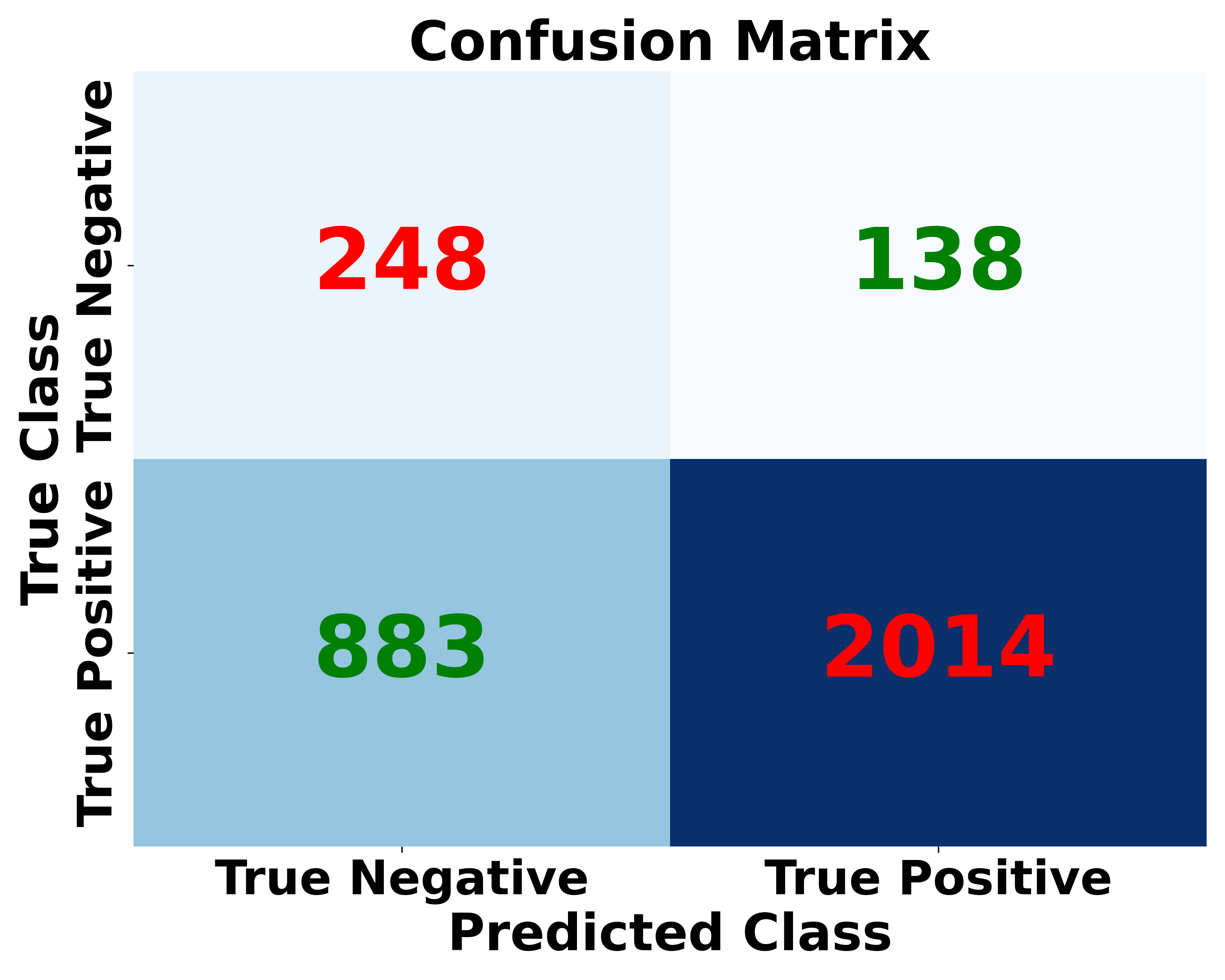} & \includegraphics[width=0.17\textwidth]{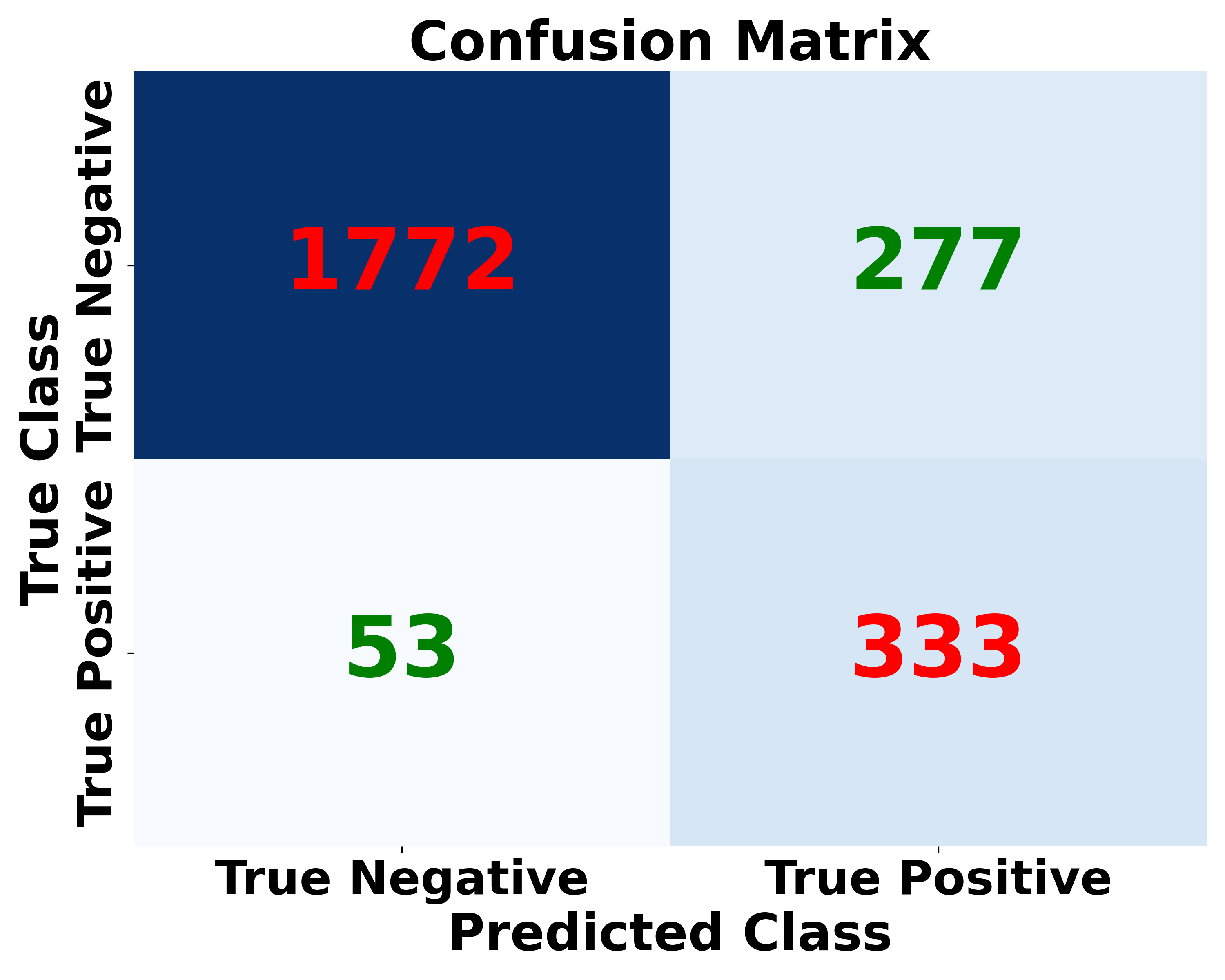} \\
    \midrule
    SVM & \includegraphics[width=0.17\textwidth]{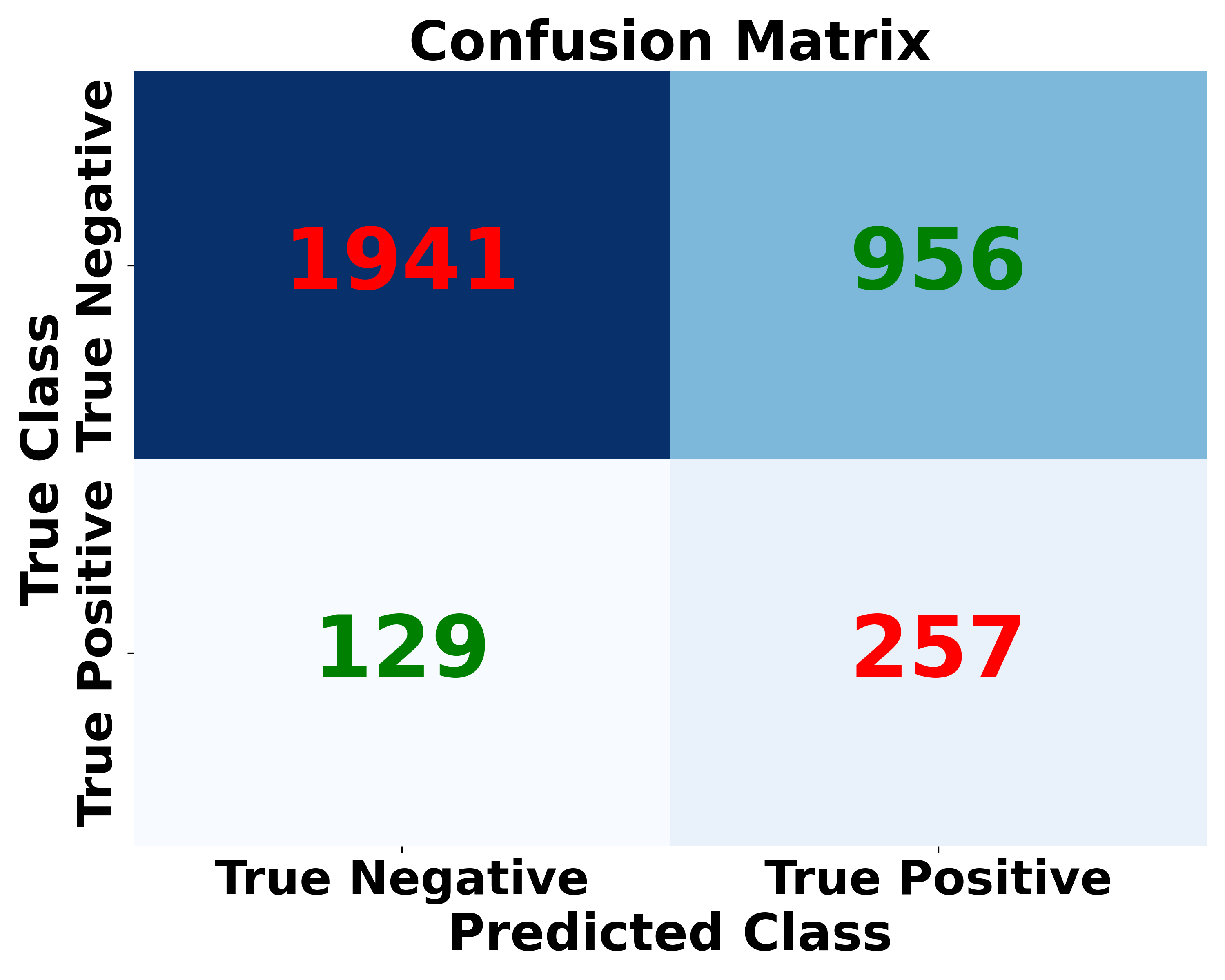} & \includegraphics[width=0.17\textwidth]{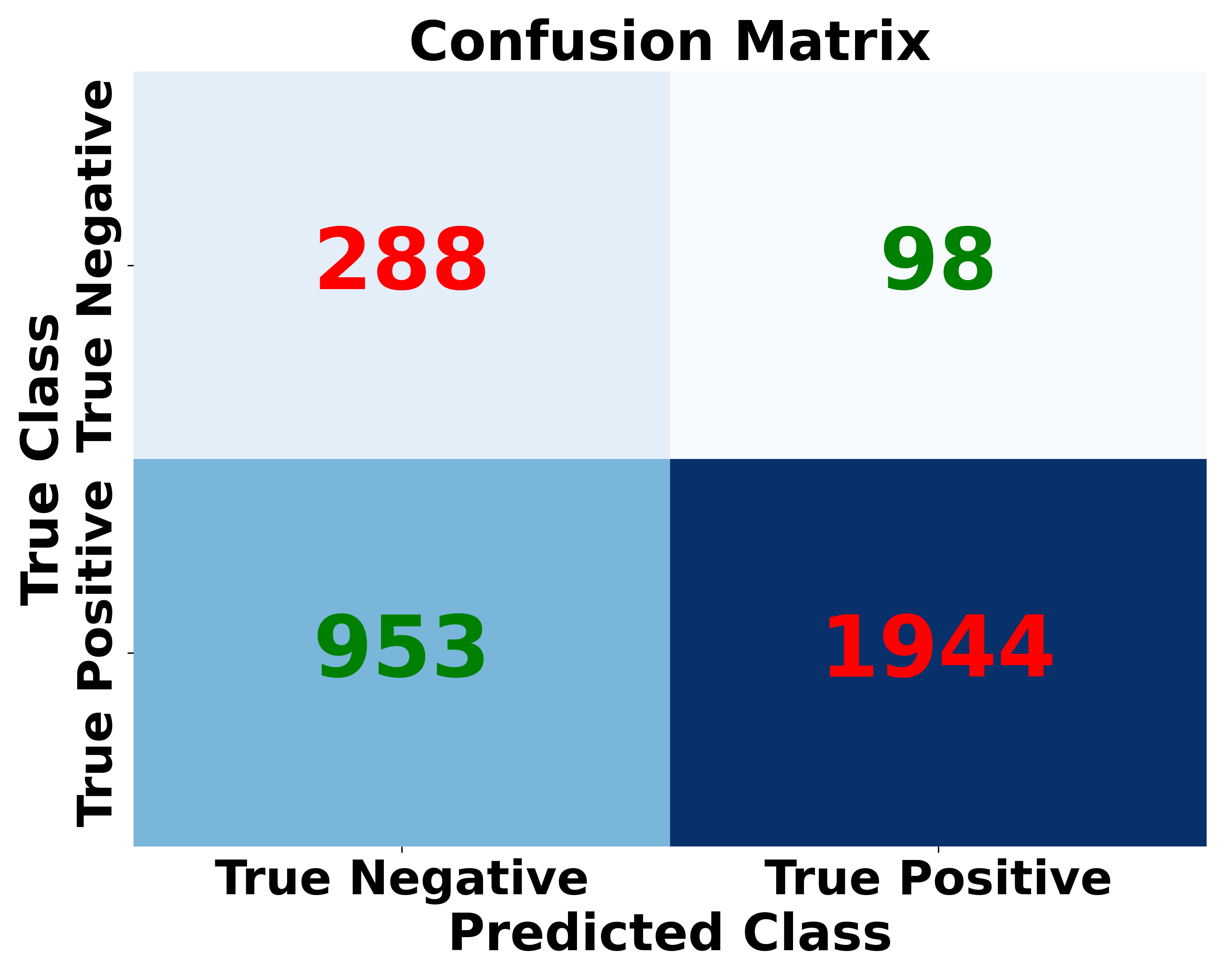} & \includegraphics[width=0.17\textwidth]{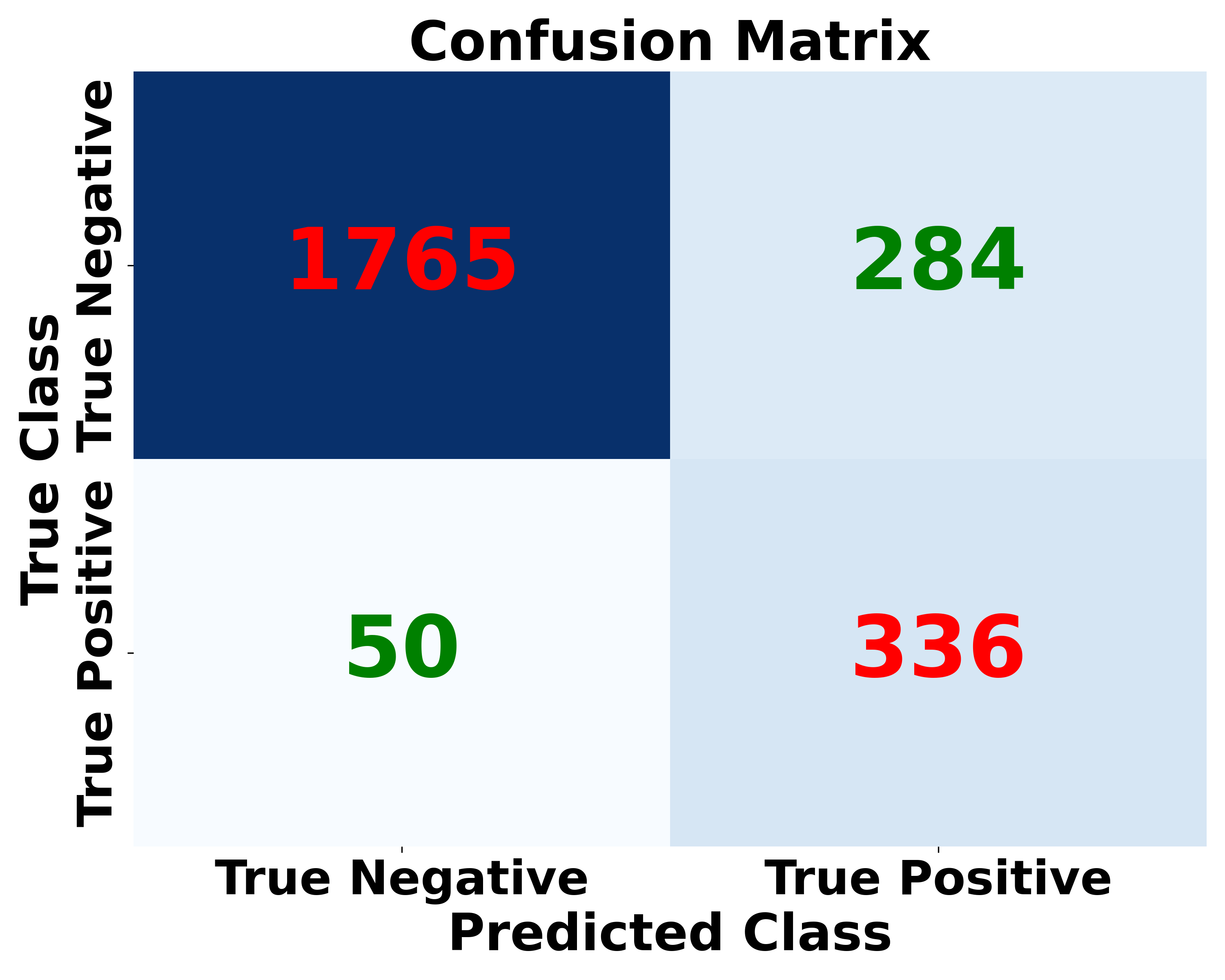} \\
    \midrule
    XGBoost & \includegraphics[width=0.17\textwidth]{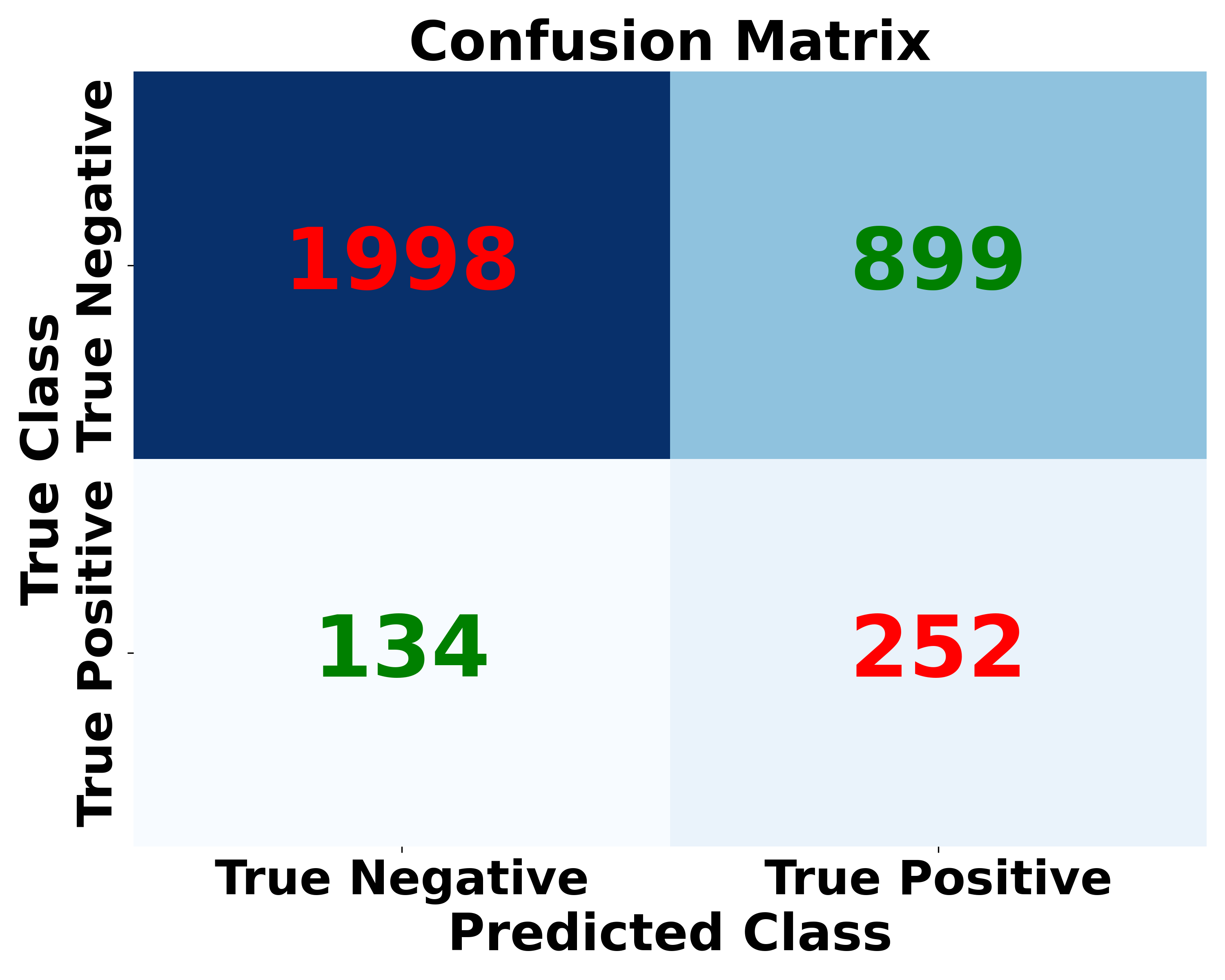} & \includegraphics[width=0.17\textwidth]{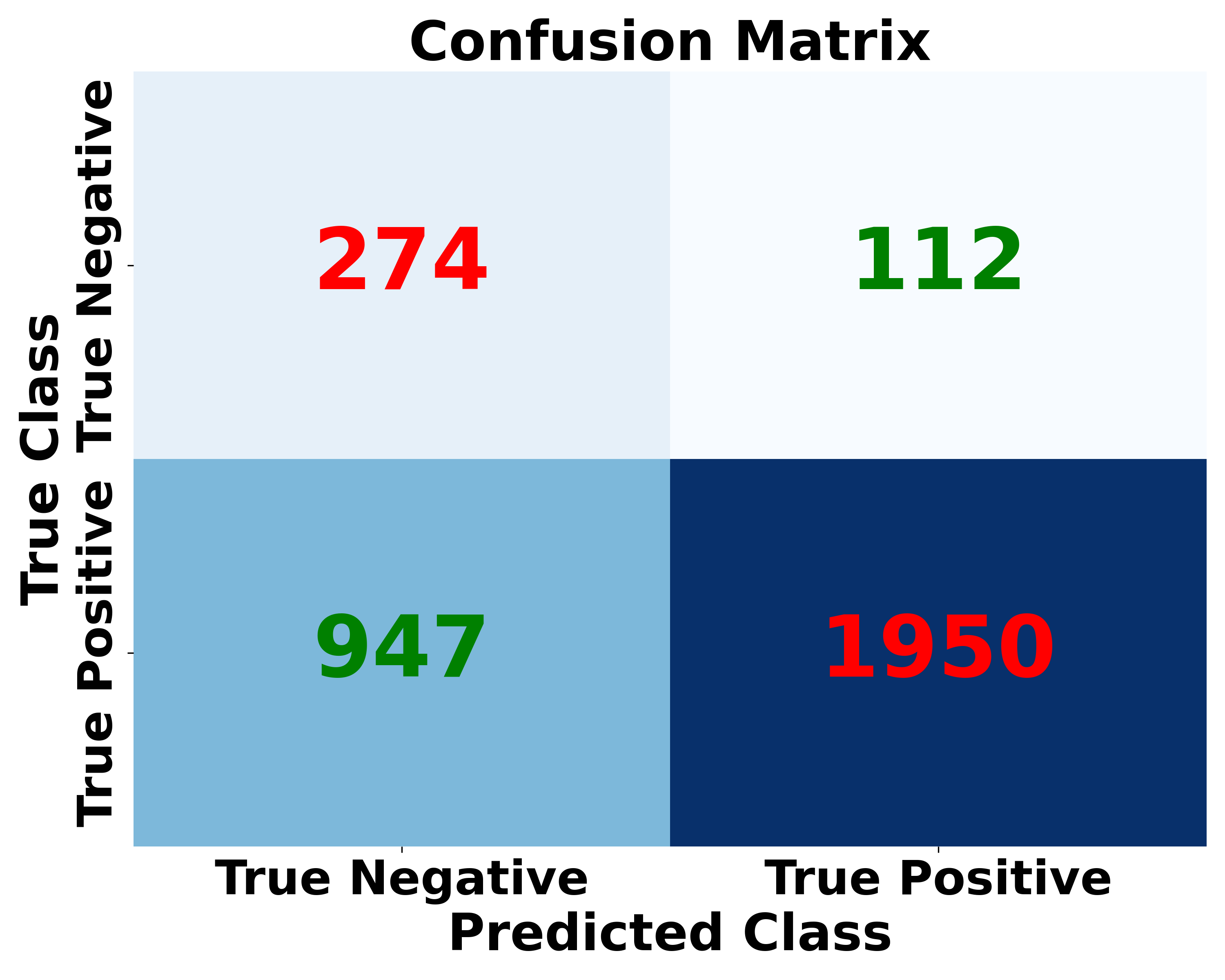} & \includegraphics[width=0.17\textwidth]{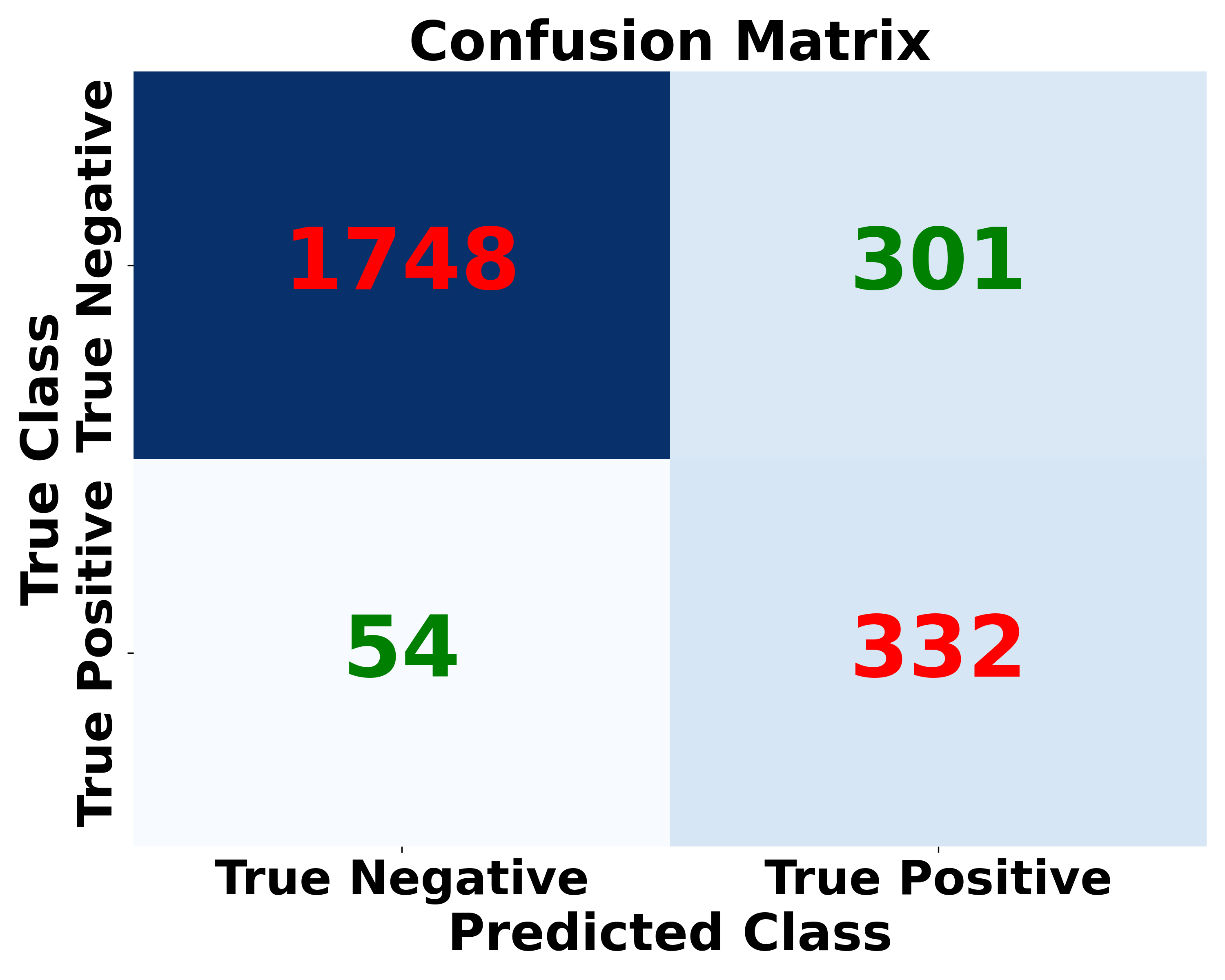} \\
    \midrule
    MLP & \includegraphics[width=0.17\textwidth]{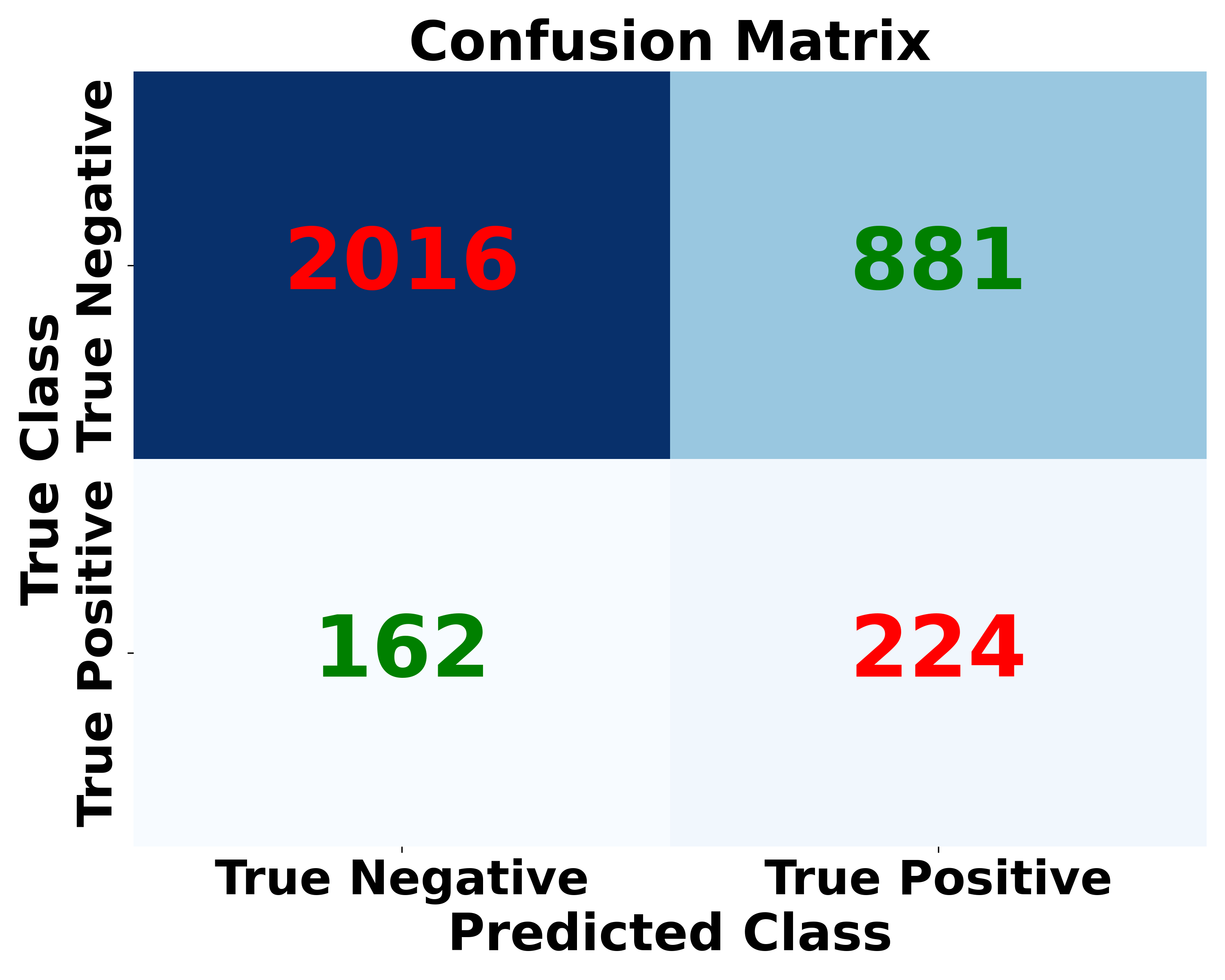} & \includegraphics[width=0.17\textwidth]{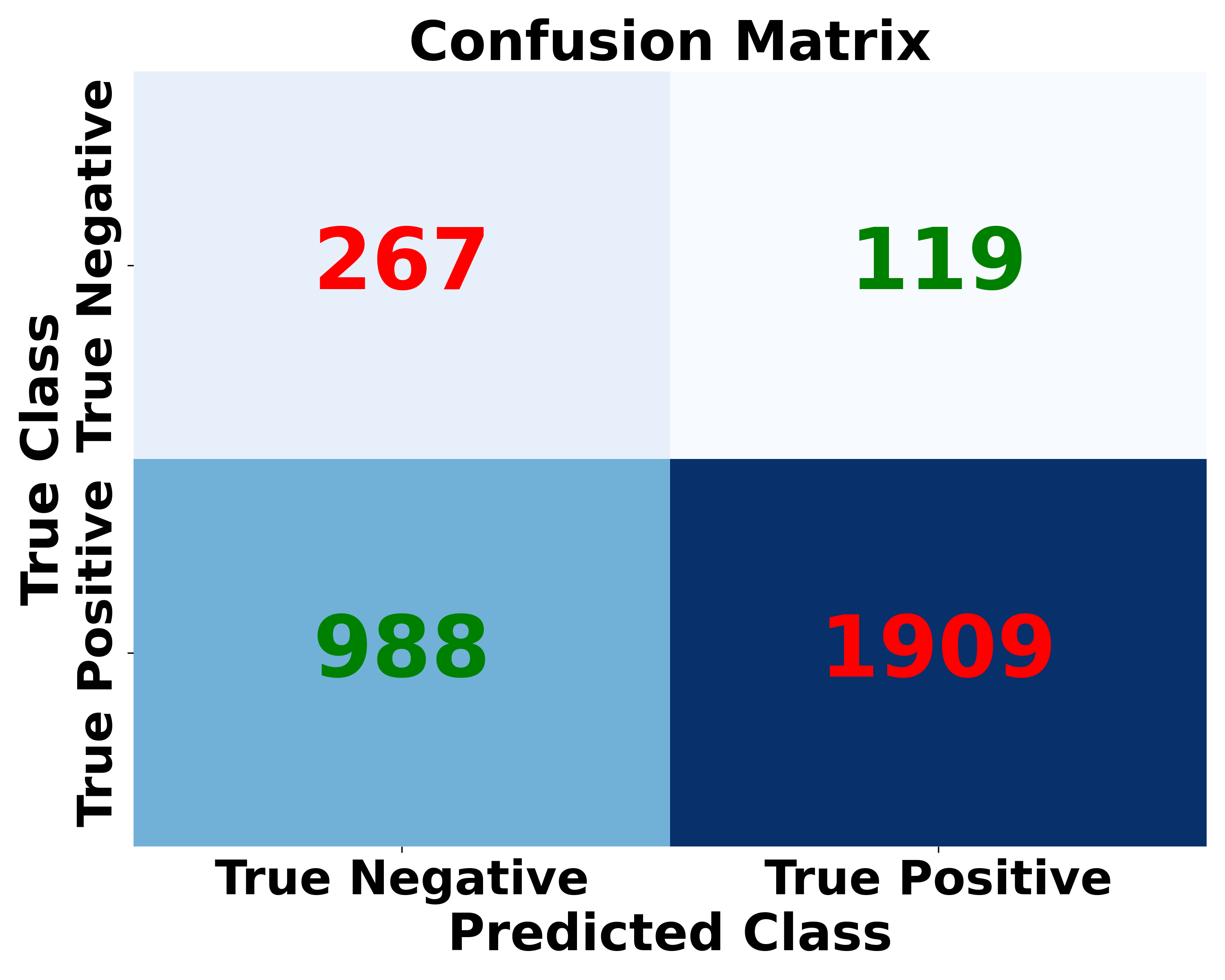} & \includegraphics[width=0.17\textwidth]{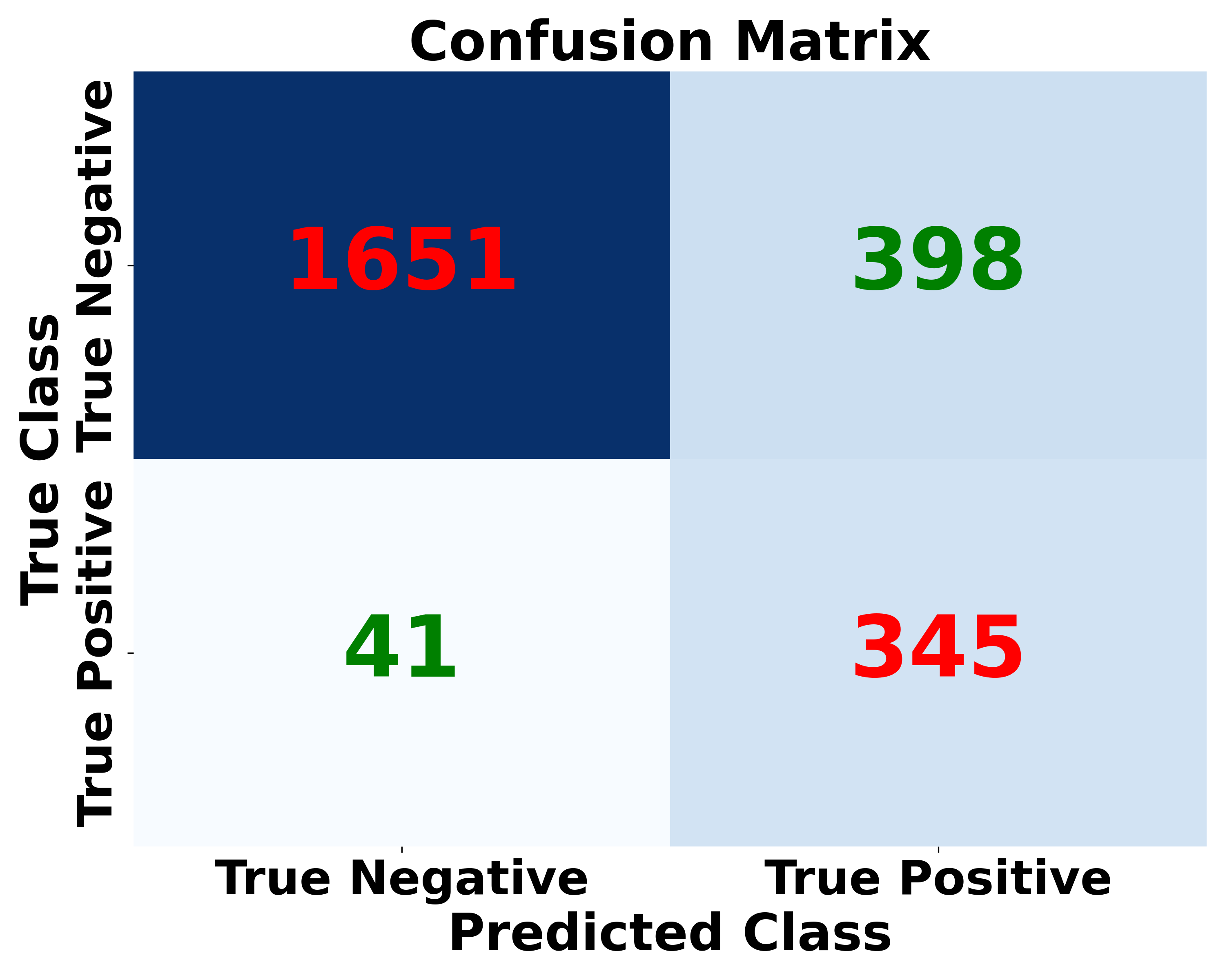} \\
    \midrule
    LightGBM & \includegraphics[width=0.17\textwidth]{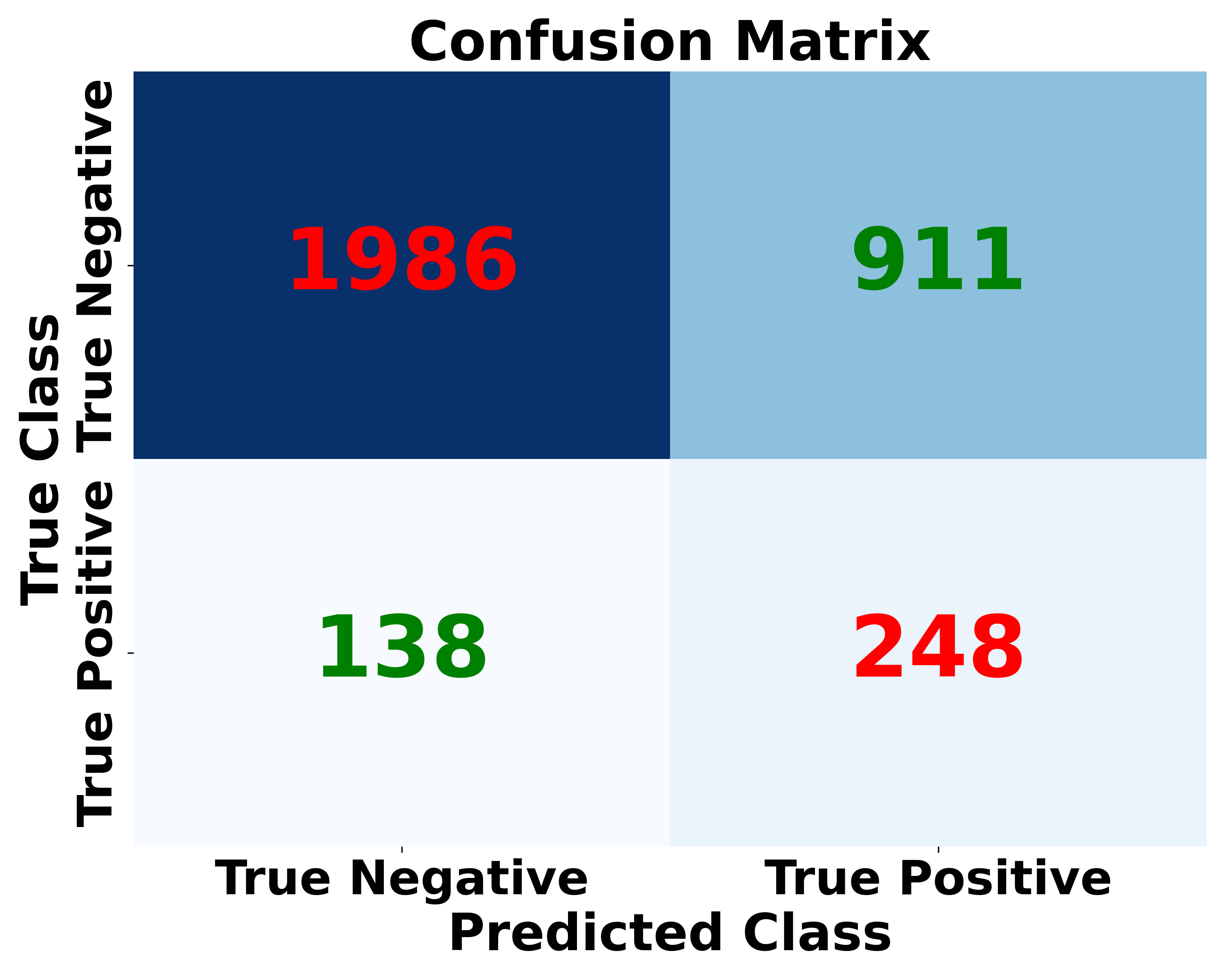} & \includegraphics[width=0.17\textwidth]{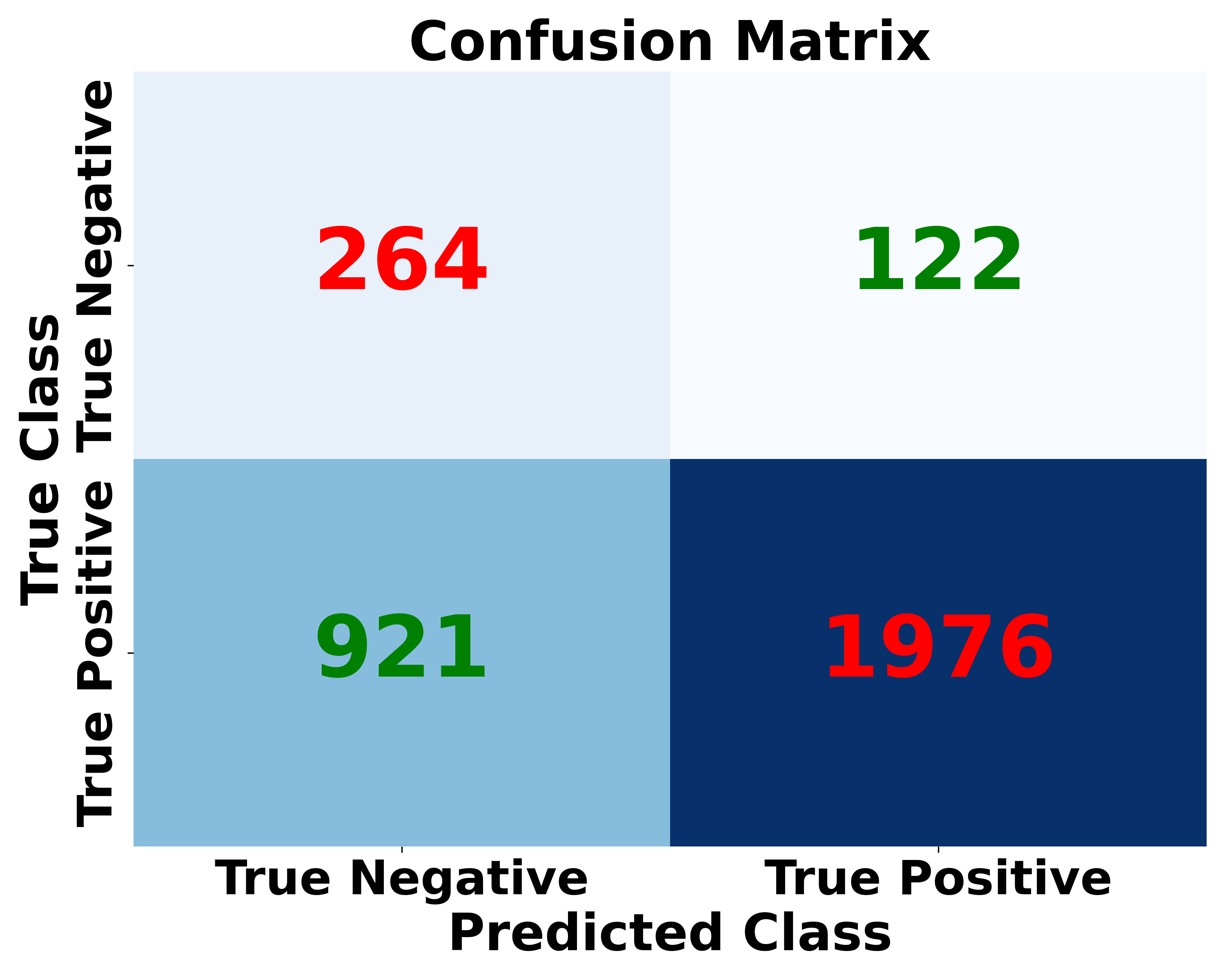} & \includegraphics[width=0.17\textwidth]{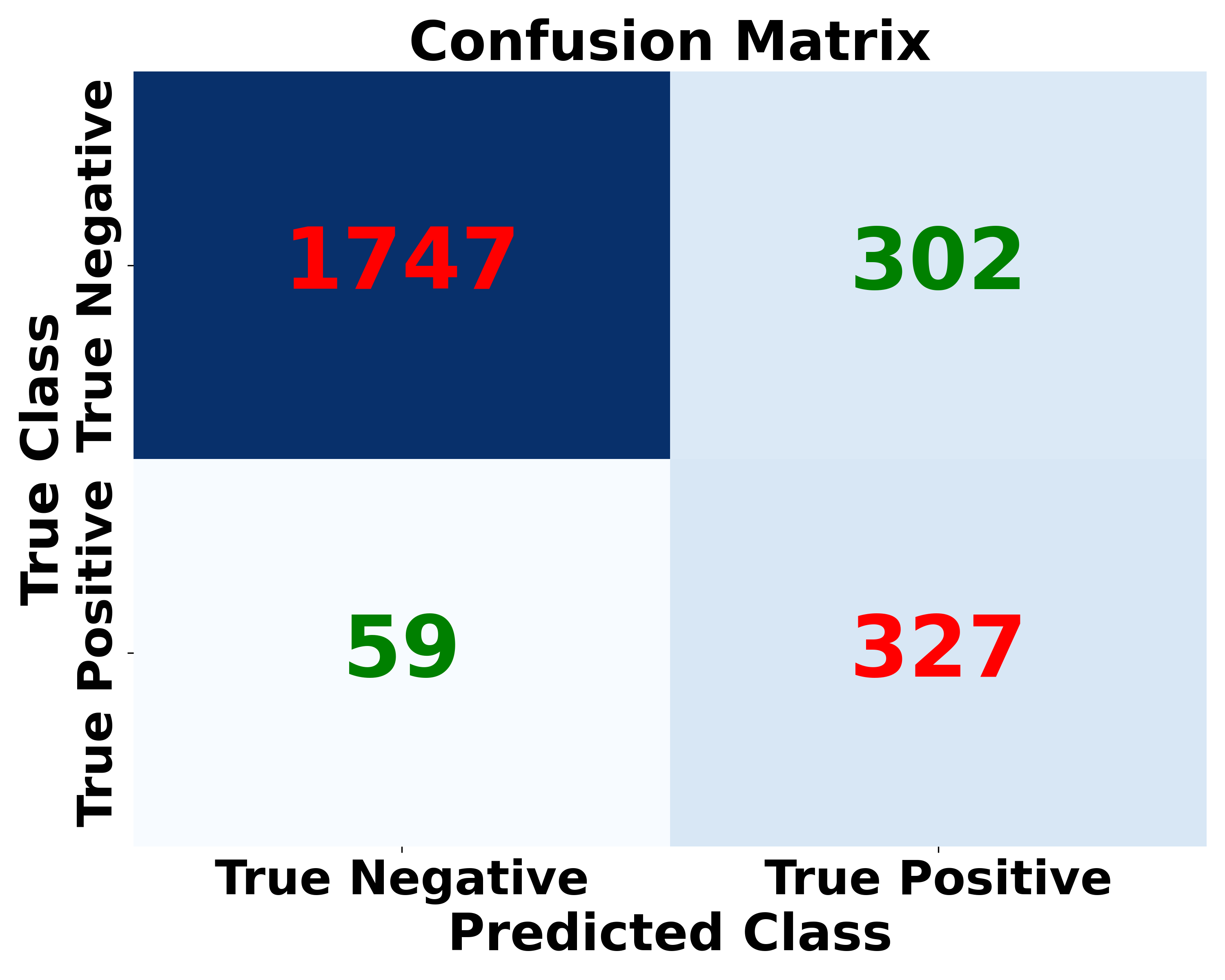} \\
    \bottomrule

  \end{tabular}
\end{table}

\end{document}